\documentclass[fleqn,usenatbib,usedcolumn]{mnras}
\usepackage[british]{babel}             

\usepackage[T1]{fontenc}
\usepackage{ae,aecompl}

\usepackage{hyperref}	
\usepackage{todonotes}  

\usepackage{graphicx}	
\usepackage{amsmath}	

\usepackage{amssymb}	
\usepackage{amsfonts}	
\usepackage{times}	

\usepackage{epsfig}
\usepackage{multirow}
\usepackage{widetext}
\usepackage{color}
\usepackage{pdflscape}
\usepackage{booktabs}
\usepackage{bm}
\usepackage{cancel}	
\usepackage{pgf,tikz}
\usepackage{mathrsfs}
\usetikzlibrary{decorations.fractals}
\usetikzlibrary{decorations.pathmorphing,backgrounds}
\usetikzlibrary{decorations.shapes}
\usetikzlibrary{decorations.footprints}
\usetikzlibrary{shapes,arrows,positioning,calc}
\tikzset{paint/.style={ draw=#1!50!black, fill=#1!50 },
    decorate with/.style=
    {decorate,decoration={shape backgrounds,shape=#1,shape size=2mm}}}
\usetikzlibrary{arrows}
 
\numberwithin{equation}{section}

\hypersetup{pdfauthor={F. Fambri, M. Dumbser, S. K\"oppel, L. Rezzolla,
    O. Zanotti}, pdftitle={ADER discontinuous Galerkin schemes for 
    general-relativistic ideal magnetohydrodynamics},
  pdfkeywords={magnetohydrodynamics, general relativity, ADER
    discontinuous Galerkin, path-conservative schemes, a posteriori
    subcell limiter, adaptive mesh refinement, local time stepping },
  bookmarksnumbered=true}
							
\newcommand{\oo}{\hphantom{\mu}}

\newcommand{\natSet}{{\rm I\!N}}

\newcommand{\be}{\begin{equation}}
\newcommand{\ee}{\end{equation}}
\newcommand{\bdm}{\begin{displaymath}}
\newcommand{\edm}{\end{displaymath}}
\newcommand{\bea}{\begin{eqnarray}}
\newcommand{\eea}{\end{eqnarray}}

\newcommand{\halb}{\frac{1}{2}}

\newcolumntype{C}[1]{>{\centering\let\newline\\\arraybackslash\hspace{0pt}}p{#1}}

\newcommand{\gammap}{\Gamma}

\newcommand{\y}{\boldsymbol{y}}
\newcommand{\x}{\boldsymbol{x}}

\renewcommand{\u}{\boldsymbol{u}}
\newcommand{\w}{\boldsymbol{w}}
\newcommand{\q}{\boldsymbol{q}}
\newcommand{\F}{\boldsymbol{F}}

\renewcommand{\v}{\boldsymbol{v}}

\newcommand{\lapse}{\alpha}
\newcommand{\shift}{\beta}
\newcommand{\Lor}{W}

\newcommand{\aposteriori}{\textit{a posteriori} }

\newfont{\numerikEleven}{ecrm1000}
\newfont{\numerikTen}{cmss10}
\newfont{\numerikNine}{cmss9}
\newfont{\numerikEight}{cmss8}
\newfont{\numerikSeven}{cmss7}

\definecolor{pergamena}{rgb}{2.44,2.15,0.67}
\definecolor{azz}{rgb}{0.85,0.90,1.00}
\definecolor{mandarino}{RGB}{237,189,101}
\definecolor{f1}{RGB}{255,204,153}
\definecolor{pera}{RGB}{237,210,61}
\definecolor{nonh}{RGB}{237,229,156}
\definecolor{menta}{RGB}{171,242,193}
\definecolor{darkspringgreen}{rgb}{0.09, 0.45, 0.27}
\definecolor{UniBlue}{RGB}{83,121,170}
\definecolor{darkgreen}{rgb}{0.0, 0.2, 0.13}

\definecolor{qqqqff}{rgb}{0.,0.,1.}
\definecolor{ffqqqq}{rgb}{1.,0.,0.}

\newcommand{\ie}{i.e.,~}
\newcommand{\eg}{e.g.,~}

\title[High-order ADER-DG schemes for GRMHD]{ADER discontinuous Galerkin
  schemes for general-relativistic ideal magnetohydrodynamics}

\author[F. Fambri, M. Dumbser, S. K\"oppel, L. Rezzolla, O. Zanotti] 
{F. Fambri$^{1}$\thanks{E-mail: francesco.fambri@unitn.it},
  M. Dumbser$^{1}$, S. K\"oppel$^{2,3}$, L. Rezzolla$^{2,3}$ ,
  O. Zanotti$^{1}$ \\
  $^{1}$Laboratory of Applied Mathematics, University of
  Trento, Via Mesiano 77, I-38123 Trento, Italy\\
  $^{2}$Institute for Theoretical Physics, Max-von-Laue-Str. 1, 60438
  Frankfurt, Germany\\
  $^{3}$Frankfurt Institute for Advanced Studies, Ruth-Moufang-Str. 1,
  60438 Frankfurt, Germany }

\pubyear{2018}

\begin{document}
\label{firstpage}
\pagerange{\pageref{firstpage}--\pageref{lastpage}}
\maketitle

\begin{abstract}
We present a new class of high-order accurate numerical algorithms for 
solving the equations of general-relativistic ideal magnetohydrodynamics
in curved spacetimes. In this paper we assume the background spacetime to
be given and static, \ie we make use of the Cowling approximation.  The
governing partial differential equations are solved via a new family of
fully-discrete and arbitrary high-order accurate path-conservative
discontinuous Galerkin (DG) finite-element methods combined with adaptive
mesh refinement and time accurate local timestepping. In order to deal
with shock waves and other discontinuities, the high-order DG schemes are
supplemented with a novel \textit{a-posteriori} subcell finite-volume
limiter, which makes the new algorithms as robust as classical
second-order total-variation diminishing finite-volume methods at shocks
and discontinuities, but also as accurate as unlimited high-order DG
schemes in smooth regions of the flow. We show the advantages of this new
approach by means of various classical two- and three-dimensional
benchmark problems on fixed spacetimes. Finally, we present a performance
and accuracy comparisons between Runge-Kutta DG schemes and ADER
high-order finite-volume schemes, showing the higher efficiency of DG
schemes.
\end{abstract}



\section{Introduction}
\label{Introduction}

Electromagnetism plays an important role in many astrophysical processes
such as compact objects and binaries consisting of black holes and
neutron stars. The general-relativistic theory of magnetohydrodynamics
(GRMHD) is a successful theory to describe these systems, combining the
fluid description of matter with a simplified theory for electromagnetic
fields in the absence of free charge carriers. Similar to
general-relativistic hydrodynamics (GRHD), first successful
(lower-dimensional) simulations of the GRMHD system date back to the
pioneering work of \citet{wilson75} more than 40 years ago
\citep[See][for recent reviews on progress in GRMHD
  simulations]{Font08,Marti2015}. In the past years, several groups
started to recast the system of GRMHD equations into a conservative form
to make use of conservative Godunov-type finite-volume schemes based on
approximate Riemann solvers and high-resolution shock-capturing schemes
(HRSC). Many GRHD and GRMHD codes have been developed over the last
decade \citep[for instance][]{Baiotti04, Duez05MHD0, Anninos05c, Anton06,
  Giacomazzo:2007ti, Anderson2008, Kiuchi2009, Bucciantini2011,
  Radice2012a, Dionysopoulou:2012pp, Radice2013b, Athena2016, Porth2017}
and applied to various topics in astrophysics. Some codes also evolve the
spacetime by feeding back the fluid and magnetic energy-momentum tensor
in the Einstein field equations, which govern the time evolution of the
metric tensor; some codes even incorporate radiation transfer like the
one proposed by \cite{Takahashi2017}, or include the full Maxwell theory
in a resistive relativistic MHD formulation
\citep[see~\eg][]{Palenzuela:2008sf, DumbserZanotti,
  Dionysopoulou:2012pp, Bucciantini2013, Bugli2014, AloyCordero}.

In this work we propose a new family of little dissipative and little
dispersive shock capturing schemes for the solution of the
general-relativistic magnetohydrodynamics equations (GRMHD), based on
high-order accurate \emph{explicit} discontinuous Galerkin (DG)
finite-element schemes on spacetime adaptive meshes (AMR) with
time-accurate local time stepping (LTS) and supplemented by a high-order
accurate \emph{a posteriori} subcell finite-volume limiter in order to
cope with shocks and discontinuities in the solution. To the best of our
knowledge, this family of algorithms has never been applied to the
GRMHD equations before.

An important and novel aspect of our approach is the interpretation of
the source terms in the GRMHD equations that account for the
gravitational field in curved spacetimes as \textit{nonconservative
  products}, instead of the usually employed algebraic source terms,
since the gravity terms in general relativity are indeed functions of the
spatial derivatives of the lapse, the shift vector and the spatial metric
tensor. In other words, given a vector of {conserved} variables
$\boldsymbol{Q}$ and the tensor of nonlinear conservative fluxes
$\boldsymbol{F} = (F^1, F^2, F^3)$, the set of GRMHD equations which are
normally written as \citep{Rezzolla_book:2013}
\begin{align} 
  \label{eq:PDE_orig}
  &\partial_t{\boldsymbol{Q}} + \nabla \cdot
        {\boldsymbol{F}}(\boldsymbol{Q}) = \boldsymbol{\mathcal{S}}(\boldsymbol{Q})
        \,, 
\end{align}
where $\boldsymbol{\mathcal{S}}$ is a generic source vector 
can, in our framework, be rewritten as
\begin{align} 
&\partial_t{\boldsymbol{Q}} + \nabla \cdot
  {\boldsymbol{F}}(\boldsymbol{Q}) +
  \boldsymbol{\mathcal{B}}(\boldsymbol{Q})\cdot \nabla \boldsymbol{Q}
  =0\,, \label{eq:PDE}
\end{align}
or in quasi-linear form,
\begin{align} \label{eq:quasi-linear-PDE}
\partial_t \boldsymbol{Q} + \boldsymbol{\mathcal{A}}(\boldsymbol{Q})
\cdot \nabla \boldsymbol{Q} = 0\,,
\end{align}
with the system matrix $\boldsymbol{\mathcal{A}}(\boldsymbol{Q}) :=
\partial \boldsymbol{F} / \partial \boldsymbol{Q} +
\boldsymbol{\mathcal{B}}(\boldsymbol{Q})$. Above, and throughout the
paper, the nabla operator {without subscript} is simply defined as
$\nabla = (\partial_x, \partial_y, \partial_z)$, and thus does not
indicate a covariant derivative. Here, $\boldsymbol{\mathcal{B}} =
(\mathcal{\boldsymbol{B}}_1, \mathcal{\boldsymbol{B}}_2,
\mathcal{\boldsymbol{B}}_3)$ is the matrix of the nonconservative product
$\boldsymbol{\mathcal{B}}(\boldsymbol{Q})\cdot \nabla \boldsymbol{Q} :=
\mathcal{\boldsymbol{B}}_1 \partial_x \boldsymbol{Q} +
\mathcal{\boldsymbol{B}}_2 \partial_y \boldsymbol{Q} +
\mathcal{\boldsymbol{B}}_3 \partial_z \boldsymbol{Q}$. The system
\eqref{eq:quasi-linear-PDE} is called hyperbolic if the matrix $\mathcal
A \cdot \boldsymbol{n}$ is diagonalizable for all normal vectors
$\boldsymbol{n} \neq 0$ with only real eigenvalues and a complete set of
bounded linearly independent eigenvectors. The hyperbolicity of the GRMHD
system has been studied in many works [see, for instance,
  \citet{Anile_book,Komissarov1999}].

This paper deals with the general-relativistic extension of the special
relativistic case presented in \citet{Zanotti2015d}. As it has been
mentioned above, the background spacetime is introduced as a
\emph{nonconservative product} in the principal part of the system on the
left hand side and is {not} treated as an algebraic source term, as it
has been conventionally treated all along in the literature so far. The
inspiration to use so-called {path-conservative} schemes for
nonconservative products has been taken from successful developments in
the context of so-called {well-balanced} numerical methods
\cite{Bermudez1994} for the solution of the shallow-water equations [see
  \cite{pares2006,Castro2006,PCS2010} for details on path-conservative
  schemes], where the bottom-slope term (which is the gradient of a known
function and accounts for gravity forces in shallow-water models) is 
discretized as a nonconservative product in the principal part of the
system rather than as a classical algebraic source term. In the shallow
water context, the family of path-conservative schemes allows to preserve
certain stationary equilibrium solutions {exactly} up to machine
precision also on the discrete level, including nontrivial equilibria,  
see \cite{GaburroDumbserSWE} and \cite{GaburroDumbserEuler}
for recent examples. At this 
stage, the development of exactly well-balanced numerical methods for the
GRMHD equations is still out of scope, but further developments in this
direction would definitely deserve attention. We also would like to stress
that the use of nonconservative products is \textit{not} related to the 
ADER-DG scheme itself. It would have been equally possible to compute the
metric derivatives analytically and discretize the gravity terms as
conventional algebraic source terms, as done in other codes for the 
GRMHD system. 

Discontinuous Galerkin methods belong to the family of finite-element
methods which consider the numerical approximation of a weak formulation
of the governing system of partial differential equations over a set of
non-overlapping elements. The discrete solution space is restricted to
the space of piecewise polynomials of maximum degree $N \geq 0$ and the
degrees of freedom (\ie the expansion coefficients) of the chosen
polynomial basis are directly evolved in time. Finite-element methods are
known also under the name of {variational-difference} or
{projection-difference}, \citep[see the original formulations
  by][]{Ritz1909,Galerkin1915}. In particular, in the DG formulation the
numerical solution is allowed to be {discontinuous} at element interfaces
[see \cite{reedhill} for the integration of the neutron transport
  equation]. It has taken nearly two decades for the DG methods to be
extended to general nonlinear hyperbolic systems, thanks to the
groundbreaking works of \cite{Cockburn1989b,Cockburn1990,CockburnShu98};
see also \cite{cockburn_2000_dg,cockburn_2001_rkd,ShuDGWENOReview} for a
review.

In the last twenty years, DG methods became increasingly popular mainly
because of four attractive properties: i) nonlinear $L_2$ stability has
been proven for general nonlinear scalar conservation laws by
\cite{jiangshu}; ii) arbitrary high order of accuracy can be easily
achieved for smooth solutions by simply increasing the polynomial degree
$N$ of the chosen basis functions; iii) high parallel scalability makes
DG methods better suited for large-scale simulations even on general
unstructured meshes when compared with high-order finite-difference or
finite-volume methods; iv) high-order DG methods are only little
dissipative and little dispersive, even when compared with high-order
finite-volume and finite-difference schemes and are thus essential for
accurate long-term simulations. The main drawback that afflicts
{explicit} DG schemes is the rather severe CFL stability condition that
constrains the timestep of the simulations to scale with approximately
$h/(2N+1)$ for hyperbolic partial differential equations, where 
$N$ is the degree  of the nodal polynomial basis used within the element 
and $h$ is the characteristic size of one DG element (not the distance 
between the individual nodal degrees of freedom). 
A way to alleviate the severe CFL timestep restriction is the use
of efficient semi-implicit DG schemes, as those proposed, for instance by
\cite{TavelliDumbser2016,FambriDumbser}.

DG methods have attracted the interest of the computational-astrophysics
community only over the last few years. In particular, the first
DG-based method for general-relativistic hydrodynamics has been developed
by \cite{Radice2011}, but it was limited to spherically symmetric
spacetimes. The first three dimensional implementation of a DG method for
relativistic flows on curved but fixed background spacetimes has been
recently presented by \cite{Bugner2016}, but without considering the
magnetic field interaction. Very recently, \cite{Miller2017} formulated
an operator-based DG method for the solution of the Einstein field
equations, while in \cite{ADERCCZ4} a high-order DG scheme for the
solution of a first-order reduction of the conformal and covariant
formulation (CCZ4) \cite{Alic:2011a} of the Z4 system of the Einstein
equations has been proposed. Also rather recently, \cite{Kidder2017}
provided a DG implementation within a task-based parallelism model for
GRMHD, while \cite{Anninos2017} presented also a DG code with
hp-refinement, and both of them complemented the high resolution of the
purely-spatial polynomials basis with multi-step high-order time
integrator, \eg Adams-Bashforth (AB3) or Runge-Kutta schemes. Indeed,
total-variation diminishing (TVD) Runge-Kutta methods are typically used
in order to reach a stable high-order time discretization of DG schemes,
\ie applying the method of lines (MOL) technique which leads to the
so-called family of RK-DG schemes.

On the other hand, the time discretization proposed in this paper is
different and is named ADER technique. The particular feature of the ADER
approach introduced by Toro and Titarev in the finite-volume context
\cite{Titarev2002, Titarev2005, Toro:2006a} is that it leads to
     {arbitrary} high-order accurate {fully-discrete} one-step schemes in
     space and time. ADER schemes have already been applied to the
     equations of relativistic MHD, both in the ideal case (see
     \citeauthor{Dumbser2008} \citeyear{Dumbser2008},
     \citeauthor{Zanotti2015} \citeyear{Zanotti2015},
     \citeauthor{Zanotti2015d} \citeyear{Zanotti2015d}b) and in the
     resistive case \citep[see][]{DumbserZanotti} and to other nonlinear
     systems of partial differential equations
     \citep[see][]{Zanotti2015c,ADERDGVisc}. Moreover, the ADER strategy
     adopted in this paper, which goes back to \cite{DumbserEnauxToro},
     applies to general systems of balance laws with conservative fluxes,
     nonconservative products and stiff or non-stiff algebraic source
     terms. In particular, it is based on a local spacetime discontinuous
     Galerkin (LSTDG) predictor step, which solves a local Cauchy problem
     in the small, based on a weak formulation of the partial
     differential equations in spacetime.

Although DG methods are proven to be nonlinearly $L_2$ stable, whenever
steep gradients or discontinuities appear in the solution, the use of an
unlimited high-order DG scheme inevitably leads to spurious oscillations
known as {Gibbs phenomenon}. In order to cope with this problem, several
attempts have been made, \eg artificial viscosity
\citep{Hartman_02,Persson_06,Feistauer4}, filtering \citep{Radice2011},
hybridisation with finite-volume/finite-difference schemes for the
selected {``troubled cells''} adopting some sort of high-order
slope-limiting procedures \citep{cbs4, QiuShu1, Qiu_2004, balsara2007,
  Zhu_2008, Zhu_13, Luo_2007, Krivodonova_2007}. Here, we employ to the
so-called {a-posteriori} finite volume {subcell limiter} (SCL) technique
recently proposed by \cite{Dumbser2014}, which is based on the so-called
multi-dimensional optimal order detection (MOOD) of \cite{CDL1} and
\cite{CDL2}. The main advantage of this approach is that the
high-resolution properties of unlimited DG methods are preserved thanks
to the introduction of a {subgrid level}, which is used {a-posteriori}
for integrating the partial differential equations in troubled cells by
means of a more robust high-order accurate finite-volume scheme [for
  completeness see also the work of \citep{CasoniHuerta1, Sonntag,
    SonntagJSC, Fechter1, MeisterOrtleb} for alternative subcell DG
  limiters]. The presented SCL has been applied to several systems of
nonlinear partial differential equations with promising results in the
work of \citeauthor{Zanotti2015c} (\citeyear{Zanotti2015c} and
\citeyear{Zanotti2015d}b) and \citet{ADERDGVisc}.

The paper is organised as follows: In Sec.
\ref{sec:Mathematical_formulation} we describe the system of governing
partial differential equations, in Sec. \ref{sec:Numerical_method} we
describe the ADER-DG scheme with the finite-volume subcell limiter and
the adaptive mesh refinement (AMR) technique. Section
\ref{sec:numerical-validation} presents the testbeds both in special and
general relativity that the scheme has passed. 
In Section \ref{sec:performance} we present strong MPI scaling results up
to 16,000 MPI ranks and performance and accuracy comparisons between
Runge-Kutta DG schemes and ADER-WENO finite-volume schemes. Finally,
Sec. \ref{sec:Conclusions} contains a summary of the results and an
outlook to future work.
Finally, Sec. \ref{sec:Conclusions} contains a summary of the results and 
an outlook to future work.

Hereafter, {Latin} indexes run from $1$ to $3$, while {Greek} indices run
from $0$ to $3$. The zeroth components refer to the {timelike} coordinate
of the corresponding tensor or vector and the {signature} of the metric
tensor is assumed to be $(-,+,+,+)$ through all the text. We use the
Einstein summation convention over repeated indexes. Wherever not
specified, the index correspondence $(V^0,V^1,V^2,V^3) =
(V^t,V^x,V^y,V^z)$ is adopted. Moreover, bold symbols are used to
indicate three-vectors (or tensors). We use units with speed of light
$c=1$ and gravitational constant $G=1$.

\section{Mathematical formulation and physical assumptions}
\label{sec:Mathematical_formulation}

The governing equations of an (ideal) fluid coupled to an
electromagnetic field and described in a curved spacetime are given by
the general-relativistic magneto-hydrodynamics equations (GRMHD). 
Following the derivation and formalism developed by \cite{DelZanna2007},
the covariant Euler-Maxwell system reads
\begin{align}
  \nabla_\mu (\rho u^\mu) = 0\,, \hspace{1cm}
  \nabla_\mu T^{\mu\nu} = 0\,,    \hspace{1cm}
  \nabla_\mu ~{}^*F^{\mu\nu} = 0\,,
\end{align}
and contains the conservation of the energy momentum tensor $T^{\mu\nu}$,
as well as the homogeneous Faraday law, with $\nabla_\mu$ being the
covariant derivative operator.

Since in most astrophysical phenomena the
electrical conductivity of the plasma is very high, the ideal-MHD
approximation (where the electrical conductivity is actually assumed to
be divergent) is a reasonable one. In this case, the electrical field
is completely determined by the fluid velocity and the magnetic field,
\begin{align}
  \label{eq:EinIMHD}
  \boldsymbol{E} = - \boldsymbol{v}\times \boldsymbol{B},
\end{align}
that is, the magnetic flux $\phi_B = \boldsymbol{B} \cdot \boldsymbol{S}$
over any surface $S$ is conserved
\begin{align}
\oint_{\partial \boldsymbol{S}} \left( \boldsymbol{E} +\boldsymbol{v}
\times \boldsymbol{B}\right) \cdot d\boldsymbol{\ell} = - \frac{d
  \phi_B}{d t} = 0\,,
\end{align}
and is advected with the fluid movement. The magnetic contribution to the
hydrodynamics equations, \ie the MHD equations, is then just a
conservation equation for the magnetic field, which we will describe in
the next Sections.

\subsection{The 3+1 split of spacetime}

The $3+1$ decomposition of spacetime is the most widely used framework to
prepare general-relativistic theories such as the GRMHD for numerical
discretization. The four-dimensional spacetime manifold is decomposed
into three-dimensional (3D) {spacelike} hypersurfaces which are
parametrized by a time coordinate $t$ and described by the 3D objects
\citep{Thorne82, Baumgarte2003, Rezzolla_book:2013}: the lapse function
$\lapse$, the spatial metric tensor $\boldsymbol{\gamma}$, the shift
vector $\boldsymbol{\beta}$, and the extrinsic curvature tensor
$\boldsymbol{K}$.

The (smooth) foliation or slicing $\Sigma_t$ defines a timelike normal
vector to the 3-hypersurface
\begin{align}
n_{\mu} = -\lapse \nabla_{\mu} t,
\quad
n_{\mu} = (-\alpha, 0_i) \,,
\quad
n^{\mu} = (  1 /\lapse , - \shift^i / \lapse ) \,,
\end{align}
which is the future-oriented {unit} ($n_\mu n^\mu = -1$) vector and can
be regarded as the four-velocity of the {Eulerian} observer, \ie {at
  rest} in the 3D hypersurface $\Sigma_t$.

Any four vector $V^{\mu}$ (or tensor) can be split into its temporal and
spatial components, respectively 
\begin{align}
  -\boldsymbol{n}\cdot \boldsymbol{V} &= 
  - n_{\mu} V^\mu\,,\\
   \boldsymbol{\gamma}\cdot \boldsymbol{V} &= 
  \left(g^{\mu}_{\nu} + n^\mu n_{\nu}\right) V^\nu\,,
\end{align}
where the relation between the purely spatial (3D) metric tensor
$\gamma_{ij}$ and the spatial-projection operator $\boldsymbol{\gamma}$
is given by
\begin{align}
&\gamma_{\mu\nu}:= 
g_{\mu\nu} + n_\mu n_{\nu}\,, \;\;
&\gamma^{\mu}_{\nu} := g^{\mu}_{\nu} + n^\mu n_{\nu} \,,
\end{align}
with the obvious property that $\boldsymbol{\gamma}\cdot \boldsymbol{n} =
0$. In this formalism, the spatial metric $\gamma_{ij}$ is used for
lowering/raising indexes of purely {spatial} vectors (or tensors).

Given a coordinate system $x^{\mu} = (t,x^i)$, where $\{x^i\}_{i=1,2,3}$
(or $\boldsymbol{x}$) are the spatial coordinates, the line element on
the foliation $\Sigma_t$ can then be expressed by the $3+1$ form of the metric,
\ie
\begin{align}
d s^2 = - \lapse^2 d t^2+\gamma_{ij} \left(d x^i + \shift^i d
t\right)\left(d x^j + \shift^j d t\right) \,.
\end{align}

\subsection{The GRMHD system}
\label{sec:derivation-grmhd}

In order to write the system of the GRMHD equations in the $3+1$
decomposition of the spacetime, we define the vector\footnote{Obviously,
  the state vector is not a physical vector but just an ordered
  collection of physical degrees of freedom (scalar, vector and tensor
  fields). One should refer to this object as a {tuple} instead.}
$\boldsymbol{V}$ of the 19~{primitive variables} as
\begin{align}
\boldsymbol{V} := \left( \rho, v_j, p, B^j, \Phi, \lapse, \shift^j,
\tilde{\gamma}_m \right)\,,\;\;\; j=1,2,3; \;\;m=1,\ldots,6\,,
\end{align}
where $\rho$ is the rest-mass density in the frame comoving with the
fluid, $\boldsymbol{v}$ the three-velocity vector, $p$ the fluid
pressure, $\boldsymbol{B}$ the magnetic field vector in the comoving
frame, $\Phi$ an artificial scalar introduced to ensure the
divergence-free constraint of the magnetic field at the discrete level
via the hyperbolic divergence-cleaning approach \citep{Dedneretal},
$\lapse$ the {lapse} function, $\bm{\shift}$, the {shift} vector, and
$\boldsymbol{\gamma}$ a vector whose components represent the six
{independent} components of the three (spatial) metric $\bm{\gamma}$, \ie
\begin{align}
\tilde{\bm{\gamma}} = \left( \gamma_{11}, \gamma_{12}, \gamma_{13},
\gamma_{22}, \gamma_{23}, \gamma_{33}\right)
\end{align}
The corresponding state vector $\boldsymbol{Q}$ of {conserved}
variables is defined as
\begin{align}
\boldsymbol{Q} := \left( \sqrt{\gamma} D, \sqrt{\gamma} S_j, \sqrt{\gamma}
\tau, \sqrt{\gamma} B^j,  \Phi, \lapse, \shift^j,
\tilde{\gamma}_{m} \right)\,.
\end{align}
Note that while $\rho$ can be seen as the rest-mass density of the fluid
evaluated by the {Lagrangian} comoving observer with four-velocity
$u^\mu$, $D$ and $v^{\mu}$ are the rest-mass density and the velocity as
measured by the {Eulerian} observer. As such, $v^{\mu}$ is a purely
spatial vector ($n_\mu v^\mu = 0$) and its norm is the one appearing in
the definition of the Lorentz factor $\Lor$. Finally, the symbol
$\gamma$ denotes instead the determinant of $\boldsymbol{\gamma}$, \ie
$\gamma = \text{det}(\gamma_{ij})$.

Also associated to the {Eulerian} frame is the (Eulerian) three-momentum
density vector $S_j$, which is related to the Lagrangian velocity $u^\mu$
through the following identities
\begin{align*}
  & u^\mu = \Lor \left( n^\mu + v^\mu\right)\,, \quad  u_\mu u^\mu=-1;\\
  & \Lor := - n_\mu u^\mu = \lapse u^t = (1-v_iv^i)^{-1/2} = (1-v^2)^{-1/2} \,,\\
  &  \boldsymbol{\gamma}\cdot  \boldsymbol{u}
  = ( g^{\mu}_{\nu} + n^\mu n_{\nu} ) u^\nu =  \Lor v^\mu\,,
  \quad
  v^i  =  u^i/\Lor + \shift^i/\lapse\,.
\end{align*}

The conserved variables $\boldsymbol{Q}(\boldsymbol{V})$ can be easily
expressed in terms of the primitive variables via
\begin{align}
  & D := \rho \Lor\,,  \\
  & \boldsymbol{S} := \rho h \Lor^2 \boldsymbol{v} +
  \boldsymbol{E}\times\boldsymbol{B}\,,\label{eq:S}\\
  & U := \rho h \Lor^2 - p + \frac{1}{2}\left(E^2 + B^2\right)\,,  \\
  & \tau := U - D\,.
\end{align}
Here, $U$ is the conserved energy density and $\tau$ the corresponding
quantity without the rest-mass energy energy density,
$h=1+\epsilon+p/\rho$ is the specific enthalpy and $\epsilon$ is the
specific internal energy \citep{Rezzolla_book:2013}.

The electric field in the Eulerian frame is indicated as $\boldsymbol{E}$
and in the ideal-MHD limit (\ie for diverging electrical conductivities)
it is determined by the simple Ohm law \eqref{eq:EinIMHD}, \ie $E_i = -
\tilde{\epsilon}_{ijk} v^j B^k$.
The cross product is given by the spatial three-Levi-Civita tensor
density $\tilde{\epsilon}$
\begin{align*}
  &
  \tilde{\epsilon}^{ijk} = \gamma^{-\frac{1}{2}} [ijk]\,, \hspace{0.5cm}
  \tilde{\epsilon}_{ijk} = \gamma^{\frac{1}{2}} [ijk]\,. \\
  &[ijk] = \left\{ \begin{array}{rl} 1 & \text{\footnotesize for even
      permutations of $(1,2,3)$}, \\
    -1 & \text{\footnotesize for odd permutations}, 
    \\
    0 &  \text{\footnotesize otherwise.}
 \end{array} \right.
\end{align*}
The (covariant) Pointing vector $\boldsymbol{E}\times\boldsymbol{B}$ in
the momentum density (\ref{eq:S}) can be written as
\begin{align*}
  \boldsymbol{E}\times\boldsymbol{B} & = \{ \tilde{\epsilon}_{ijk} E^j
  B^k\} = \{ -\tilde{\epsilon}_{ijk} \tilde{\epsilon}^{jmn} v_m B_n B^k \}
  \\ &= \{ v_i \left(B_k B^k \right) - B_i \left( v_k B^k\right) \}
  = \boldsymbol{v} B^2 - \boldsymbol{B} \left(\boldsymbol{v}\cdot \boldsymbol{B}\right) .
\end{align*}

Given all these definitions, the system of partial differential equations
for ideal GRMHD can be written in the very compact {nonconservative}
homogeneous form \eqref{eq:PDE}, where the conservative fluxes
$\boldsymbol{F}$ and the nonconservative product
$\boldsymbol{\mathcal{B}} (\boldsymbol{Q})\cdot \nabla \boldsymbol{Q}$
are given by
\begin{equation}
\boldsymbol{F} := \gamma^{\frac{1}{2}}\left( \begin{array}{c}
 \lapse v^i D - \shift^i D \\
 \lapse T^i_j - \shift^i S_j\\ 
\lapse \left( S^i - v^i D\right) - \shift^i \tau \\ 
 \left( \lapse v^i - \shift^i \right) B^j    - \left( \lapse v^j - \shift^j \right) B^i  \\ 
0 \\
   0 \\ 0\\ 0
  \end{array}\right)
\,,\label{eq:fluxes}  
\end{equation}
\begin{equation} 
\boldsymbol{\mathcal{B}}(\boldsymbol{Q})\cdot \nabla \boldsymbol{Q} :=
\left( \begin{array}{c} 
0 \\ 
\gamma^{\frac{1}{2}}\left(U\partial_j\lapse  -\frac{1}{2}\lapse T^{ik}\partial_j \gamma_{ik} - S_i \partial_j \shift^i\right) \\ 
\gamma^{\frac{1}{2}}\left( S^j\partial_j \lapse - \frac{1}{2} T^{ik} \shift^j \partial_j \gamma_{ik} - T_i^j \partial_j \shift^i \right) \\ 
-\shift^j \partial_i \left(\gamma^{\frac{1}{2}} B^i\right) + \lapse  \gamma^{\frac{1}{2}} \gamma^{ji} \partial_i \Phi   \\ 
\gamma^{-\frac{1}{2}}\lapse c_h^2 \partial_j \left(\gamma^{\frac{1}{2}} B^j\right) - \shift^j \partial_j \Phi  
\\ 0 \\ 0  \\ 0\end{array}\right)\,,
 \label{eq:NCP} 
\end{equation}
and where $T^{ij}$ denotes the spatial stress-energy tensor 
\begin{align}\label{eq:EMtensor}
T^{ij} &:= \rho h \Lor^2 {v^i v^j} - {E^i E^j} - {B^i B^j} + \left[ p +
  \frac{1}{2} \left(E^2 + B^2\right) \right] \gamma^{ij} \nonumber \\ &=
{S^i v^j} + p_\text{tot} \gamma^{ij} - \frac{{B^i B^j}}{\Lor^2} - ({B}_k
{v}^k) {v^i B^j}\,,
\end{align}
with the total pressure comprising both the fluid and the magnetic
pressure, \ie
\begin{equation}
p_\text{tot}=p + p_\text{mag} = p + \frac{1}{2} \left[ B^2 / \Lor^2 + (B\cdot v
  )^2 \right] \,.
\end{equation}

Since we are here interested in static spacetimes (Cowling approximation), 
the system of equations does not contain explicitly the extrinsic-curvature 
tensor $\bm K$, which can be expressed simply in terms of metric functions 
\citep{MTW1973, York79, Gourgoulhon2012}
\begin{align}
\lapse T^{ij} K_{ij} := \frac{1}{2} T^{ik} \shift^j \partial_j
\gamma_{ik} + S^j_{\oo i} \partial_j\shift^i\,.
\end{align} 

As mentioned above, the divergence-free constraint of the magnetic field
is here taken into account at the discrete level through the so-called
{hyperbolic} Generalized Lagrangian Multiplier approach (GLM, and also
known as "divergence-cleaning") proposed by \citet{Dedneretal}, \ie by
{augmenting} the GRMHD system with an additional {auxiliary} equation for
an {artificial} scalar field $\Phi$, in order to propagate away numerical
errors in the divergence-free constraint of the magnetic field
\begin{equation}
 \partial_i \left( \sqrt{\gamma} B^i \right) =0\,.
\end{equation}  

In order to achieve a more efficient divergence cleaning, we also  
allow the characteristic velocity of the divergence cleaning $c_h$ to be 
larger than the speed of light, \ie $c_h \geq 1$ in (\ref{eq:fluxes}). 
Typical values for the cleaning speeds are chosen in the range 
$c_h \in [1,2]$.

\subsection{Equation of state, primitive recovery, characteristic speeds}

For the closure of the GRMHD equations, an equation of state
$p=p(\rho,\epsilon)$ has to be chosen. With the aim of simplicity, we
here consider the ideal-fluid (or ``Gamma-law'') equation of state
(Sec.~\ref{sec:numerical-validation})
\begin{align}
p=\rho\epsilon(\Gamma-1)
\end{align}
where $\Gamma$ is the polytropic index. In the same spirit, for the
recovery of primitive variables $\boldsymbol{V}(\boldsymbol{Q})$, we
employ a standard approach corresponding to the third option reported in
Sect. 3.2 of \cite{DelZanna2007}; possible alternatives for performing
the inversion of system (\ref{eq:S}) are discussed by \cite{NGMD2006}.

For the characteristic wave speeds in GRMHD we usually employ the
standard {magnetosonic approximation} for the wave speeds \citep[as
  in][]{Gammie03}, but accounting also for the possibility $c_h > 1$.
Note that this choice of eigenvalues corresponds to the standard choice
when $c_h=1$ and is also valid in the general relativistic hydrodynamics
limit. 

\section{Numerical methods}
\label{sec:Numerical_method}

\subsection{ADER discontinuous Galerkin schemes}

As mentioned in the Introduction, the numerical scheme that we adopt is
the ADER discontinuous Galerkin (DG) scheme supplemented with an
\emph{a-posteriori} finite-volume subcell limiter approach with AMR,
presented in the series of papers \citeauthor{Zanotti2015c}
(\citeyear{Zanotti2015c} and \citeyear{Zanotti2015d}b) and
\citet{ADERDGVisc} in the context of the Euler equations of compressible
gas dynamics, ideal MHD, special relativistic RMHD, but also compressible
Navier-Stokes and viscous and resistive MHD equations. A brief overview
of the numerics is given in the following.
 
After choosing a mesh partition $\Omega_h = \{ \Omega_i \}$, which is
Cartesian and {spacetime} adaptive through a \emph{cell-by-cell}
approach \citep[see][]{Khokhlov98}, with the property
\begin{align}
\Omega = \bigcup \limits_{i=1,\ldots N_E} \Omega_i, \hspace{1cm} \bigcup
\limits_{i\neq j; \;\;i,j=1,\ldots N_E} \Omega^{\circ}_i \cap
\Omega^{\circ}_j = \varnothing \label{eq:mesh}
\end{align}
with $\Omega$ being the computational domain, $N_E$ the total number of
spatial elements and ``$\left.\right.^{\!\!\circ}$'' denoting the interior
operator. The weak formulation of the governing equations
(\ref{eq:PDE}) is then written in the form
\begin{align}
& \int \limits_{t^n}^{t^{n+1}} \int\limits_{\Omega_i } \phi_k \left(
  \partial_t \boldsymbol{Q}  + \nabla \cdot
  \boldsymbol{F}(\boldsymbol{Q})+ \boldsymbol{\mathcal{B}}(\boldsymbol{Q}) \cdot\nabla
  \boldsymbol{Q}\right) \,d\boldsymbol{x}\,dt = 0\,, \label{eq:weakPDE}
\end{align}
where $\phi_k\in \mathcal{U}_h^N$ is a generic basis element for the
vector space $\mathcal{U}_h^N$ of {piecewise} polynomials of maximum
degree $N\geq 0$ defined over $\Omega$ and which are allowed to be
{discontinuous} across the element interfaces $\partial \Omega_i$.
In this work, the set of basis and test functions $\{\phi_k\}$ has been
chosen as the set of Lagrange interpolation polynomials of maximum degree
$N$ over $\Omega_i$ with the property
\begin{align}
\phi_k(\x_{\text{GL},i}^m) = \left\{ \begin{array}{rl} 1 & \text{if}\;\;
  k=m; \\0 & \text{otherwise}; \end{array}\right.\hspace{0.7cm}
k,m=1,\ldots,(N+1)^{\text{d}}
\end{align}
with $\{\x_{\text{GL},i}^m\}$ being the set of the Gauss-Legendre (GL)
quadrature points in $\Omega_i \subset \mathbb{R}^d$ [see \cite{stroud}
for a detailed discussion of multidimensional quadrature]. For this
reason, the chosen polynomial basis is said to be a \emph{nodal} basis
with respect to the Gauss-Legendre quadrature points.

Since the chosen AMR grid is locally Cartesian, the spatial integrals of
Eq. (\ref{eq:ADER-DG}) can be evaluated in a
\emph{dimension-by-dimension} fashion in $x$, $y$ and $z$ direction, and
the corresponding nodal test and basis functions are defined after
rescaling the domain of integration $\Omega_i$ to the unit element
$[0,1]^d$. Therefore, we only need the tensor product of the GL
quadrature points in the unit interval $[0,1]$, denoted by
$\{\xi_{\text{GP}}^m\}_{m=1,\ldots,N+1}$ in the following. Note that the
total number of GL quadrature points $\{\x_{\text{GP}}^m\}$ in
$\Omega_i$, as well as the total number of basis elements $\{ \phi_k\}$,
is $(N+1)^d$.

After integration by parts of the flux-divergence term,
Eq.~(\ref{eq:weakPDE}), can be rewritten as
\begin{align}
&\int \limits_{t^n}^{t^{n+1}} \int\limits_{\Omega_i } \phi_k\, \partial_t
  \boldsymbol{Q} \,d\boldsymbol{x}\,dt + \int \limits_{t^n}^{t^{n+1}}
  \int\limits_{\partial \Omega_i } \phi_k \,
  \boldsymbol{F}(\boldsymbol{Q}) \cdot \boldsymbol{n} \, dS\,dt
  \nonumber\\
&- \int \limits_{t^n}^{t^{n+1}} \int\limits_{\Omega_i } \nabla \phi_k
  \cdot \boldsymbol{F}(\boldsymbol{Q}) \,d\boldsymbol{x}\,dt + \int
  \limits_{t^n}^{t^{n+1}} \int\limits_{\Omega_i } \phi_k \,
  \boldsymbol{\mathcal{B}}(\boldsymbol{Q}) \cdot \nabla \boldsymbol{Q}
  \,d\boldsymbol{x}\,dt= 0\,. \label{eq:DGPDE}
\end{align}

After restricting the space of the solutions to the set of piecewise
polynomials $\boldsymbol{u}_h (\boldsymbol{x}, t) \in \mathcal{U}_h^N$,
\ie
\begin{align}
\boldsymbol{u}_h(\boldsymbol{x},t^n) = \phi_k (\boldsymbol{x})\;
\hat{\boldsymbol{u}}^n_k, \;\;\;k=1,\ldots,(N+1)^d, \quad \boldsymbol{x} \in
\Omega_i, \nonumber
\end{align}
the following higher order accurate \emph{path-conservative} ADER-DG
scheme is obtained for the so-called {degrees of freedom} of
$\boldsymbol{u}_h$, or expansion coefficients, $\hat{\boldsymbol{u}}_k^n$:
\begin{align}
&\left( \, \int\limits_{\Omega_i} \phi_k \phi_l \, d\boldsymbol{x}\right)
  \left( \hat{\boldsymbol{u}}_l^{n+1} - \hat{\boldsymbol{u}}_l^{n} \, \right) +
  \nonumber \\ & \int \limits_{t^n}^{t^{n+1}} \!\! \int\limits_{\partial
    \Omega_i } \!\! \phi_k \mathcal{G}\left(\q_h^-, \q_h^+ \right) \cdot
  \boldsymbol{n} \, dS \, dt + \int \limits_{t^n}^{t^{n+1}} \!\!
  \int\limits_{\partial \Omega_i } \!\! \phi_k \mathcal{D}\left(
  \q_h^-,\q_h^+ \right) \cdot \boldsymbol{n} \, dS \, dt \nonumber \\ & -
  \int \limits_{t^n}^{t^{n+1}} \!\! \int\limits_{\Omega_i } \!\!\! \nabla
  \phi_k \cdot \boldsymbol{F}(\q_h) \,d\boldsymbol{x}\,dt + \int
  \limits_{t^n}^{t^{n+1}} \!\! \int\limits_{\Omega_i^{\circ} } \phi_k
  \boldsymbol{\mathcal{B}}(\boldsymbol{q}_h) \cdot \nabla \boldsymbol{q}_h
  \,d\boldsymbol{x}\,dt = 0\,, \nonumber \\
\label{eq:ADER-DG}
\end{align}
where an element-local {spacetime predictor} solution $\q_h(\x,t)$
has been introduced and the details related to its computation are given
in the next section.

Due to the discontinuous character of the solution $\q_h$ at the element
interfaces $\partial \Omega_i$, the surface integral of the fluxes is
computed by means of an approximate Riemann solver $\mathcal{G}$
depending on the boundary extrapolated data $\q_h^-$ and $\q_h^+$
evaluated at the left and right of an element interfaces,
respectively. In this paper we mainly use the simple Rusanov flux
\citep{Rusanov1961a}
\begin{equation}
  \mathcal{G}\left(\q_h^-, \q_h^+ \right) \cdot \boldsymbol{n} = \frac{1}{2}
  \left( \boldsymbol{F}(\q_h^+) + \boldsymbol{F}(\q_h^-) \right) \cdot \boldsymbol{n}
  - \frac{1}{2} s_{\max} \left( \q_h^+ - \q_h^- \right)\,,
	\label{eq.rusanov} 
\end{equation} 
where $s_{\max}$ denotes the maximum signal speed computed in $\q_h^-$
and $\q_h^+$. Any other monotone numerical flux function could be used 
equally well, see \cite{toro-book} for an overview of different Riemann
solvers. 
On the other hand, the {jump term} of the
nonconservative product has been approximated with a so-called
{path-conservative} scheme \citep{pares2006,Castro2006} of the
form
\begin{align}
\mathcal{D}\left( \q_h^-,\q_h^+ \right) \cdot \boldsymbol{n} = & \frac{1}{2}
\left(\int \limits_{0}^{1} \boldsymbol{\mathcal{B}} \left(
\bm{\psi}(\q_h^-,\q_h^+,s) \right)\cdot\boldsymbol{n} \, ds
\right)\cdot\left(\q_h^+ - \q_h^-\right)\,, \label{eq:PC}
\end{align}
that is based on the theory of \cite{DLM1995} on hyperbolic partial
differential equations with nonconservative products and which must obey
the generalized Rankine-Hugoniot or {consistency} condition
\begin{align}
  &
  \mathcal{D}\left( \q_h^-,\q_h^+ \right) \cdot \boldsymbol{n} -
  \mathcal{D}\left( \q_h^+,\q_h^- \right) \cdot \boldsymbol{n} =
  \nonumber \\
  & \hskip 2.0cm
  \int \limits_{0}^{1} \boldsymbol{\mathcal{B}} \left( \bm{\psi}(\q_h^-,\q_h^+,s)
  \right)\cdot\boldsymbol{n} \, \partial_s \bm{\psi} \, ds\,.
\end{align}
The path $\bm{\psi}=\bm{\psi}(\q_h^-,\q_h^+, s)$ is a Lipschitz
continuous function with $0 \leq s \leq 1$, $\bm{\psi}(\q_h^-,\q_h^+, 0)
= \q_h^-$ and $\bm{\psi}(\q_h^-,\q_h^+, 1) = \q_h^+$. We
here use the simplest possible path, \ie the straight-line segment path
\begin{align}
\bm{\psi} = \bm{\psi}(\q_h^-, \q_h^+, s) = \q_h^- + s \left( \q_h^+ -
\q_h^- \right)\,, && s \in [0,1]\,, 
\end{align}
and the line integral in (\ref{eq:PC}) is easily evaluated by
sufficiently accurate numerical quadrature rules [see
  \citet{ADERNC,OsherNC} for details].

Notice that the combination of (\ref{eq.rusanov}) with (\ref{eq:PC})
represents the extension of the Rusanov (or local Lax-Friedrichs) flux to
the nonconservative case. Indeed, other more sophisticated schemes may be
used with the aim of reducing the numerical dissipation [see \eg the
{HLLEM}-type version of \citet{NCP_HLLEM}, which is an extension of the
HLLEM flux of \citet{HLLEM}, or the path-conservative Osher schemes
forwarded in \citet{OsherNC}].

Note also that the choice made here to interpret the gravity terms as a
nonconservative product makes them appear not only in the volume integral
in Eq. (\ref{eq:ADER-DG}), but also also in the {Riemann solver} via
Eq. (\ref{eq:PC}) above. This contribution to the Riemann solver is not
present in classical discretisations as purely algebraic source
term. However, the main advantage of path-conservative schemes is that
they allow at least in principle the construction of well-balanced
numerical schemes that are able to preserve particular steady-state
solutions of the governing partial differential equations
exactly. Although the development of exactly well-balanced schemes for
the GRMHD equations is beyond the scope of this work, it represents an
interesting extension of the formalism presented here.

As a concluding remark in this Section we note that the ADER-DG scheme
(\ref{eq:ADER-DG}) is $(N+1)$-th order accurate for smooth
solutions. Since the final algorithm is a purely {explicit} DG scheme, a
CFL-type stability condition on the time step holds in the form
\begin{align}
\Delta t_{\text{DG}} < \text{CFL}\frac{h_{\text{min}} }{d
  \left(2N+1\right)} \frac{1}{|\lambda_{\text{max}}|}, \label{eq:CFL}
\end{align}
where $h_{\text{min}}$ is the minimum characteristic mesh-size, $d$ is
the number of spatial dimensions, $\lambda_{\text{max}}$ is the maximum
signal velocity of the system of partial differential equations, and CFL
is a constant coefficient such that $0 < \text{CFL} < 1$. If not
stated otherwise, the standard value for the tests presented  
in this paper is $\text{CFL}=0.9$. 
{For the results of a numerical 
von Neumann stability analysis of ADER-DG schemes, see \eg 
\cite{dissdumbser,QiuDumbserShu,Dumbser2008}.}

\subsection{Spacetime discontinuous Galerkin predictor}
\label{sec:STDG}

First introduced in Eq. (\ref{eq:ADER-DG}), the spacetime {predictor}
$\q_h$ is an \emph{``interior''} solution of the partial differential
equations within each element, based on the following weak formulation of
(\ref{eq:DGPDE}) in spacetime:
\begin{align}
 \int \limits_{t^n}^{t^{n+1}} \!\! \int\limits_{\Omega_i^{\circ} }
 \theta_k \, \partial_t \q_h \,d\boldsymbol{x} \, dt + \int
 \limits_{t^n}^{t^{n+1}} \!\! \int\limits_{\Omega_i^{\circ} } \theta_k \,
 \nabla \cdot \boldsymbol{F}(\q_h) \,d\boldsymbol{x}\,dt &\nonumber \\
 + \int
 \limits_{t^n}^{t^{n+1}} \!\! \int\limits_{\Omega_i^{\circ} } \theta_k
 \boldsymbol{\mathcal{B}}(\q_h ) \cdot \nabla \q_h \,d\boldsymbol{x}\,dt&=
 0\,, \label{eq:predictor}
\end{align}
where the spatial domain of integration has been reduced to only the
{interior} of the space elements $\Omega^{\circ}_i$, \ie without
integration by parts of the space-integrals. As a result, one obtains a
system of $N_E$ {independent} (element-local) equation systems.

Note the introduction here of the new basis set $\{\theta_k\}$ for the
vector space $\mathcal{Q}_h^N$ of piecewise {spacetime} polynomials of
maximum degree $N$, and the discrete solution $\q_h(\x,t)$ is represented
in terms of the basis functions $\theta_k$ as
\begin{equation}
  \q_h(\boldsymbol{x},t) = \theta_k(\boldsymbol{x},t) \, {\hat \q}_k\,.
\label{eq.stdof}
\end{equation}
Also in this case, a {nodal} basis is used, based on the Gauss-Legendre
quadrature points referring to the spacetime element $\Omega_i\times
[t^n, t^{n+1}]$. After integration in time by parts of the first term in
(\ref{eq:predictor}) and after invoking the \emph{causality principle}
(upwinding in time) then the following $N_E$ {independent systems} of
$(N+1)^{(d+1)}$ nonlinear equations in the spacetime degrees of freedom
$\hat{\q}_k$ are obtained:
\begin{align}
   \int\limits_{\Omega_i^{\circ}} \theta_k(\boldsymbol{x},t^{n+1})
   \q_h(\boldsymbol{x},t^{n+1}) \, d\boldsymbol{x} -
   \int\limits_{\Omega_i^{\circ}} \theta_k(\boldsymbol{x},t^{n})
   \boldsymbol{u}_h(\boldsymbol{x},t^{n}) \, d\boldsymbol{x} &\nonumber\\ - \int
   \limits_{t^n}^{t^{n+1}} \!\!  \int\limits_{\Omega_i^{\circ} } \!\!\!
   \partial_t \theta_k \q_h(\boldsymbol{x},t) \,d\boldsymbol{x}
   \, dt + \int \limits_{t^n}^{t^{n+1}} \!\!
   \int\limits_{\Omega_i^{\circ} } \!\!\! \theta_k \nabla \cdot
   \boldsymbol{F}(\q_h) \,d\boldsymbol{x}\,dt &\nonumber\\ + \int
   \limits_{t^n}^{t^{n+1}} \!\!  \int\limits_{\Omega_i^{\circ} } \theta_k
   \boldsymbol{\mathcal{B}}(\q_h) \cdot \nabla\q_h \,d\boldsymbol{x}\,dt &= 0\,, \nonumber
   \\ i=1,2,\ldots,N_E;\quad
   k=1,2,\ldots,(N+1)^{(d+1)}. \quad\quad\quad& \label{eq:DOFpredictor}
\end{align}
The system of equations (\ref{eq:DOFpredictor}) can be solved via a simple
discrete Picard iteration for each element $\Omega_i$, without needing
any communication with neighbour elements \citep{Dumbser2008}.

{We should stress that the choice of an appropriate
  initial guess $\q_h^0(\x,t)$ for $\q_h(\x,t)$ is crucial to obtain a
  computationally efficient scheme.  One can either use an extrapolation
  of $\q_h$ from the previous time interval $[t^{n-1},t^n]$, as suggested
  in \cite{ADERPrim}, or a second-order accurate MUSCL-Hancock method, as
  suggested in \cite{Hidalgo2011}. For the initial guess, one can write a
  Taylor series expansion in time and then simply needs to compute
  approximations to the time derivatives of $\q_h$ at time $t^n$, where
\begin{equation}
\boldsymbol{\mathcal{L}}(\u_h(\x,t^n)) = - \nabla \cdot \F(\u_h(\x,t^n)) - 
\boldsymbol{\mathcal{B}}\left(\u_h(\x,t^n)\right) \cdot  \nabla \u_h(\x,t^n)
\end{equation} 
is used as an abbreviation in the following. A second-order accurate 
MUSCL-type initial guess for for $\q_h(\x,t)$ is given by 
\begin{equation} 
  \q^0_h(\x,t) = \u_h(\x,t^n) + \left(t - t^n\right)  \boldsymbol{\mathcal{L}}(\u_h(\x,t^n)), 
\end{equation} 
while a third-order accurate initial guess for $\q_h(\x,t)$ reads 
\begin{equation}
\q^0_h(\x,t) = \u_h(\x,t^n) + \left(t - t^n\right)  \boldsymbol{k}_1 + 
 \frac{1}{2} \left(t - t^n\right)^2 \frac{\left( \boldsymbol{k}_2 - \boldsymbol{k}_1 \right)}{\Delta t}, 
\end{equation} 
where $\boldsymbol{k}_1 := \boldsymbol{\mathcal{L}}\left(\u_h(\x,t^n)
\right)$ and $\boldsymbol{k}_2 :=
\boldsymbol{\mathcal{L}}\left(\u_h(\x,t^n) + \Delta t \boldsymbol{k}_1
\right)$.  For an even higher-order accurate initial guess, one can
employ the continuous extension Runge-Kutta (CERK) schemes proposed in
\cite{OwrenZennaro}. For the use of CERK schemes as time integrators of
explicit discontinuous Galerkin schemes, see \cite{Gassner2011a}.  If an
initial guess of the order $N$ is chosen, it is sufficient to use
\emph{one single} Picard iteration in order to solve
\eqref{eq:DOFpredictor}.  } It is also useful to remark once again that
one-step ADER schemes are particularly well suited for AMR with
time-accurate local-timestepping (LTS) and allow a consistent reduction
of MPI communications compared to classical Runge-Kutta time stepping
schemes, see \cite{AMR3DCL,AMR3DNC}, and \cite{Zanotti2015c},
\cite{Zanotti2015d}, \cite{ADERDGVisc} for details.

{Furthermore, in ADER schemes for nonlinear hyperbolic PDE, limiters  
need to be applied only once per time step, while in Runge-Kutta based 
method of lines schemes, the limiter needs to be applied in each Runge-Kutta 
stage again.}   

{For a detailed comparison of Runge-Kutta and ADER
  finite-volume schemes, see \cite{dumbser_diffapprox} and
  \cite{Balsara2013}, while Runge-Kutta DG and Lax-Wendroff DG schemes
  (the latter are very similar to ADER-DG schemes) have been compared in
  \cite{QiuDumbserShu}, also concerning computational performance. We
  also show a detailed computational performance comparison between
  ADER-DG schemes and RKDG schemes for GRMHD at the end of this paper in
  Section \ref{sec:performance}.}

\subsection{\textit{A-posteriori} subcell finite-volume limiter}
\label{sec:limiter}

The ADER-DG scheme (\ref{eq:ADER-DG}) is formally of order $N+1$ for
smooth solutions, hence the method must be {oscillatory} for $N>0$ in the
presence of discontinuities, since the scheme is {linear} in the sense of
\cite{Godunov59}, thus inevitably generating spurious oscillations (this
is also known as the {``Gibbs phenomenon''} in the context of signal
analysis). In order to cope with this problem, a special treatment is
needed \emph{wher}ever and \emph{when}ever the solution is discontinuous,
or the gradients in the discrete solution are sufficiently steep. In our
specific implementation, this problem is handled as follows: After
evaluating the predictor solution $\q_h(\x,t)$ via Eq.
(\ref{eq:DOFpredictor}), a so-called {candidate} solution
$\u_h^*(\x,t^{n+1})$ is computed through the {unlimited} one-step ADER-DG
scheme (\ref{eq:ADER-DG}). Next, the candidate solution
$\u_h^*(\x,t^{n+1})$ is checked against mathematical and physical
admissibility criteria, which are collectively referred as the relaxed
\emph{discrete maximum principle} (DMP). These criteria are: the absence
of {floating point errors} (NaNs), the {positivity} of pressure and
density of the fluid, the velocity being lower than the light speed and a
possible (successful) conversion from conservative to primitive variables
$\boldsymbol{V}=\boldsymbol{V}(\boldsymbol{Q})$ [see
  \cite{ADER_MOOD_14,Zanotti2015d}]. Typical scenarios that may
potentially violate the cited admissibility criteria are: the vicinity of
steep-gradients or discontinuities, under-resolved flow features, as well
as very low pressure and density conditions, \eg atmospheres around
compact objects or vacuum regions.

For the subcell finite-volume limiter we introduce the notation
$\v_h(\x,t^n) = \mathcal{P}\left( \u_h(\x,t^n) \right)$ as the
$L_2$-projection of $\u_h$ onto the space of piecewise {constant}
functions on a given {sub-grid} defined within $\Omega_i$, where the
individual cells of the subgrid are denoted by $\Omega_{i,s}$ with
$\bigcup \Omega_{i,s} = \Omega_i$. Following \citet{Dumbser2014},
\citeauthor{Zanotti2015c} (\citeyear{Zanotti2015c},
\citeyear{Zanotti2015d}b), and \citet{ADERDGVisc}, each element
$\Omega_i$ is divided into $N_s^d$ equidistant subgrid cells
$\Omega_{i,s}$ with $N_s \geq N+1$. If we denote by $\bar{\v}_{i,s}^n$
the individual subcell averages within each subcell $\Omega_{i,s}$, then
the projection $\mathcal{P}$ reads
\begin{equation}
   \bar{\v}_{i,s}^n := \frac{1 }{| \Omega_{i,s} |} \int
   \limits_{\Omega_{i,s}} \u_h(\x,t^n) d\x \,.
	\label{eq.subcellaverage} 
\end{equation}  

In practice, the relaxed DMP used in this paper reads:
\begin{equation} 
\min \limits_{\y\in {\cal{V}}_i} (\v_h(\boldsymbol{y},t^n))-\delta \leq
\v_h^*(\boldsymbol{x},t^{n+1})\leq \max \limits_{\boldsymbol{y}\in
  {\cal{V}}_i}(\v_h(\y,t^n))+\delta
\,, 
\label{eq:DMP}
\end{equation}
where ${\cal{V}}_i$ is the set containing the space-element $\Omega_i$
and its Voronoi neighbours that share a common node with $\Omega_i$.
Here, the parameter $\delta$ in (\ref{eq:DMP}) is chosen as
\begin{equation}
	\delta =\max \left( \delta_0\,, \epsilon \times \left( \max
        \limits_{y\in {\cal{V}}_i}(\boldsymbol{u}_h(\boldsymbol{y},t^n))- \min
        \limits_{y\in {\cal{V}}_i}(\boldsymbol{u}_h(\boldsymbol{y},t^n))\right)\,
        \right)\,,
\end{equation}
with $\delta_0=10^{-8}$ and $\epsilon=10^{-7}$, which is more restrictive
than what used in previous work [\citep{Dumbser2014}, \citeauthor{Zanotti2015d}].
 
If the candidate solution $\u_h^*$ violates {any} of the criteria of the
relaxed DMP (\ref{eq:DMP}), then it is {locally} rejected and the cell
$\Omega_i$ is flagged as a troubled cell and a limiter status flag
$\tilde{\beta}^{n+1}_i$ is set to $\tilde{\beta}^{n+1}_i=1$; conversely,
it is set to $\tilde{\beta}^{n+1}_i=0$ if all admissibility criteria are
satisfied in cell $\Omega_i$ at time $t^{n+1}$. For a troubled cell
$\Omega_i$, the numerical solution is then {recomputed}, starting again
from the old time level $t^n$, but using now a more robust numerical
scheme than the high-order ADER-DG scheme.

We have here selected as numerical scheme on the subgrid level a
second-order accurate MUSCL-Hancock TVD finite-volume scheme with MinMod
slope limiter \citep{toro-book}, mostly because of its proven robustness
in the presence of shock waves and low density atmospheres. For cells
which were unlimited at the old time (\ie $\beta_i^n=0$), it is easy to
compute the necessary subcell averages via the projection
(\ref{eq.subcellaverage}), while for limited cells at time $t^n$, the
subcell averages are already available from the previous time step. As an
alternative, a higher accurate ADER-WENO finite-volume schemes can be
used [see \citet{Dumbser2014,AMR3DCL}], bearing in mind that the WENO
approach does not clip local extrema, in contrast to the chosen
second-order TVD method. However, for the GRMHD system considered here we
have found the subcell TVD limiter to be much more robust than the WENO
scheme.

Formally, we can write both the second-order MUSCL-Hancock scheme, as
well as a high-order ADER-WENO scheme, as
\begin{align}
& \bar{\boldsymbol{v}}_{i,s}^{n+1} - \bar{\boldsymbol{v}}_{i,s}^{n} +
  \int 
  \limits_{t^n}^{t^{n+1}} \!\! \!\! \int\limits_{\partial \Omega_{i,s} } \!\!\!\!
  \mathcal{G}\left(\q_h^-, \q_h^+ \right) \cdot \boldsymbol{n} \, dS \, dt
  \nonumber \\ & + \!\!\int \limits_{t^n}^{t^{n+1}} \!\!\!\!
  \int\limits_{\partial \Omega_{i,s} } \!\!\!\!  \mathcal{D}\left(
  \q_h^-,\q_h^+ \right) \cdot \boldsymbol{n} \, dS \, dt + \int
  \limits_{t^n}^{t^{n+1}} \!\!\!\!\!\! \int\limits_{\Omega_{i,s}^{\circ} }\!\!\!\!
  \boldsymbol{\mathcal{B}}(\boldsymbol{q}_h) \cdot \nabla \boldsymbol{q}_h
  \,d\boldsymbol{x}\,dt = 0\,, \nonumber \\
\label{eq:ADER-FV}
\end{align}
which is very similar to the ADER-DG scheme (\ref{eq:ADER-FV}).

High order in space, together with non-oscillatory properties, are
achieved in Eq. (\ref{eq:ADER-FV}) via a {nonlinear} reconstruction of
piecewise polynomials from the known cell averages
$\bar{\boldsymbol{v}}_{i,s}^n$ using either a TVD or a WENO
reconstruction. Denoting by $\boldsymbol{w}_h(\x,t^n)$ the result of
this reconstruction, it can then be used to compute the predictor
$\q_h(\x,t)$, either via Eq. (\ref{eq:DOFpredictor}), where
$\u_h(\x,t^n)$ is simply replaced by $\w_h(\x,t^n)$ and the control
volume of the spacetime integration is replaced by $\Omega_{i,s} \times
[t^n, t^{n+1}]$, or via the simple MUSCL-Hancock evolution step to the
half-time level [see \cite{toro-book} for details].

From Eq. (\ref{eq:ADER-FV}), a new piecewise constant solution $\v_h(\x,
t^{n+1})$ given by the cell averages $\bar{\boldsymbol{v}}_{i,s}^{n+1}$
is obtained, from which we can then reconstruct the final, {limited} DG
polynomial as $\u_h(\x,t^{n+1}) = \mathcal{R}\left( \v_h(\x,t^n) \right)
$, where $\mathcal{R}$ is the reconstruction operator associated with the
projector $\mathcal{P}$, so that $\mathcal{R} \circ \mathcal{P} =
\mathcal{I}$, with $\mathcal{I}$ the identity operator [see
  \cite{Dumbser2014} for details].
For the subcell finite-volume scheme a different CFL stability condition
applies and takes the form
\begin{align}
\Delta t_{\text{FV}} < \text{CFL}\frac{h_{\text{min}}}{d \, N_s}
\frac{1}{|\lambda_{\text{max}}|}, \label{eq:CFLweno}
\end{align}
with $h_{\min}$ the minimum cell size referred to the DG control volumes
$\Omega_i$. Choosing $N_s \geq N+1$ is a natural requirement that allows
to reconstruct the of degrees of freedom of $\u_h$ from the piecewise
constant solution $\v_h$ via $\mathcal{R}$. Following \cite{Dumbser2014}
we choose $N_s = 2N + 1$ so that $\Delta t_{\text{FV}} = \Delta
t_{\text{DG}}$. This choice allows us to maximise the resolution
properties of the chosen subcell finite-volume scheme and to run it at
its maximum possible CFL number. 
For alternative higher order ADER-WENO finite-volume schemes for the 
relativistic MHD equations with reconstruction in primitive variables, 
the reader is referred to \cite{BalsaraKim} and \cite{ADERPrim}.

\subsection{Adaptive Mesh Refinement}

The ADER-DG algorithms with subcell finite-volume limiter described above
has been here implemented on spacetime adaptive Cartesian meshes. Detains
on our AMR algorithm have been described in \citet{AMR3DCL} and in
\citeauthor{Zanotti2015c} (\citeyear{Zanotti2015c}, as well as
\citeyear{Zanotti2015d}b) and \citet{ADERDGVisc}, and we refer the
interested reader to these works. The AMR strategy adopted here is named
``{cell-by-cell}'' refinement and consists in providing a space-tree data
structure \citep[see][for details]{Khokhlov98, Peano1, Peano2, AMR3DCL},
whose ``leaves'' correspond to the spatial elements $\Omega_i$ used by
the numerical scheme described before. The main alternative to a
space-tree data structure is the use of so-called ``\emph{patches}'',
\citep[see][]{Berger-Oliger1984, berger85, Berger-Colella1989}, where a
set of independent overlaying Cartesian sub-grid domains, or ``patches'',
is introduced and activated when necessary. In our AMR approach the
numerical solution is checked independently along every single
space-element for an eventual recursive refining or recoarsening
process.

\begin{figure*}
  \centering
  \includegraphics[width=0.3\textwidth]{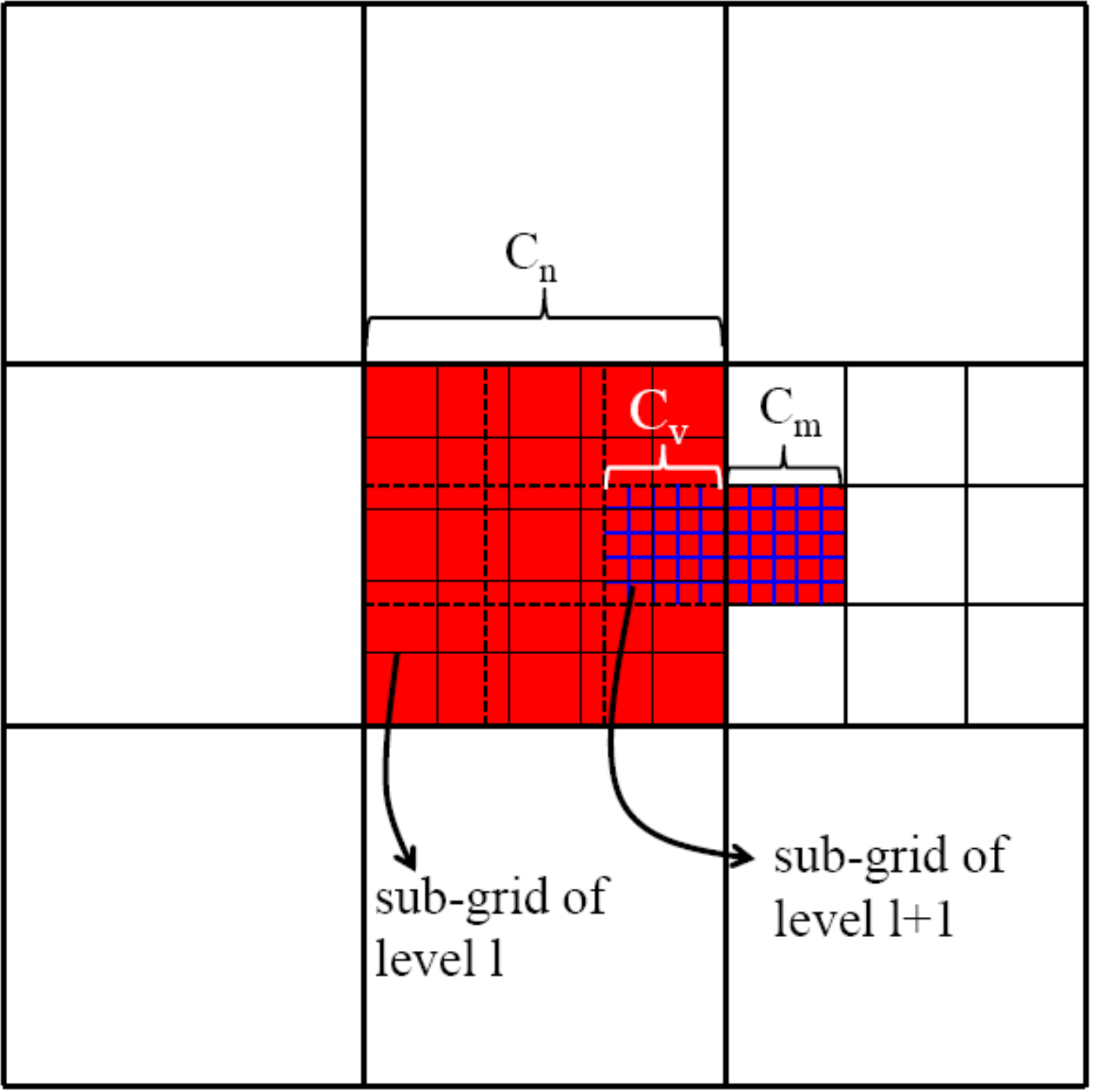}
  \hspace{3cm}
  \includegraphics[width=0.4\textwidth]{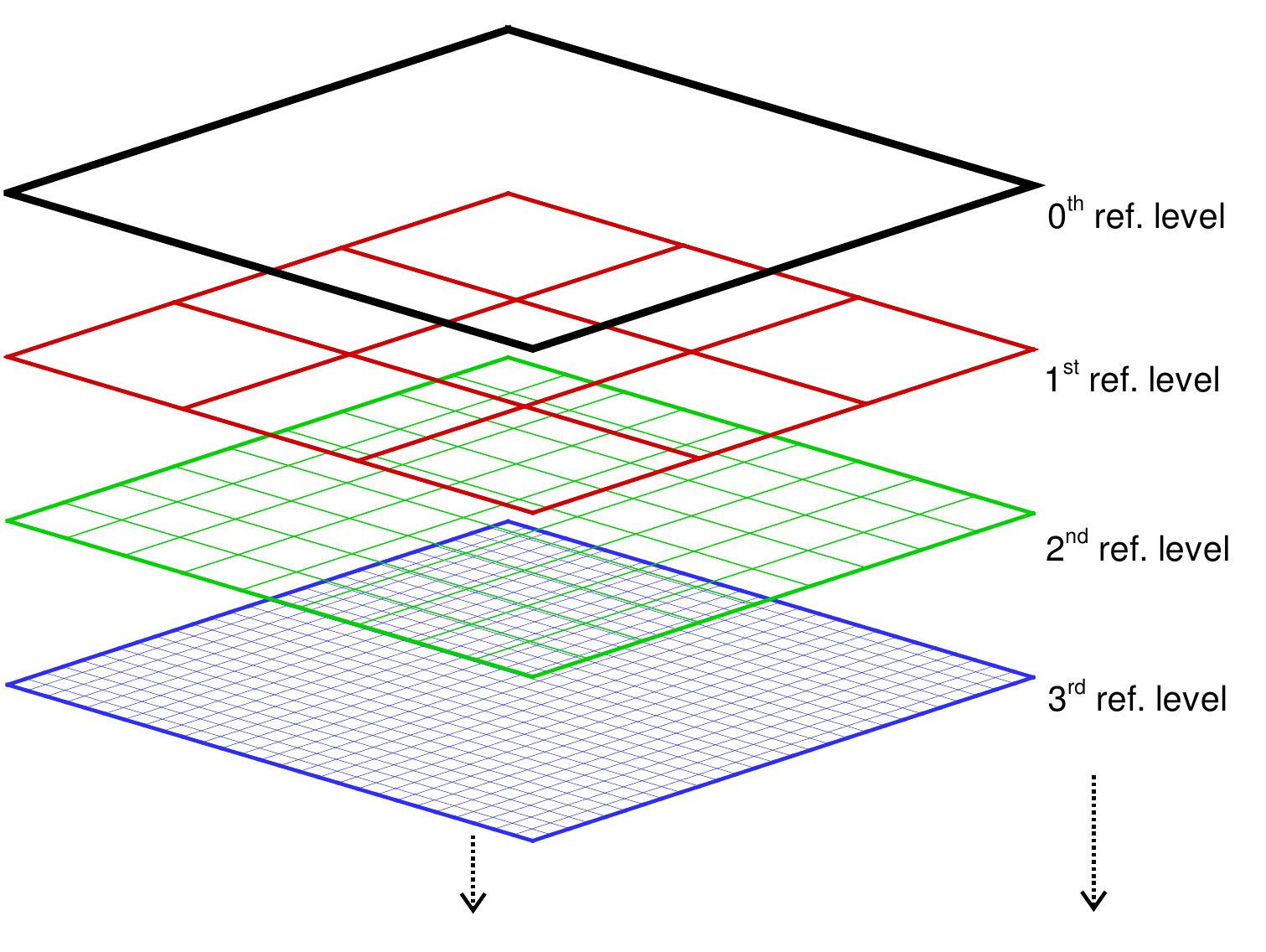}
  \caption{On the left, an example of combination of AMR and DG subcell
    reconstruction is shown. The limited cells ($\beta=1$)
    $\mathcal{C}_n$ and $\mathcal{C}_m$ are highlighted in red. The
    simplest way for the polynomial reconstruction between
    $\mathcal{C}_n$ and $\mathcal{C}_m$ elements is: (i) project the
    piecewise constant solution from $\mathcal{C}_n$ to the virtual
    child-element $\mathcal{C}_v$ [see Fig. \ref{fig:AMRmaps}]; (ii) do
    polynomial reconstruction along the same refinement level, between
    $\mathcal{C}_v$ and $\mathcal{C}_m$. At the right, the
    \emph{space-tree} structure of the refinement levels for a single
    element at the coarsest level $\ell_0$ is shown, corresponding to the
    choice $\mathcal{R}=3$. }
  \label{fig:AMR}
\end{figure*}

In practice, starting from an initial Cartesian grid of refinement level
$\ell=\ell_0=0$, which is the basic mesh without refinement, the
tree-type infrastructure of finer {refinement levels} is made
accessible. The refinement levels $\ell>0$ are built according to the so
called {refinement factor} $\mathcal{R}$ which is the number of smaller
space-elements per space-direction in which a coarser element is broken
in a refinement process, or which are merged in a recoarsening
stage. Note that choosing a refinement factor $\mathcal{R}=2$ would
generate the well known ``quadtrees'' in two-dimensional (2D) meshes and
``octrees'' in 3D meshes. For an arbitrary refinement factor
$\mathcal{R}$, general space-trees are obtained [see also
  \cite{Peano1,Peano2}].

For practical purposes, a finite number of refinement levels is provided,
\ie from the coarser $\ell = \ell_0$ to a finest possible refinement
level $\ell = \ell_{\text{max}}\in\natSet^+_0$. The
refinement/recoarsening process is driven by a prescribed
{refinement-estimator function} ${\chi} =
{\chi}\left(\u_h(\x,t^n)\right)$, which is a function of discrete
gradients and second derivatives of a scalar \emph{indicator function}
$\varphi$, and by two thresholds $\chi^+$ and $\chi^-$. Elements are
marked for refinement whenever $\chi > \chi^+$ and for recoarsening
whenever $\chi < \chi^-$ [for details on the definition of $\chi$ see
  \cite{Loehner1987,Zanotti2015d,ADERDGVisc}]. In general, the indicator
function $\varphi$ can be chosen to be any mathematical quantity of
physical interest that varies in the computational domain and time, \eg
the local rest-mass density, the pressure, or a function of the
state-variables and of their gradients, \eg the Lorentz factor or the
vorticity, but also the limiting-status $\beta_i^n$. Hereafter and unless
stated otherwise, we have simply used the rest-mass density as indicator
function, \ie $\varphi = \rho$. An alternative choice of $\chi$ that deserves
investigations in the future would consist in the evaluation of the 
numerical production of entropy as both error and smoothness indicator [see
\citet{PS:entropy, SCR:CWENOquadtree, CS:epsweno} for details], but has 
not yet been used in the present paper. 

To simplify the AMR algorithm, two neighbour elements are allowed to
belong either to the same level $\ell$ or to an adjacent refinement level
$\ell \pm 1$. To each element in the tree we assign a basic {element
  status} which is
\begin{align}
\sigma_i &= \left\{\begin{array}{rcl} 
-1\,, & & \text{for the so-called \emph{parent cells}} \\
\phantom{-}0\,, & & \text{for \emph{active elements}} \\
+1\,, & & \text{for the so-called \emph{virtual children}}
\end{array}\right. \nonumber \\
 i&=1,\ldots, N_{\text{tot}},
\end{align}
where $N_{\text{tot}}$ is the total number of space-elements present in
the tree. Note that $N_{\text{tot}}$ should be distinguished from the
total number of {active} elements $N_E$, which are the leaves of the tree
that define the $\Omega_i$ used in the numerical scheme, and for which
$N_{\text{tot}}>N_E$ holds in general. The so-called {parent cells}
($\sigma_i=-1$) are those tree elements which contain active elements on
a higher level and finally a {virtual child cell} ($\sigma_i=+1$) is a
tree element which is {contained} within an active cell that belongs to a
{lower} and {adjacent} refinement level $\ell-1$.

\begin{figure*} 
\input{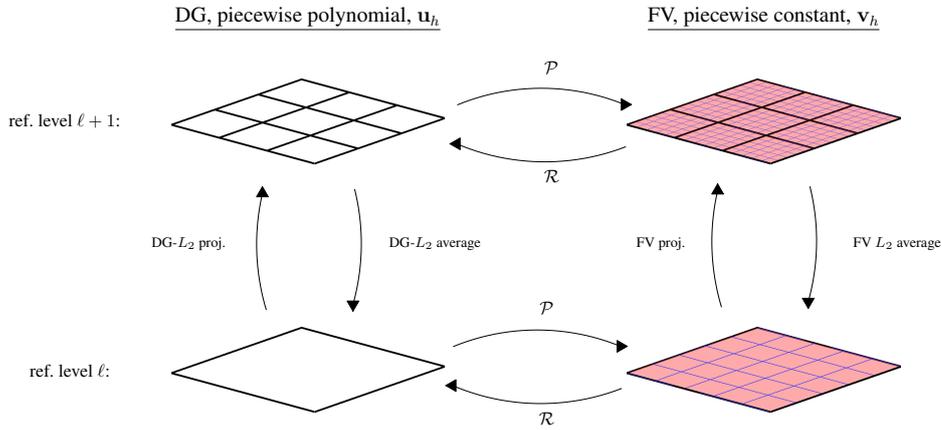} 
\caption{Mapping of the numerical solution between the piecewise
  polynomials $\u_h$ of the DG scheme and the piecewise constant data
  $\v_h$ of the finite-volume scheme as well as between two different
  AMR-levels $\ell$ and $\ell+1$.}
\label{fig:AMRmaps}
\end{figure*}

Apart from the storage of flux contributions from neighbour cells within
our high-order time-accurate local time stepping (LTS) algorithm
\cite{AMR3DCL}, virtual cells are also needed for high-order
finite-volume schemes to provide the necessary data for polynomial
reconstructions (TVD, WENO) on a given refinement level if two adjacent
active cells belong to {different} refinement levels; this is illustrated
schematically in Fig. \ref{fig:AMR}. This strategy produces a locally
uniform grid around each cell and greatly simplifies reconstruction. Our
strategy of generating a locally uniform grid around each cell is very
different from the approach based on genuinely multidimensional CWENO
reconstructions proposed by \cite{SCR:CWENOquadtree}.

The dynamics of the numerical solution on virtual elements is given by
standard $L_2$ projection (for virtual children) or averaging (for parent
cells), as depicted in Fig. \ref{fig:AMRmaps}, where the mapping between
the chosen solution spaces, piecewise polynomial (unlimited) or
piecewise constant (limited), and between two adjacent refinement levels
$\ell$ and $\ell+1$ is depicted.

Finally, due to the possibility of handling a large range of spatial
scales within the same domain, corresponding to very different CFL time
restrictions, a time-accurate and fully conservative {local time
  stepping} (LTS) has been implemented in order to use the smallest
admitted timestep only where necessary, and a large timestep where it is
allowed \citep[see][]{AMR3DCL}. A flow diagram illustrating the main
stages of the final algorithm presented in this Section can be found in
\cite{Dumbser2014}.

\section{Numerical validation}
\label{sec:numerical-validation}

 In the following sections, we will present a series of numerical
 validations of the numerical algorithms introduced so far and
 characterised by a path-conservative ADER-DG scheme supplemented by an
 {a-posteriori} finite-volume limiter applied to AMR grids. The tests
 have been performed both in special and in general relativity, employing
 either two or three spatial dimensions. Furthermore, the validations
 will be distinguished in ``smooth flows'' (Sec. \ref{sec:smooth}), for
 which we will be able to measure the actual convergence order of the
 scheme, and ``non-smooth flows'' (Sec. \ref{sec:discontinuous_sr} and
 \ref{sec:discontinuous_gr}), for which we will illustrate the ability of
 our approach to handle accurately shocks and large gradients.

All of these tests share a number of common properties which we list
below and have been employed unless stated otherwise: i) the adiabatic
index has been chosen equal to $\Gamma=4/3$; ii) the refinement
factor has been chosen as $\mathcal{R}=3$; iii) a second-order
MUSCL-Hancock TVD finite-volume method with reconstruction in primitive
variables on the subgrid-level has been employed as subcell finite-volume
limiter for the ADER-DG $\mathbb{P}_N$ method; iv) the Rusanov (or local
Lax-Friedrichs) approximate Riemann solver has been used; v) problems in
curved spacetimes have been solved employing Kerr-Schild (KS)
coordinates, either spherical or Cartesian.

\subsection{Smooth general-relativistic flows} 
\label{sec:smooth}

We first test the high order of convergence of our ADER-DG schemes
against three different scenarios in curved spacetimes given respectively
by: i) the Michel accretion of gas onto a black hole in KS {spherical}
(KSS) coordinates and in 2D); ii) a stationary non-selfgravitating fluid
torus in equilibrium around a black hole, again in 2D; iii) the Michel
accretion with a radial magnetic field in KS {Cartesian} (KSC)
coordinates and in 3D.

To ensure that the flow is actually smooth, in the following tests we will
restrict our computational domain to regions that are fully filled with
fluid. In this way, after successively refining the mesh, we evaluate the
 $L_2$ and $L_\infty$ error norms at different DG polynomial
degrees and mesh resolutions so as to to measure the convergence order of
our numerical implementation and compare it with the expected mathematical
one. Anticipating what will be shown in more detail in the following
sections, the numerical results {confirm} the high order of accuracy of
the presented numerical scheme. Indeed, using the results shown in Tables
\ref{tab:CTMichel2D}-\ref{tab:CTBondi3D} we can conclude that the ADER-DG
$\mathbb{P}_N$ method reaches its design accuracy $N+1$ in most cases.

{For all these convergence tests, since the reference
  solutions are stationary in time, we used the initial condition as the
  external state vector in the chosen approximate Riemann solver whenever
  evaluating the fluxes \emph{only at the boundary interfaces}
  $\boldsymbol{x}\in \partial \Omega$.}

\subsubsection{\textbf{2D Michel accretion onto a Schwarzschild black hole}}
\label{sec:conv}

As a first test of a smooth flow with an analytical solution we consider
the spherical transonic accretion of an isentropic fluid onto a
nonrotating black hole is known as {Michel solution}
\citep{Michel72}. For the sake of
completeness we give the explicit expressions of the lapse, the shift and
the spatial metric of a Kerr black hole with mass $M$ and spin $a$ in
{Cartesian} Kerr-Schild coordinates $(x,y,z)$
\begin{equation*}
  \alpha = S^{-\halb}\,, \quad \beta^i = \frac{2 H}{S} l_i\,, \quad H = M
  \frac{r^3}{r^4 + a^2 z^2}\,, \quad S = 1 + 2 H\,,
\end{equation*}
\begin{equation} 	
	\gamma_{ij} = \left( \begin{array}{ccc} 1 + 2 H l_x^2 & 2 H l_x
          l_y & 2 H l_x l_z \\ 2 H l_x l_y & 1 + 2 H l_y^2 & 2 H l_y l_z
          \\ 2 H l_x l_z & 2 H l_y l_z & 1 + 2 H l_z^2
	\end{array} \right)\,,
	\label{eq.kerr.metric}
\end{equation}
with
\begin{equation*}
  l_x := \frac{r x + a y}{r^2 + a^2}\,, \qquad l_y := \frac{r y - a
    x}{r^2 + a^2}\,, \qquad l_z := \frac{z}{r}\,,
\end{equation*}
and
\begin{equation*}
  r =
  \sqrt{ \frac{x^2 + y^2 + z^2 - a^2}{2} + \sqrt{\left(\frac{x^2 + y^2 + z^2 -
      a^2}{2}\right)^2 + z^2 a^2} }\,.
\end{equation*}
Conversely, The Kerr metric in {spherical} Kerr-Schild coordinates
$(r,\theta,\phi)$ is given by \citep{Komissarov04}
\begin{equation*}
  \alpha = (1+z)^{-\frac{1}{2}}, \quad \beta^i = \left( \frac{z}{1+z},0,0
  \right) , \quad
\end{equation*}
\begin{equation} 	
	\gamma_{ij} = \left( \begin{array}{ccc} 1 + z & 0 & - a \sin^2
          \theta (1 + z) \\ 0 & \rho^2 & 0 \\ - a \sin^2 \theta (1 + z) &
          0 & {\Sigma} \sin^2 \theta/{\rho^2}
	\end{array} \right)\,,
	\label{eq.kerr.metric.spherical}
\end{equation}
with
\begin{equation*}
	 \rho^2 := r^2 + a^2 \cos^2 \theta\,, \qquad z := \frac{2 r}{\rho^2}\,,
\end{equation*}
\begin{equation*} 
	\Delta := r^2 + a^2 - 2 M r\,, \qquad \Sigma = (r^2 + a^2)^2 - a^2
        \Delta \sin^2 \theta\,.
\end{equation*}

\begin{figure}
  \begin{center} 
    \includegraphics[width=0.46\textwidth]{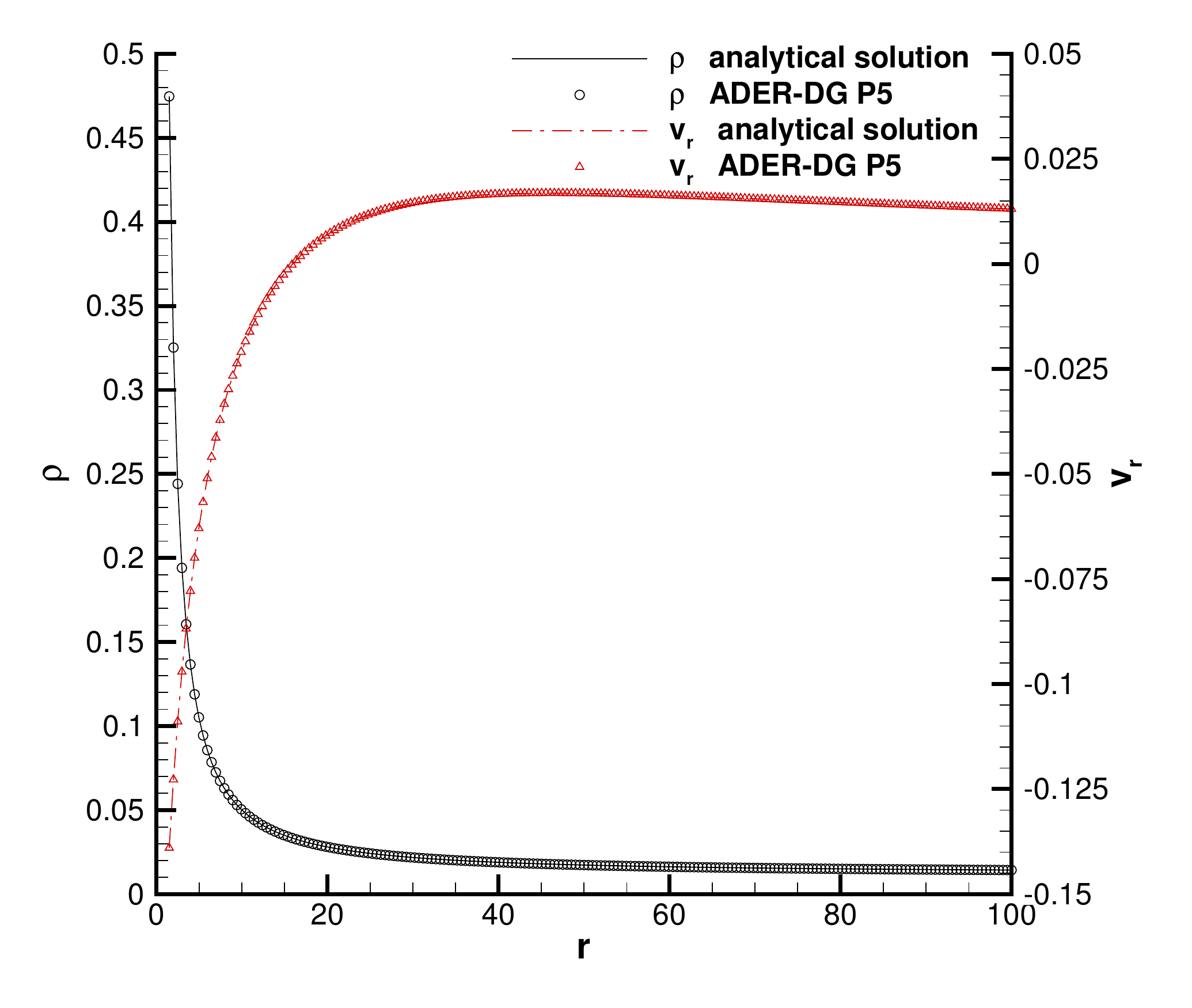}
    \caption{ \label{fig:Michel} Numerical solution for the
      two-dimensional Michel accretion test in KSS coordinates obtained
      with our ADER-DG $\mathbb{P}_5$ at $t=100$. The numerical solution of
      density (black) and radial velocity  (red) interpolated along $200$ points at
      $\theta=1.5$ are plotted. The numerical domain is $(r,\theta) \in
      \Omega = [1.5,100]\times[0.15, 3.0]$.}
  \end{center}
\end{figure}

After taking the metric \eqref{eq.kerr.metric.spherical} with $a=0$ and
defining the values of the free parameters of the problem, \ie the mass
of the black hole $M=1$, the critical radius $r_c=8\,M$ and the critical
density $\rho_c M^2=1/16$, the Michel solution can be determined
analytically [see, \eg \cite{Rezzolla_book:2013}].

We have performed this test in spherical KS coordinates with a spatial
domain $(r,\theta)\in\Omega=[1.5,100]\times[0.15, 3.0]$, discretized with
a uniform mesh of $200\times 32$ elements and solved with our ADER-DG
$\mathbb{P}_5$ scheme. A graphical representation of the numerical
results and their comparison with the analytic solution is shown in
Fig. \ref{fig:Michel}, while the results of the convergence study are
provided in Table \ref{tab:CTMichel2D}. Clearly, we can note an
excellent agreement between analytical and numerical solution and that
the latter converges at the expected and high order.

\begin{table}
  \centering
  \begin{tabular}{ c c  cccc c }
   \hline
    & $N_x$ &  $\mathcal{E}_{L_2}$ & $\mathcal{E}_{L_\infty}$ & $L_2$ & $L_\infty$   &
   Exp. \\ 
   \hline
   \multirow{4}{*}{\rotatebox{90}{{DG-$\mathbb{P}_1$}}}
& 10	& 5.51E-05	& 9.95E-05	& --	&  --				& \multirow{4}{*}{2}	 \\
& 20	& 1.26E-05	& 2.35E-05	&  2.13	& 2.08	&\\
& 40	& 3.01E-06	& 5.70E-06	&  2.06	& 2.05	&\\
& 80	& 7.39E-07	& 1.40E-06	&  2.03	& 2.02	&	\\
   \cline{2-6}
   \hline
	 \multirow{4}{*}{\rotatebox{90}{{DG-$\mathbb{P}_2$}}}
&6	&  2.53E-05	& 3.26E-05		&   ---	&  ---	&\multirow{4}{*}{3}\\
&12	&  3.32E-06	& 4.46E-06		&  2.93	& 2.87	&\\
&18	&  1.01E-06	& 1.37E-06		&  2.93	& 2.91	&\\
&30	&  2.26E-07	& 3.07E-07		&  2.94	& 2.93	 &\\
   \cline{2-6}
   \hline
	 \multirow{4}{*}{\rotatebox{90}{{DG-$\mathbb{P}_3$}}}
&4	&  3.93E-07	& 1.11E-06		&   ---	&  ---	&\multirow{4}{*}{4}\\
&6	&  5.95E-08	& 1.72E-07		&  4.65	& 4.59	& \\
&8	&  1.77E-08	& 4.70E-08		&  4.21	& 4.51	&\\
&12	&  3.55E-09	& 8.05E-09		&  3.96	& 4.35	&\\
   \cline{2-6}
   \hline
	 \multirow{4}{*}{\rotatebox{90}{{DG-$\mathbb{P}_4$}}}
&2	& 3.83E-06	& 5.36E-06	&  ---	&  ---	&\multirow{4}{*}{5}\\
&3	& 4.10E-07	& 5.16E-07	& 5.51	& 5.77	& \\
&4	& 9.13E-08	& 1.23E-07	& 5.22	& 4.97	&\\
&5	& 2.88E-08	& 4.11E-08	& 5.18	& 4.93	&\\
   \cline{2-6}
   \hline
	 \multirow{4}{*}{\rotatebox{90}{{DG-$\mathbb{P}_5$}}}
&2	&  6.33E-08	& 3.30E-08		&   ---	&  ---	&\multirow{4}{*}{6}\\
&3	&  4.22E-09	& 2.36E-09		&  6.68	& 6.50	& \\
&4	&  6.88E-10	& 3.85E-10		&  6.31	& 6.30	&\\
&5	&  1.70E-10	& 1.05E-10		&  6.27	& 5.83	 &\\ 
 \cline{2-6}
   \hline
	 \multirow{4}{*}{\rotatebox{90}{{DG-$\mathbb{P}_6$}}}
&2	& 1.08E-08	& 4.67E-09	&   ---	&  ---	&\multirow{4}{*}{7}\\
&3	& 4.56E-10	& 2.51E-10	&  7.81	& 7.21	& \\
&4	& 5.38E-11	& 3.62E-11	&  7.43	& 6.72	&\\
&5	& 1.04E-11	& 8.11E-12	&  7.37	& 6.71	&\\
   \cline{2-6}
   \hline	
  \end{tabular}
  \caption{ \label{tab:CTMichel2D} $L_2$ and $L_\infty$ errors and
    convergence rates for the 2D Michel accretion in spherical
    Kerr-Schild coordinates for the ADER-DG-$\mathbb{P}_N$ scheme. We
    report the convergence results for the rest-mass density $\rho$ at
    $t=10$ up to $N=6$, and contrast the results with the expected
    rate. The domain has been chosen different (enlarged) for the cases
    $N=5$ and $N=6$ in order to keep away the numerical error from the
    machine limit. Similar results have also been obtained for all other
    flow variables.}
\end{table}

\subsubsection{\textbf{2D torus interior around a Schwarzschild black hole}}
\label{sec:2D_torus_interior}

Next, we consider the numerical convergence study of a stationary
solution of a thick disk (or {axisymmetric test-fluid torus}) orbiting
around a Schwarzschild black hole ($a=0$) of mass $M=1$ in 2D spherical
KS coordinates. The theory of the equilibrium of these
non-selfgravitating fluids in GRHD has been first proposed by
\cite{Abramowicz78, Kozlowski1978} and has been the subject of a vast
literature. 
{ For completeness, we give in appendix \ref{sec:TorusIC} a brief
description of the setup of the primitive
variables of this test problem, referring the interested
reader to \cite{Font02a} or to Chap. 11 of \cite{Rezzolla_book:2013}, but also to \cite{Anton06,DelZanna2007} for details about a more
general configuration of the fluid, depending on the selected values of
physical parameters.
 
}

\begin{figure*}
  \begin{center} 
    \includegraphics[width=0.49\textwidth]{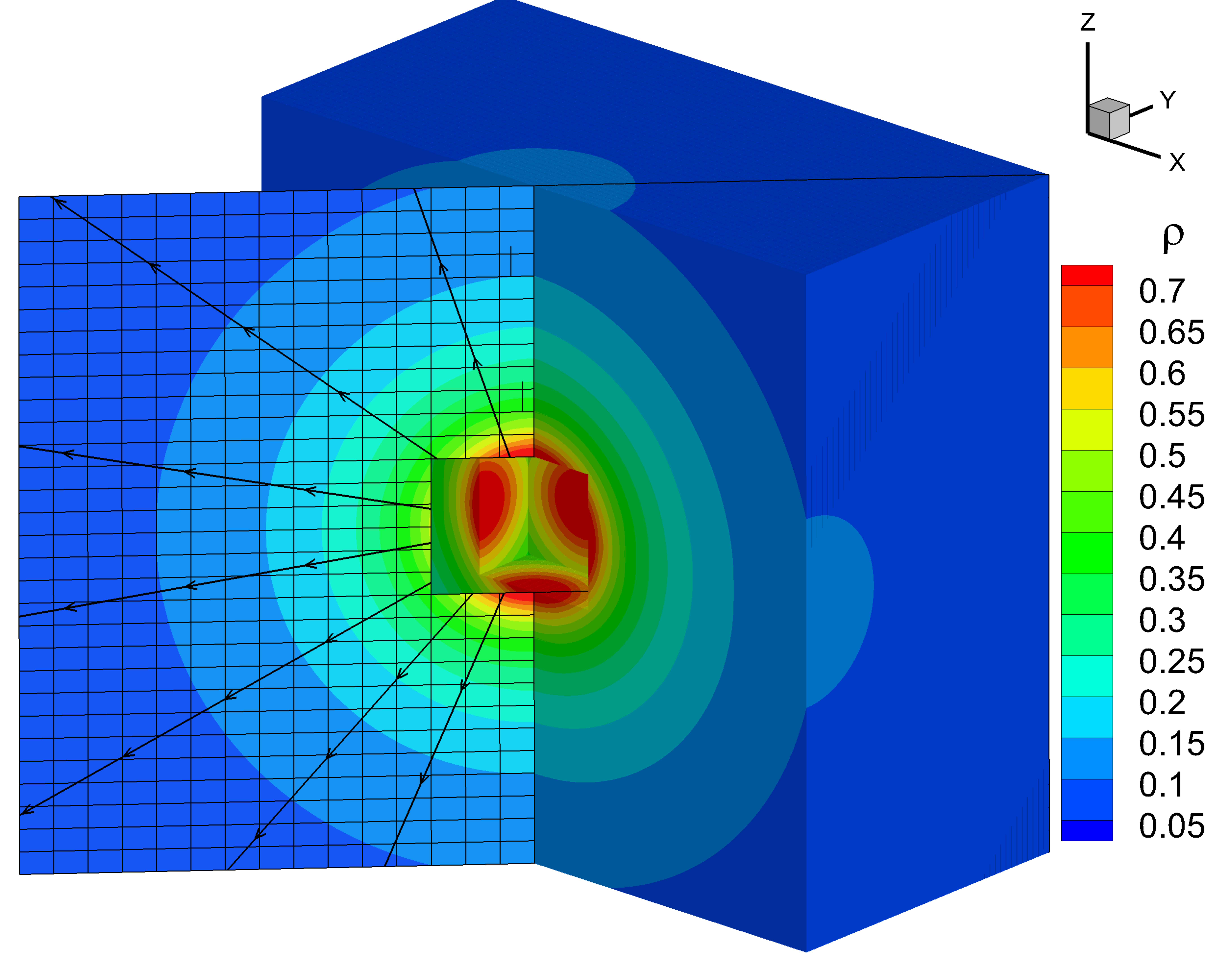}
    \includegraphics[width=0.49\textwidth]{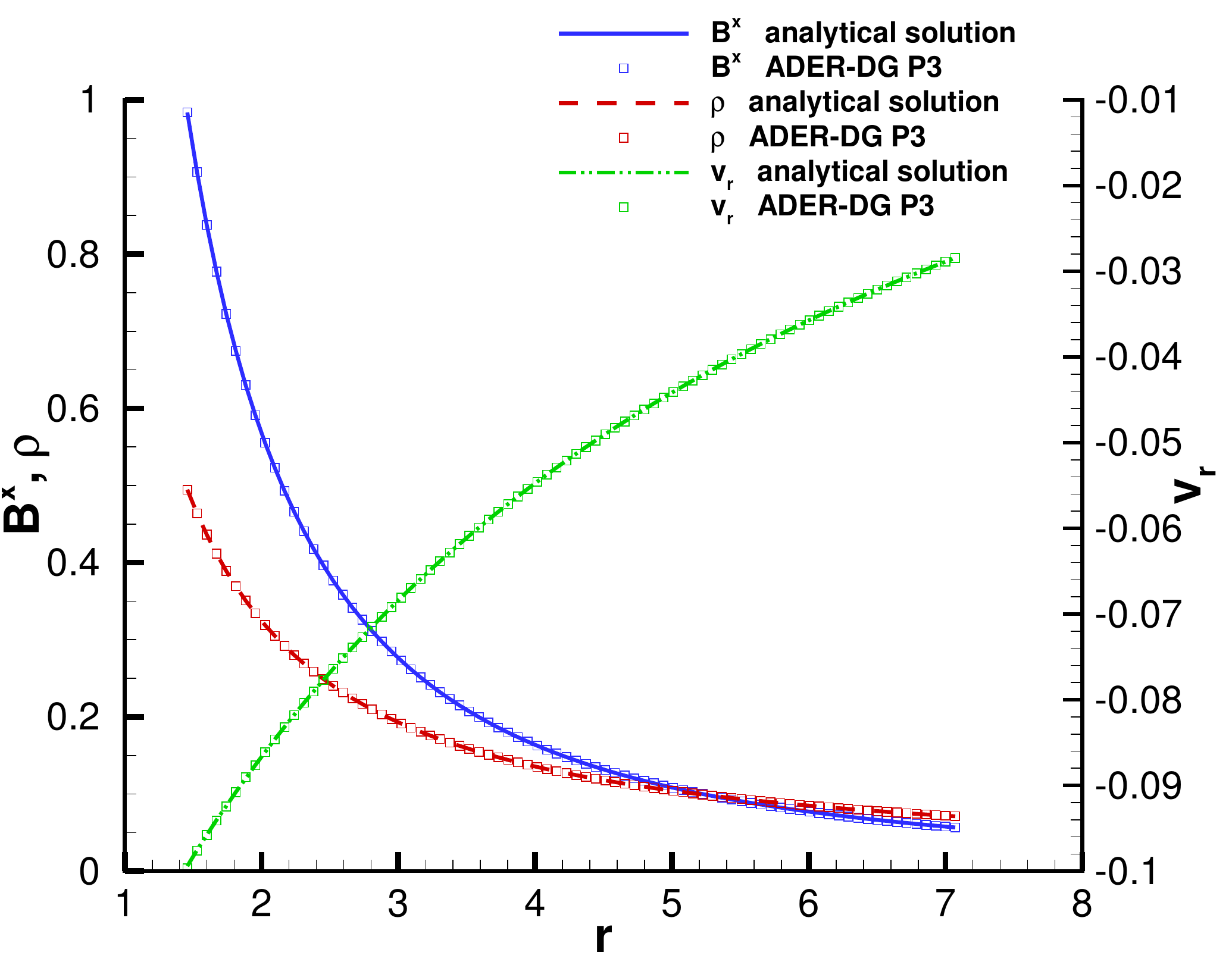}
    \caption{ \label{fig:Michel3D} Numerical solution for the 3D Michel
      accretion test with radial magnetic field in KSC coordinates
      obtained with our ADER-DG $\mathbb{P}_3$ at $t=20$. \textit{Left
        panel:} 3D visualization of the numerical solution and mesh: the
      space elements at $y<0$ are artificially blanked (not-visible), at
      $y>0$ are coloured by the rest-mass density. Moreover, the computed
      density is shown also along the 2D cut-plane $y=x\leq0$ together
      with the stream-traces of the magnetic field. \textit{Right panel:} numerical
      solution interpolated along $200$ points at $z=0$ and $y=x$ for the
      rest-mass density (red), the $x$ component of the velocity (green) and
      magnetic field (blue) vectors are plotted next to the analytical
      solution. The numerical domain is $\bm{x} \in \Omega = [-5,
        5]^3$. }
  \end{center}
\end{figure*}

The free parameters of the problem have been chosen to be a specific
angular momentum of $\ell_0 = 3.8$, a potential gap $\Delta W = -
10^{-3}$ (inside and nearly filling its {Roche lobe}). The polytropic
constant and exponent have been chosen equal to $K=0.0496$ and
$\gammap=4/3$, respectively.

Also in this case, for a rigorous testing of the convergence order we
have simulated only an inner portion of the torus which is fully filled
by fluid, namely, the one covered by the coordinate patch
$(r,\theta)\in\Omega=[7,10.5]\times[1.47,1.67]$. The corresponding
measured convergence order after evolving the set of the GRHD equations
in spherical KS coordinates are reported in Table \ref{tab:CTTorus}, once
again showing the expected high order of convergence of our ADER-DG
scheme. We conclude this test by remarking that torus simulations where
the torus is fully contained in the computational domain, which therefore
includes also a region set to {atmosphere}, will be presented in
Sec. \ref{sec:full_torus}.

\begin{table}
  \centering
  \begin{tabular}{ c c  cccc c }
    \hline
    & $N_x$ &  $\mathcal{E}_{L_2}$ & $\mathcal{E}_{L_\infty}$ & $L_2$ & $L_\infty$   &
    Exp. \\ 
    \hline
    \multirow{4}{*}{\rotatebox{90}{{DG-$\mathbb{P}_1$}}}
& 10	& 5.05E-07	& 2.67E-06	 &  ---	&  ---	  &\multirow{4}{*}{2}	 \\
& 20	& 1.58E-07	& 9.14E-07	 	& 1.68	& 1.55	 &\\
& 30	& 7.52E-08	& 4.34E-07	 	& 1.83	& 1.84	 &\\
& 40	& 4.40E-08	& 2.50E-07	 	& 1.86	& 1.92	 &\\
   \cline{2-6}
   \hline
	 \multirow{4}{*}{\rotatebox{90}{{DG-$\mathbb{P}_2$}}}
&10	&   5.29E-08	& 1.42E-07	 &  ---	&  ---	  &\multirow{4}{*}{3}\\
&15	&   1.81E-08	& 5.22E-08	 & 2.65	& 2.47	  &\\
&20	&   8.45E-09	& 2.35E-08	 & 2.64	& 2.78	  &\\
&30	&   2.83E-09	& 7.84E-09	 	& 2.70	  & 2.70  &\\
   \cline{2-6}
   \hline
	 \multirow{4}{*}{\rotatebox{90}{{DG-$\mathbb{P}_3$}}}
&8	&   3.65E-09	& 1.97E-08	 &  ---	&  ---	 &\multirow{4}{*}{4}\\
&10	&   1.60E-09	& 9.77E-09	 & 3.69	& 3.13	 & \\
&15	&   3.12E-10	& 2.10E-09	 & 4.04	& 3.79	 &\\
&20	&   1.08E-10	& 7.36E-10	 	& 3.69	 & 3.64 &\\
   \cline{2-6}
   \hline
	 \multirow{4}{*}{\rotatebox{90}{{DG-$\mathbb{P}_4$}}}
&2	&   1.03E-07	& 3.60E-07	 	&  ---	&  ---	 &\multirow{4}{*}{5}\\
&3	&   1.07E-08	& 3.96E-08	 	& 5.57	& 5.44	 &\\
&4	&   3.30E-09	& 1.15E-08	 	& 4.10	& 4.29	 &\\
&5	&   1.33E-09	& 5.85E-09	 	& 4.08	& 3.04	 &\\
   \cline{2-6}
   \hline	
  \end{tabular}
  \caption{ \label{tab:CTTorus} $L_2$ and $L_\infty$ errors and
    convergence rates for the 2D torus-interior problem in spherical Kerr-Schild
    coordinates for the ADER-DG-$\mathbb{P}_N$ scheme. We report the
    convergence results for the rest-mass density $\rho$ at $t=10$ up to
    $N=4$, and contrast the results with the expected rate. Similar
    results have also been obtained for all other flow variables. }
\end{table} 

\subsubsection{\textbf{3D Michel accretion with radial magnetic field}}
\label{sec:3D_Michel}

This is the 3D version of the similar test presented in
Sec. \ref{sec:2D_torus_interior}, with the addition of one spatial
dimension (corresponding to the azimuthal Killing vector) and of a
{radial magnetic field}. Although such a magnetic field is unphysical,
since it leads to a nonzero divergence and hence to the presence of a
{magnetic monopole}, it is nevertheless widely used for testing GRMHD
codes [see, \eg \cite{Etienne:2010ui}]. Here, we use it to test the
convergence order of our high-order method by considering also the
magnetic component of the set of partial differential equations. In
addition, to stress-test our numerical infrastructure, we have employed
for this test 3D Cartesian KS coordinates, so that the magnetic field
lines are not aligned with any of the coordinate axis.

The chosen contravariant components of the radial magnetic field
takes the form
\begin{align}
  B^i(\boldsymbol{x},t) = \gamma^{-\frac{1}{2}} M^2 B_0 \frac{x^i}{r^2}\,,
  \qquad B_0 =
  \frac{2.688}{M}\left(\frac{b^2}{\rho}\right)_{\text{hor}}^{\frac{1}{2}}\,,
\end{align}
where the black-hole mass is again set to $M=1$ and $b^\mu$ is the
magnetic field measured by the Lagrangian observer comoving with the
fluid, \ie
\begin{align}
b^\mu := \frac{ \left(\delta^\mu_\nu + u^\mu u_\nu\right) B^\nu}{ -n_\nu
  u^\nu}\,.
\end{align}

The spatial domain is in this case given by $(x,y,z)\in\Omega=[-5,+5]^3$
and is partitioned with a uniform mesh of $30^3$ elements, where we have
employed a very simple {cubic excision} to avoid the singularities at the
coordinates' origin location of the black hole as shown in the left panel
of Fig. \ref{fig:Michel3D}. At the excision boundary, we impose the exact 
solution of the problem as boundary condition in all variables. 

After adopting a ratio $(b^2/\rho)_{\text{hor}}=4$ at the horizon, the
results of the convergence study are presented in Table
\ref{tab:CTBondi3D}, while graphical representation of the numerical
results is offered in the right panel of Fig. \ref{fig:Michel3D}, which
reports the numerical solution interpolated along $200$ points at $z=0$
and $y=x$ for the rest-mass density and the $x$-component of the velocity
and of the magnetic field vectors as plotted against to the analytical
solutions. Clearly, also in this case the numerical solution is shown to
converge at the expected order of accuracy, confirming the validity of
our implementation in the presence of a magnetic field and of a
nontrivial coordinate mapping.

\begin{table}
  \centering
  \begin{tabular}{ ccccccc }
   \hline
   & $N_x$ &  $\mathcal{E}_{L_2}$ & $\mathcal{E}_{L_\infty}$ & $L_2$ & $L_\infty$   &
   Exp. \\ 
   \hline
   \multirow{4}{*}{\rotatebox{90}{{DG-$\mathbb{P}_1$}}}
& 10	&	6.588E-04	&	2.163E-04	  &---  &--- 	& \multirow{4}{*}{2}\\
& 20	&	1.69E-04	&	8.258E-05		&	1.96	&	1.39&\\
& 30	&	7.56E-05	&	4.193E-05		&	1.99	&	1.67&\\
& 40	&	4.26E-05	&	2.490E-05		&	2.00	&	1.81&	\\
   \cline{2-6}
   \hline
	 \multirow{4}{*}{\rotatebox{90}{{DG-$\mathbb{P}_2$}}}
&10&	3.75E-05	&	2.39E-05	&---	&--- 	&\multirow{4}{*}{3}\\
&15&	1.35E-05	&	7.40E-06		&	2.53	&	2.90&\\
&20&	6.62E-06	&	3.61E-06		&	2.47	&	2.49&\\
&30&	2.44E-06	&	1.35E-06		&	2.46	&	2.42 &\\
   \cline{2-6}
   \hline
	 \multirow{4}{*}{\rotatebox{90}{{DG-$\mathbb{P}_3$}}}
&8	&	1.73E-06	&	1.15E-06	&---	&--- 	&\multirow{4}{*}{4}\\
&10	&	6.38E-07	&	3.72E-07		&	4.47	&	5.05& \\
&15	&	1.10E-07	&	6.67E-08		&	4.35	&	4.24&\\
&20	&	3.25E-08	&	1.83E-08		&	4.22	&	4.50&\\
   \cline{2-6}
   \hline
	 \multirow{4}{*}{\rotatebox{90}{{DG-$\mathbb{P}_4$}}}
&6	&	4.45E-07	&	4.17E-07	&---	&--- 	&\multirow{4}{*}{5}\\
&8	&	1.04E-07	&	9.78E-08		&	5.05	&	5.04& \\
&12	&	1.35E-08	&	1.19E-08		&	5.03	&	5.19&\\
&16	&	3.20E-09	&	2.58E-09		&	5.01	&	5.33&\\
   \cline{2-6}
   \hline
	 \multirow{4}{*}{\rotatebox{90}{{DG-$\mathbb{P}_5$}}}
&4	&	1.90E-07	&	3.92E-07	&---	&---  &\multirow{4}{*}{6}\\
&6	&	1.32E-08	&	3.65E-08		&	6.57	&	5.85& \\
&8	&	2.37E-09	&	6.29E-09		&	5.98	&	6.12&\\
&10	&	6.42E-10	&	1.60E-09		&	5.85	&	6.14&\\ 
 \cline{2-6}
   \hline
	 \multirow{4}{*}{\rotatebox{90}{{DG-$\mathbb{P}_6$}}}
&6	&	1.26E-06	&	1.77E-06	&---	&--- 	&\multirow{4}{*}{7}\\
&8	&	1.72E-07	&	3.96E-07		&	6.93	&	5.20& \\
&10	&	4.13E-08	&	1.09E-07		&	6.39	&	5.78&\\
&12	&	1.34E-08	&	3.65E-08		&	6.18	&	5.99&\\
   \cline{2-6}
   \hline	
 \end{tabular}
 \caption{ \label{tab:CTBondi3D}$L_2$ and $L_\infty$ errors and
   convergence rates for the 3D Michel accretion with radial magnetic
   field in Cartesian Kerr-Schild coordinates for the
   ADER-DG-$\mathbb{P}_N$ scheme. We report the convergence results for
   the magnetic field component $B^x$ at $t=10$ up to $N=6$, and contrast
   the results with the expected rate. Similar results have also been
   obtained for all other flow variables. }
 \end{table} 

\subsection{Non-smooth special-relativistic flows}
\label{sec:discontinuous_sr}

\begin{table} 
\begin{center} 
\begin{tabular}{ccccccccc} %
\hline
    & $\rho$ &$v_x$&$v_y$&$v_z$ & $p$ & $B^x$&$B^y$&$B^z$ \\
\hline
\hline
RP1, $x > 0$    & 0.125 & 0.0 & 0.0 &0.0 & 0.1 & 0.5&-1.0&0.0 \\
RP1, $x \leq 0$ & 1.0   & 0.0 & 0.0 &0.0 & 1.0 & 0.5& 1.0&0.0 \\ 
\hline
\hline
RP2, $x > 0$ & 1.0      &  -0.45 & -0.2 & 0.2 & 1.0  & 2.0&-0.7&0.5 \\
RP2, $x \leq 0$ & 1.08  &   0.40 &  0.3 & 0.2 & 0.95 & 2.0& 0.3&0.3 \\ 
\hline
\end{tabular} 
\caption{
\label{tab:RP1D}
Initial conditions of the MHD variables for the Riemann problems.}
\end{center}
\end{table} 

The tests considered in this section are considerably different from
those discussed so far in that they do not involve smooth flows and allow
therefore for the presence of nonlinear waves, either in the form of
shocks or of steep gradients as those present at the fluid interface with
an atmosphere.

\subsubsection{\textbf{Riemann problems}}

We start by considering two standard Riemann (or shock-tube) problems,
here referred to respectively as RP1 and RP2, and originally proposed in
the context of {special} relativistic MHD by \cite{BalsaraRMHD}. Although
these tests are solved on flat spatial hypersurfaces, \ie
$\gamma_{ij}=\delta_{ij}$, where $\delta_{ij}$ is the identity
three-matrix, they employ different setups for the gauge variables, the
lapse function and the shift vector. In particular, Table \ref{tab:RP1D}
provides all the considered initial conditions for the MHD variables of
RP1 and RP2, while the lapse, the $x$-component of the shift and the
final time are chosen to be $(\alpha,\beta_x, t_{\rm final})=\{0.5, 0.0,
0.8\},\, \{1.0, 0.0,0.4\},\, \{1.0, 0.4, 0.16\},\, \{2.0, 0.0,0.2\}$. The
adiabatic index for RP1 and RP2 has been set to be $\Gamma=2$ and 
$\Gamma=5/3$, respectively.

\begin{figure*}
  \begin{center} 
    \includegraphics[width=0.49\textwidth]{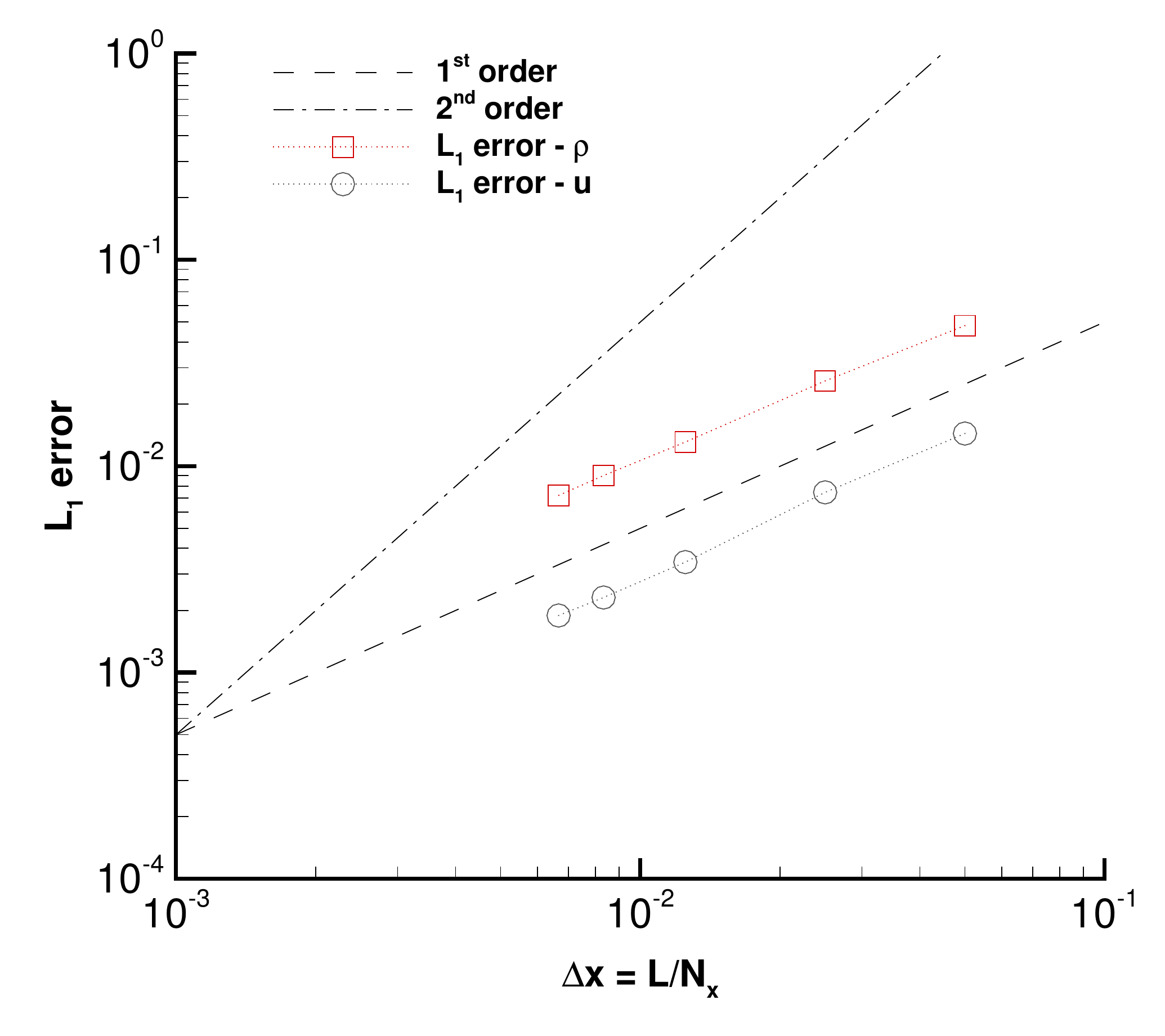} 
    \caption{ \label{fig:CTRP} Convergence study against Riemann problem
      RP2 of table \ref{tab:RP1D}. $L_1$ errors are plotted against the
      discretization step $\Delta x=L/N_x$, with $L=1$ being the length
      of the one-dimensional domain, $N_x$ the discretization number, \ie
      the number of high-order space-elements in the $x$-direction. These
      tests have been performed with the fourth-order accurate ADER-DG-P3
      scheme supplemented by our second-order subcell finite-volume
      limiter.}
  \end{center}
\end{figure*}

\begin{figure*}
  \begin{center} 
    \includegraphics[width=0.49\textwidth]{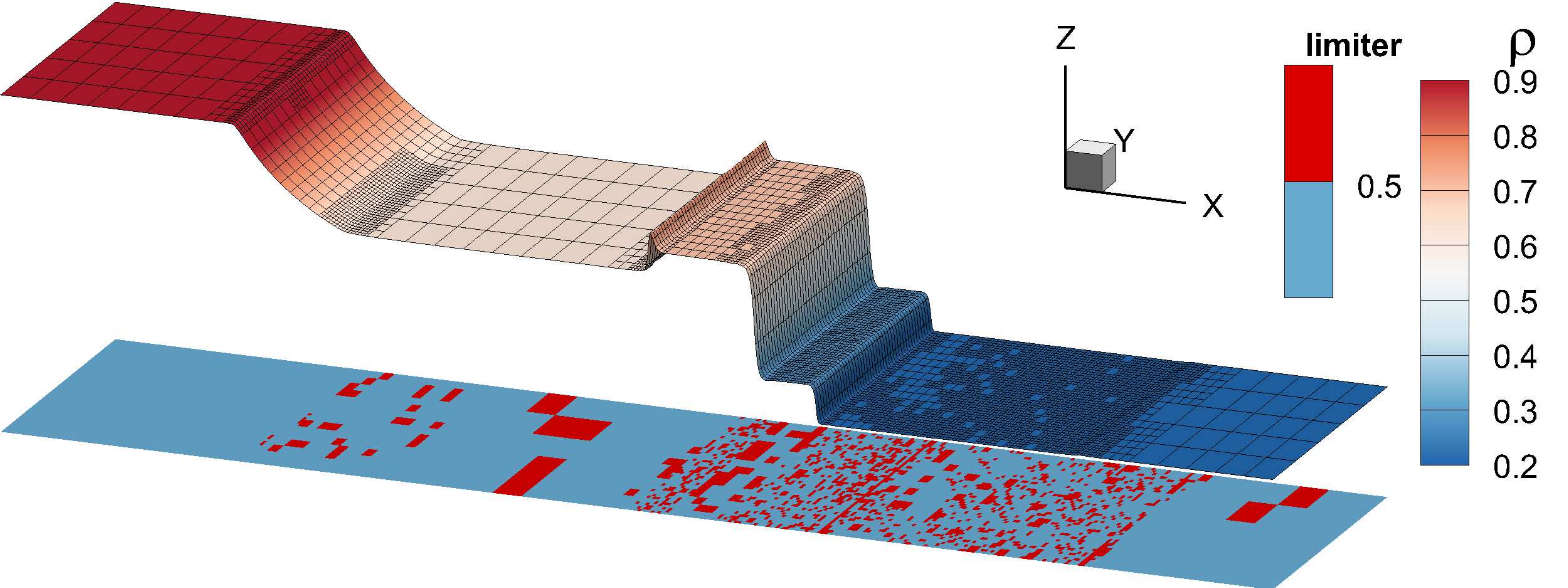} 
    \includegraphics[width=0.49\textwidth]{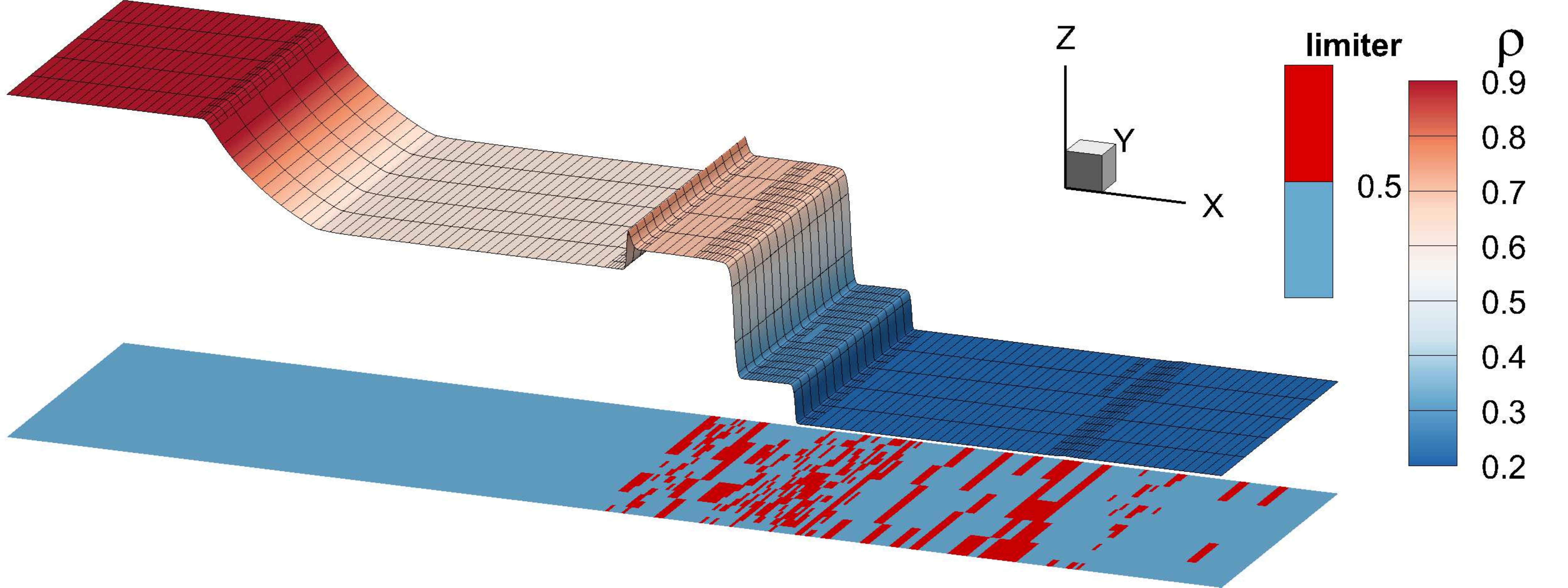} 
    \includegraphics[width=0.49\textwidth]{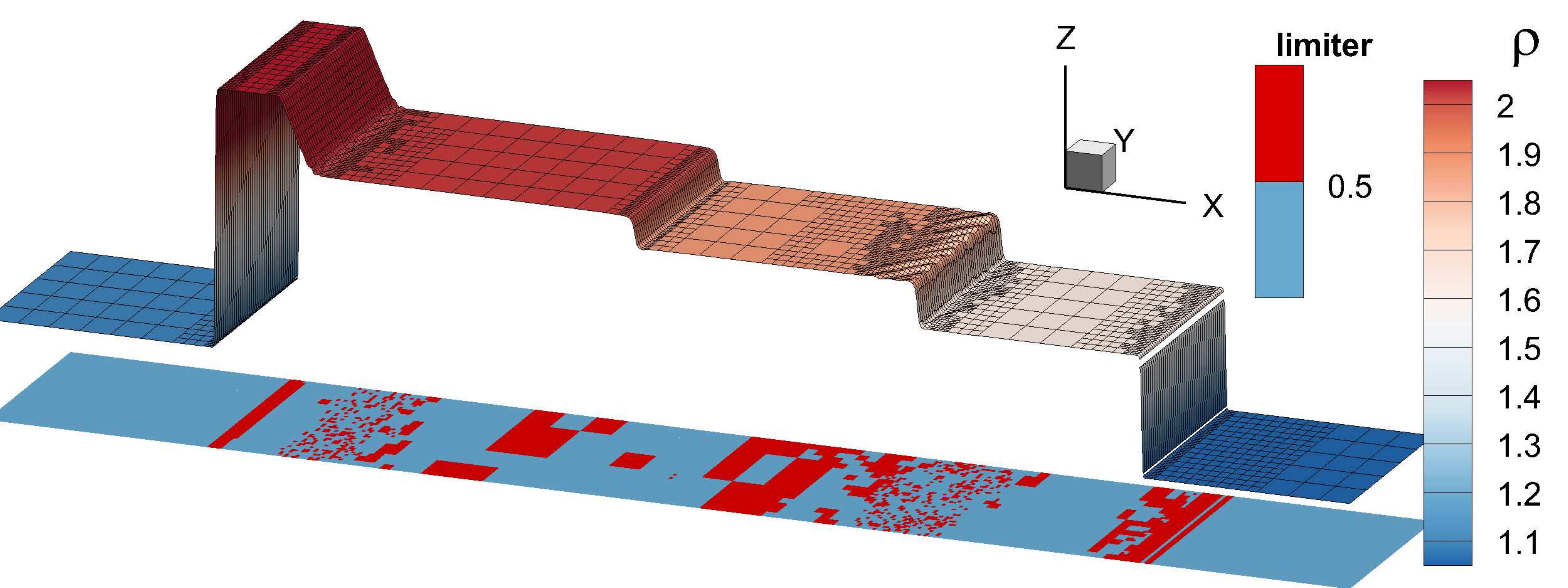}
    \includegraphics[width=0.49\textwidth]{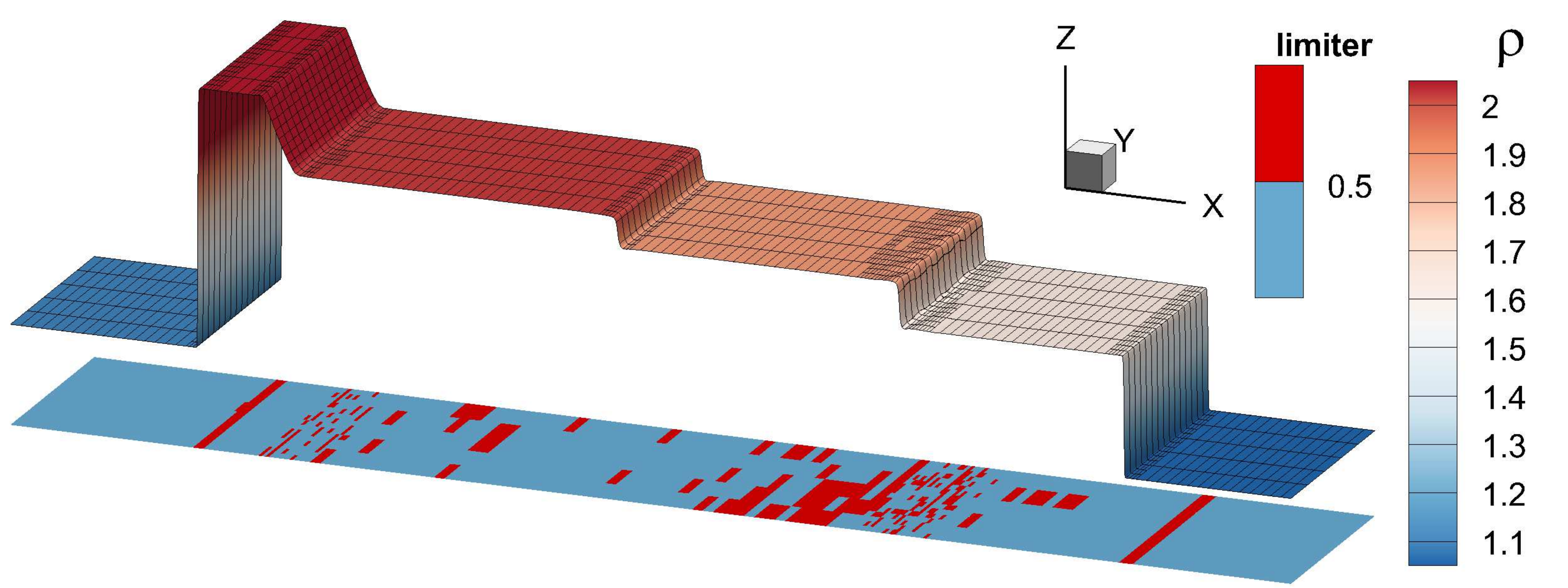}
    \caption{ \label{fig:Balsara3D} 3D view of the rest-mass density, the
      corresponding AMR grid and, on the horizontal plane, the
      corresponding limiting status, obtained with our ADER-DG
      $\mathbb{P}_N$ with finite-volume subcell limiting. From the top
      panel to the bottom, from left to right: i) RP1 at $t_{\text{final}}=0.2$ with
      $\alpha=2$, $\mathbb{P}_3$, with a coarsest grid of $40\times5$
      elements, $\ell_{\text{max}} = 2$; ii) RP1 at
      $t_{\text{final}}=0.2$ with $\alpha=2$, $\mathbb{P}_5$, with a
      coarsest grid of $120\times5$ elements, $\ell_{\text{max}} = 1$;
      iii) RP2 at $t_{\text{final}}=0.275$ with $\alpha=2$,
      $\mathbb{P}_3$, with a coarsest grid of $40\times5$ elements,
      $\ell_{\text{max}} = 2$; iv) RP2 at $t_{\text{final}}=0.275$ with
      $\alpha=2$, $\mathbb{P}_5$, with a coarsest grid of $120\times5$
      elements, $\ell_{\text{max}} = 1$. The limited cells, using the
      subcell ADER-TVD finite-volume scheme, are highlighted in red along
      the horizontal plane below the 3D plot of the rest-mass density
      $\rho$, while unlimited DG-$\mathbb{P}_N$ cells are highlighted in
      blue.}
  \end{center}
\end{figure*}

\begin{figure*}
  \begin{center} 
    \includegraphics[width=0.4\textwidth]{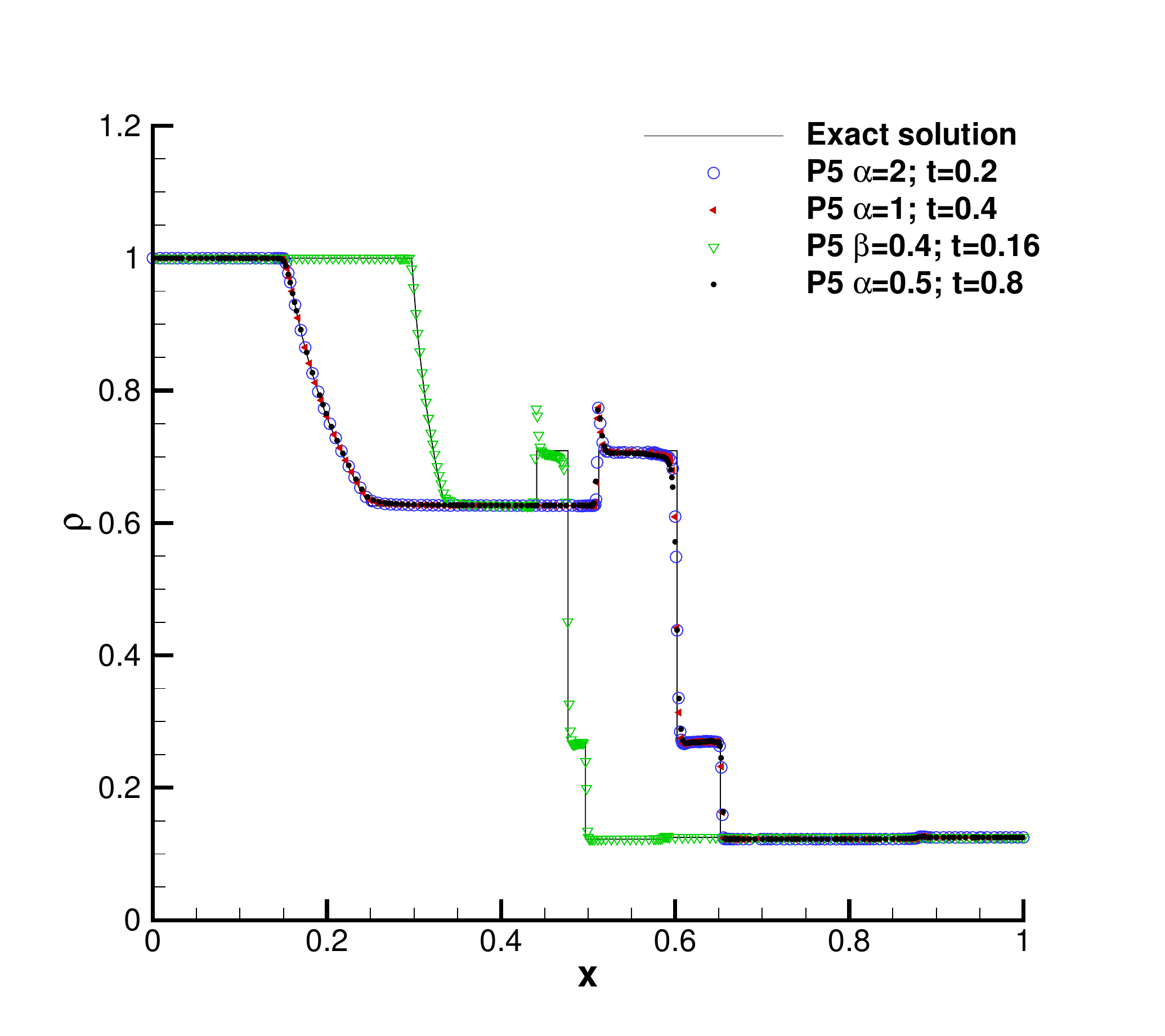}
    \includegraphics[width=0.4\textwidth]{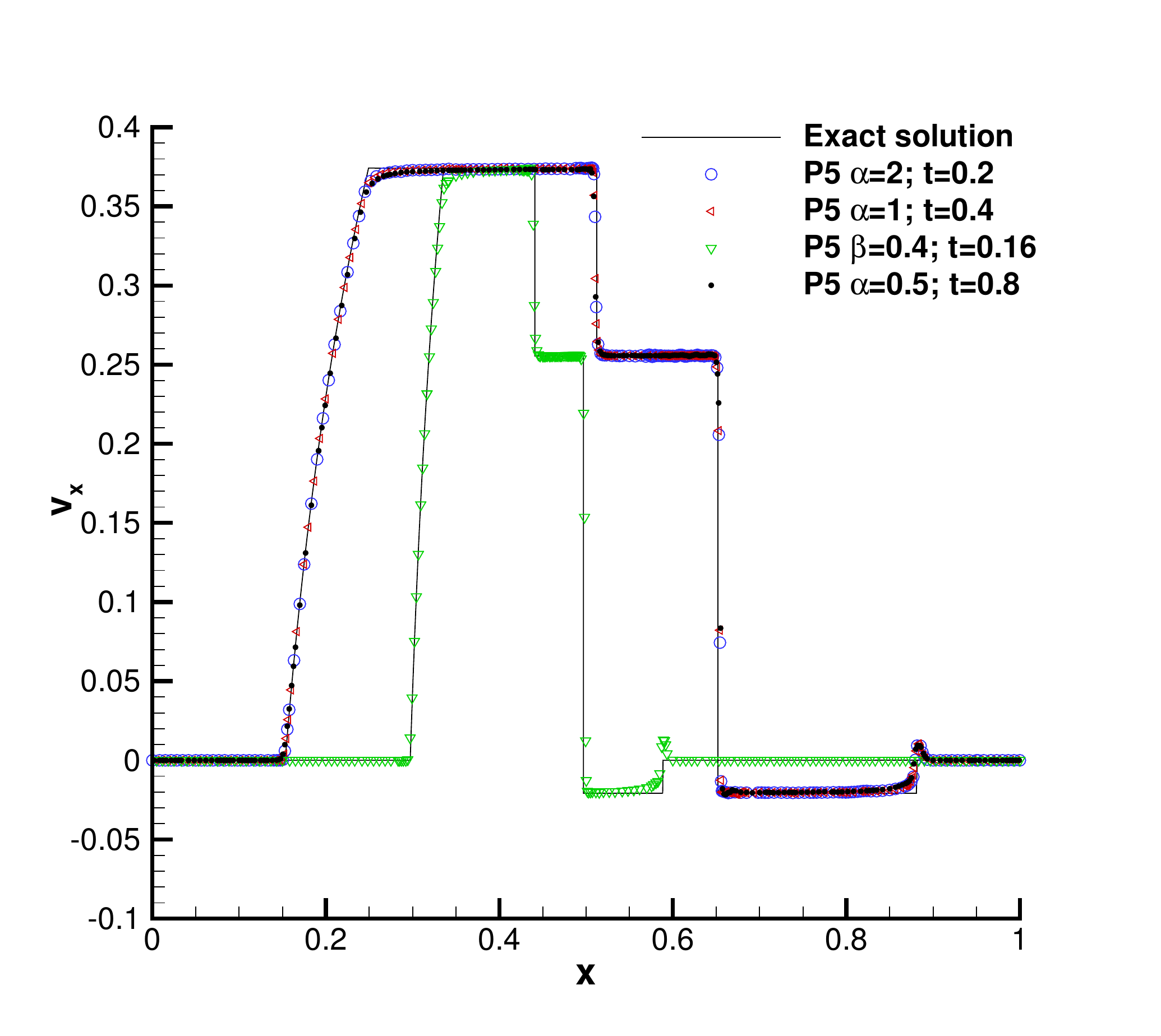}
    \includegraphics[width=0.4\textwidth]{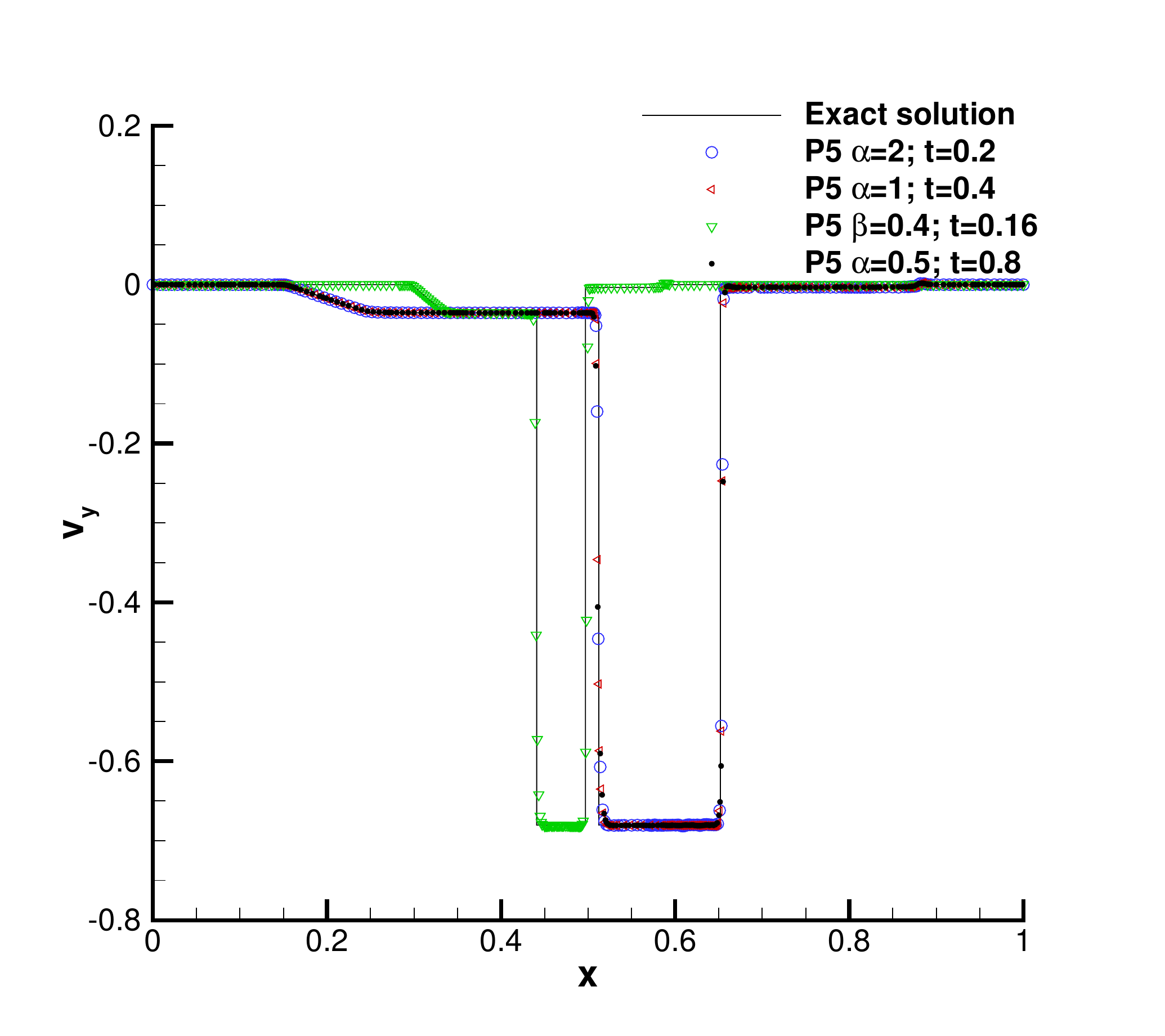} 
    \includegraphics[width=0.4\textwidth]{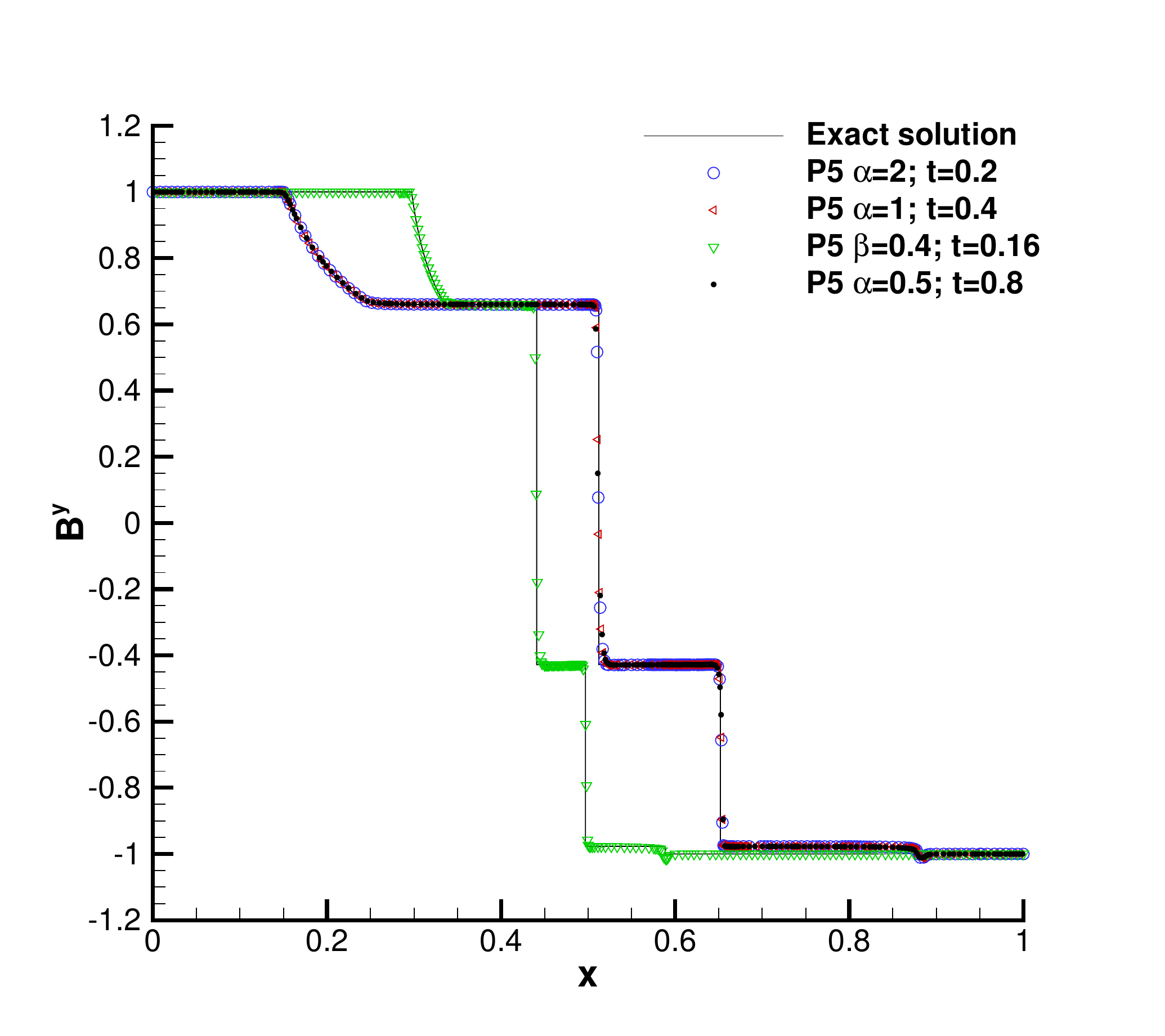} 
   \caption{ \label{fig:Balsara1P5} Riemann Problem 1 (RP1): the
     different panels show the various physical variables interpolated
     along a 1D cut, starting from a coarsest grid of $120\times5$
     elements by using the ADER-DG-$\mathbb{P}_5$ scheme supplemented with
     the \aposteriori ADER-TVD subcell and one single refinement level
     $\ell_{\text{max}} = 1$. Shown with solid lines are the
     corresponding solutions from an exact Riemann solver.}
  \end{center}
\end{figure*}

For these tests, the HLL approximate Riemann solver has been used. Figure
\ref{fig:Balsara3D} offers a 3D view of the rest-mass density variable
for the proposed shock-tube problems and the corresponding AMR grid and
limiting status, for the case $\lapse=2$, obtained with our
ADER-DG-$\mathbb{P}_3$ scheme using a level-zero mesh of $40\times5$
space-elements onto with $\ell_{\text{max}}=2$ maximum refinement levels
are added, and an ADER-DG-$\mathbb{P}_5$ scheme on a level-zero grid of
$120\times5$ elements with one single refinement level
$\ell_{\max}=1$. The corresponding one-dimensional (1D) cuts relative to
the $\mathbb{P}_5$ solutions are presented instead in
Fig. \ref{fig:Balsara1P5} relatively to the test configurations listed in
Table \ref{tab:RP1D}; shown with solid lines are the corresponding
solutions from the exact Riemann solver of \cite{Giacomazzo:2005jy}.
{
In the presence of moving discontinuities, the expected order of convergence 
of any shock capturing method is at most one. In Fig. \ref{fig:CTRP} 
we show the results of a numerical convergence study for RP2, indicating 
that the numerical method converges indeed with the expected order of one 
for flows with shocks and discontinuities. 
}

Overall, the results of these tests confirm the high-resolution
shock-capturing capability, but also the robustness, of the new class of
ADER-DG $\mathbb{P}_N$ schemes. In addition, they show that the
{a-posteriori} finite-volume {sub-grid} limiter is activated only in very
small portions of the domain and, in the case of genuine shocks, it is
very narrowly concentrated near the discontinuity.


\begin{figure*}
\begin{center}
  \includegraphics[width=0.45\textwidth]{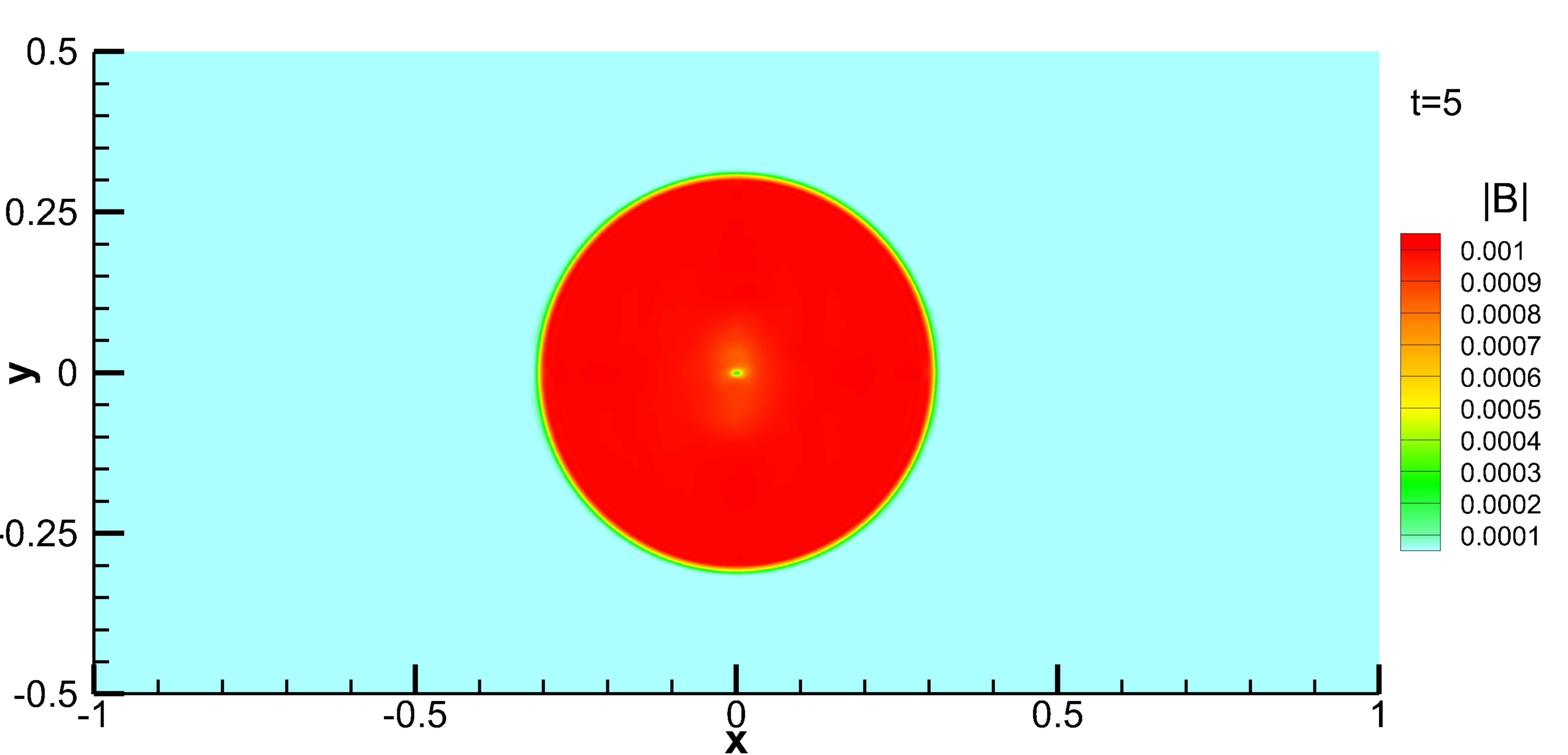}    
  \includegraphics[width=0.45\textwidth]{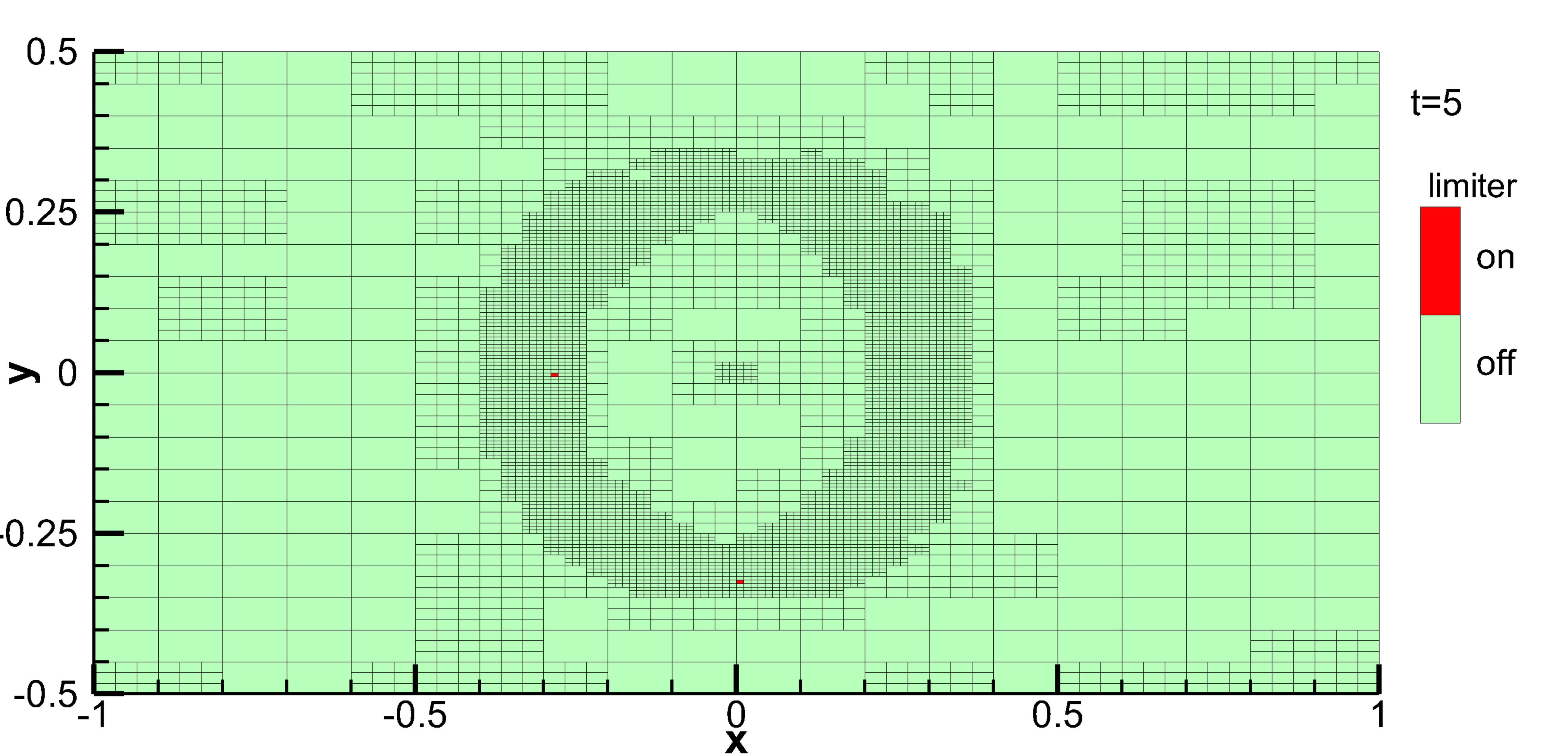}   \\  
  \includegraphics[width=0.45\textwidth]{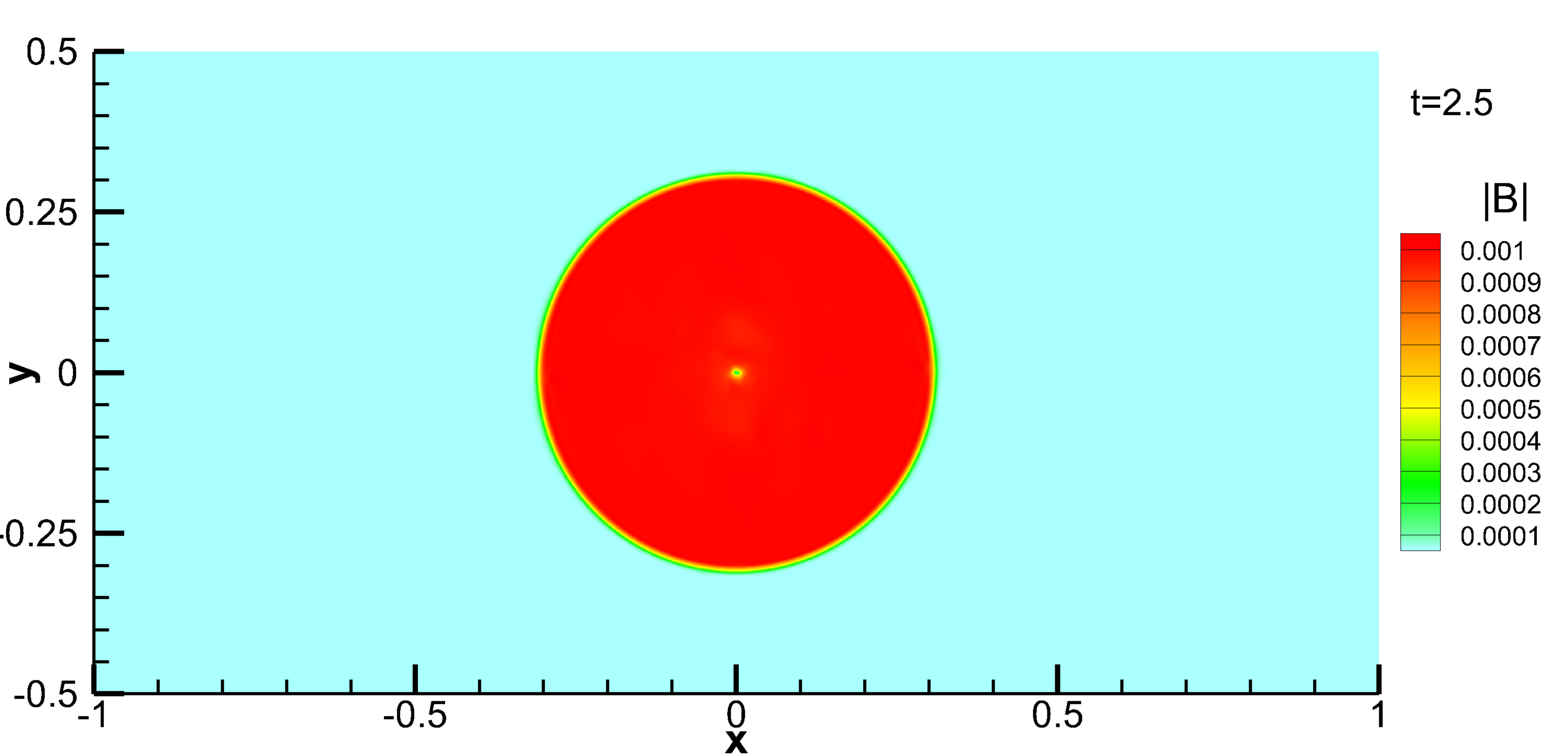}  
  \includegraphics[width=0.45\textwidth]{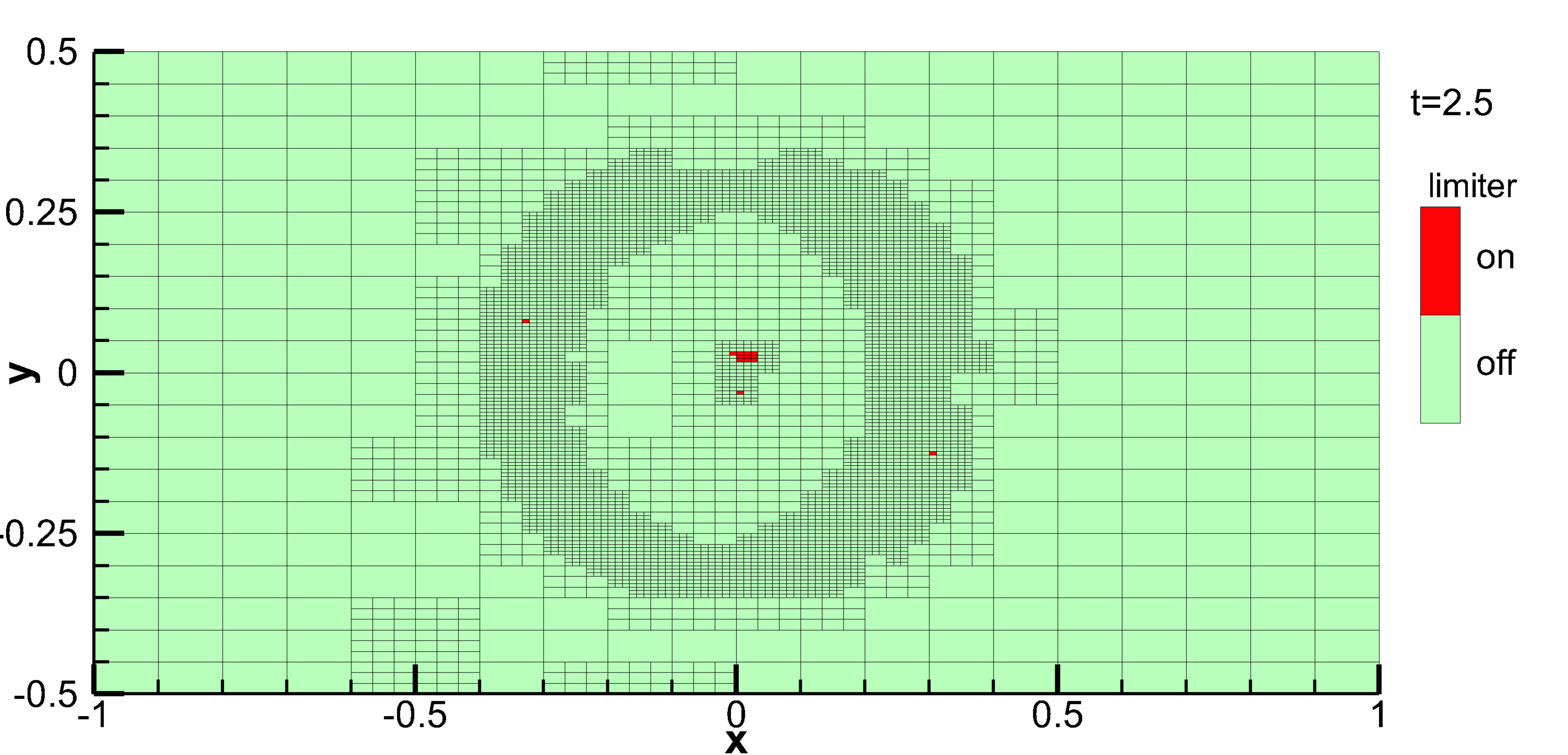}    
  \caption{Advected magnetic field loop problem (SRMHD) obtained with the ADER-DG-$\mathbb{P}_4$ 
  	scheme supplemented 
    with the \aposteriori TVD subcell limiter. 
    Left panels: $\mathbb{P}_4$-solution obtained for the magnetic
    field magnitude; right panels: AMR-grid, troubled cells (red) and
    unlimited cells (green). Solution at time $t=5.0$ with lapse function
    $\alpha=1.0$ (top row) and at time $t=2.5$ with lapse function
    $\alpha=2$ (bottom row).}
\label{fig:ML_B}
\end{center}
\end{figure*}
\begin{figure*}
\begin{center}
  \includegraphics[width=0.45\textwidth]{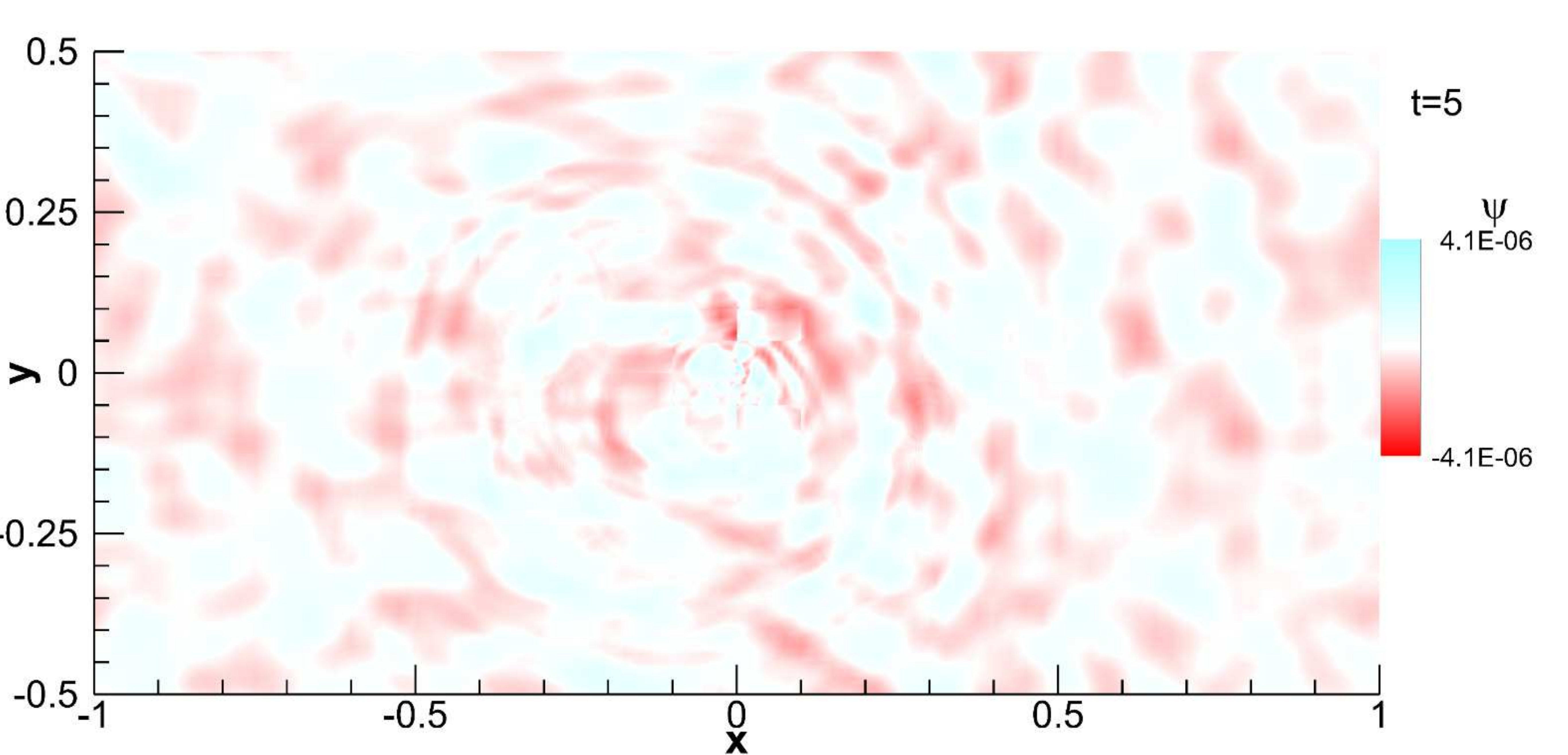}    
  \includegraphics[width=0.45\textwidth]{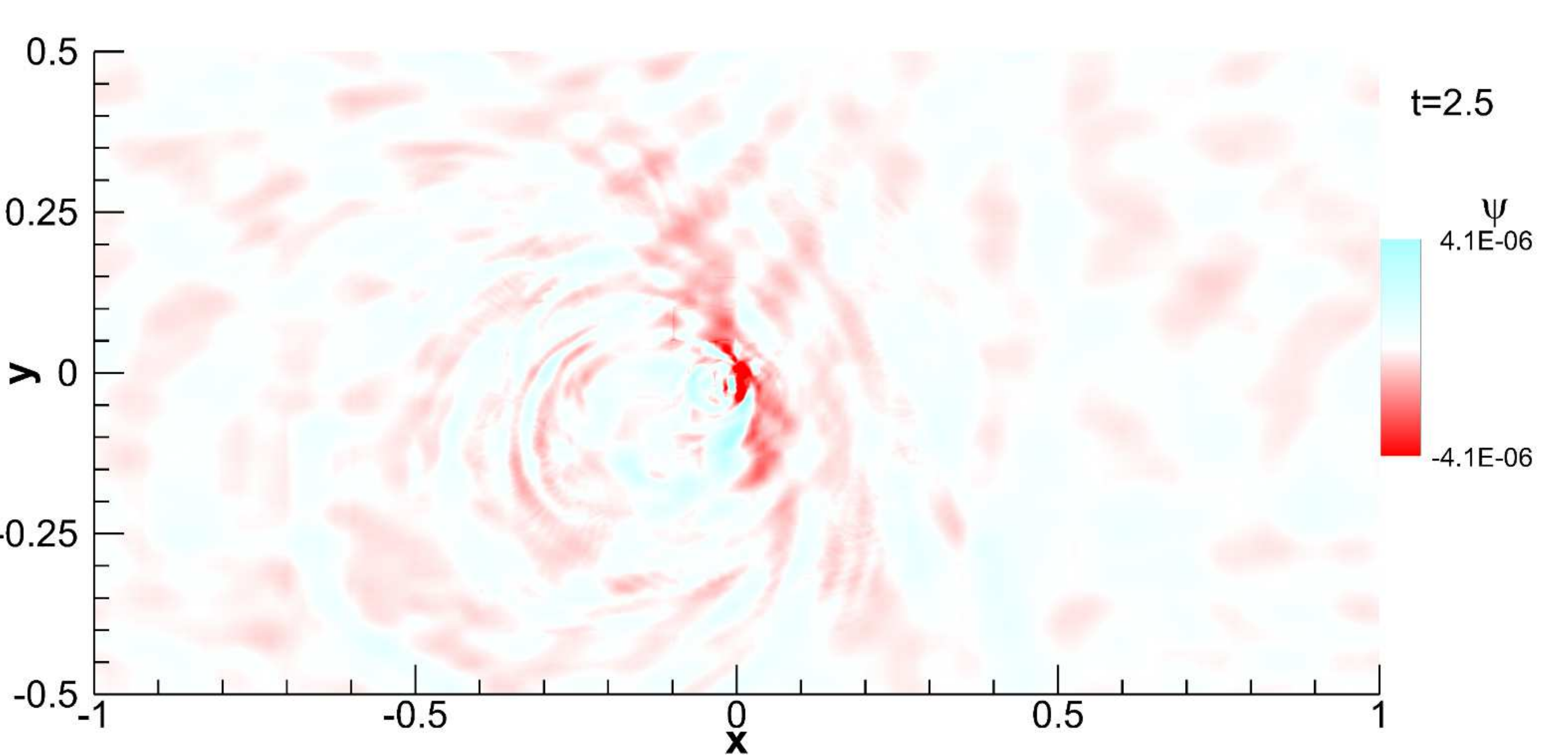}      
  \caption{Advected magnetic field loop problem (SRMHD) obtained with the ADER-DG-$\mathbb{P}_4$ 
  	scheme supplemented 
    with the \aposteriori TVD subcell limiter: results for the divergence cleaning scalar $\psi$ at time $t=5.0$ with lapse function
    $\alpha=1.0$ (left) and at time $t=2.5$ with lapse function $\alpha=2$ (right).}
\label{fig:ML_B_psi}
\end{center}
\end{figure*}
\begin{figure*}
\begin{center}
\includegraphics[width=0.33\textwidth]{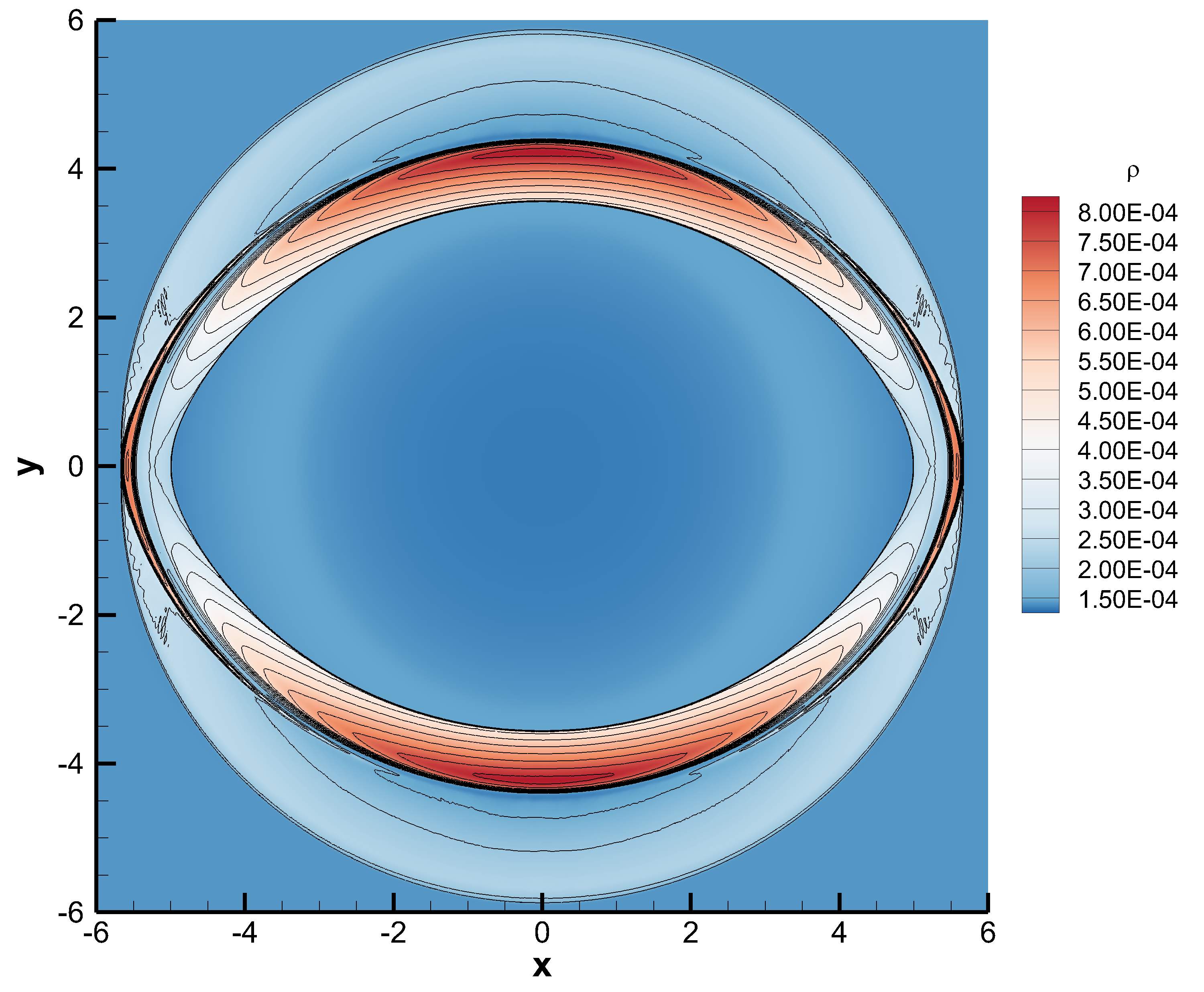}     
\includegraphics[width=0.33\textwidth]{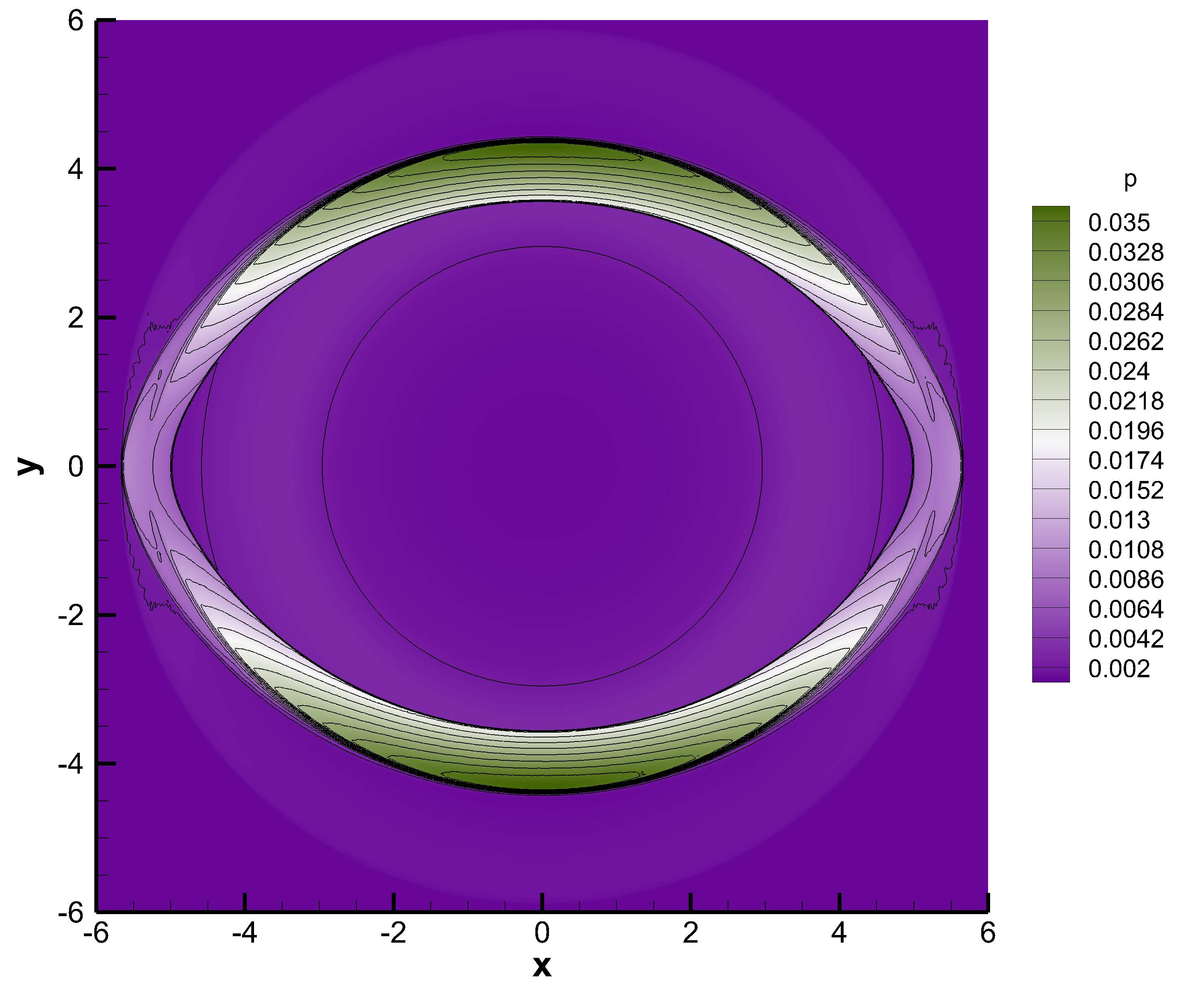}    
\includegraphics[width=0.33\textwidth]{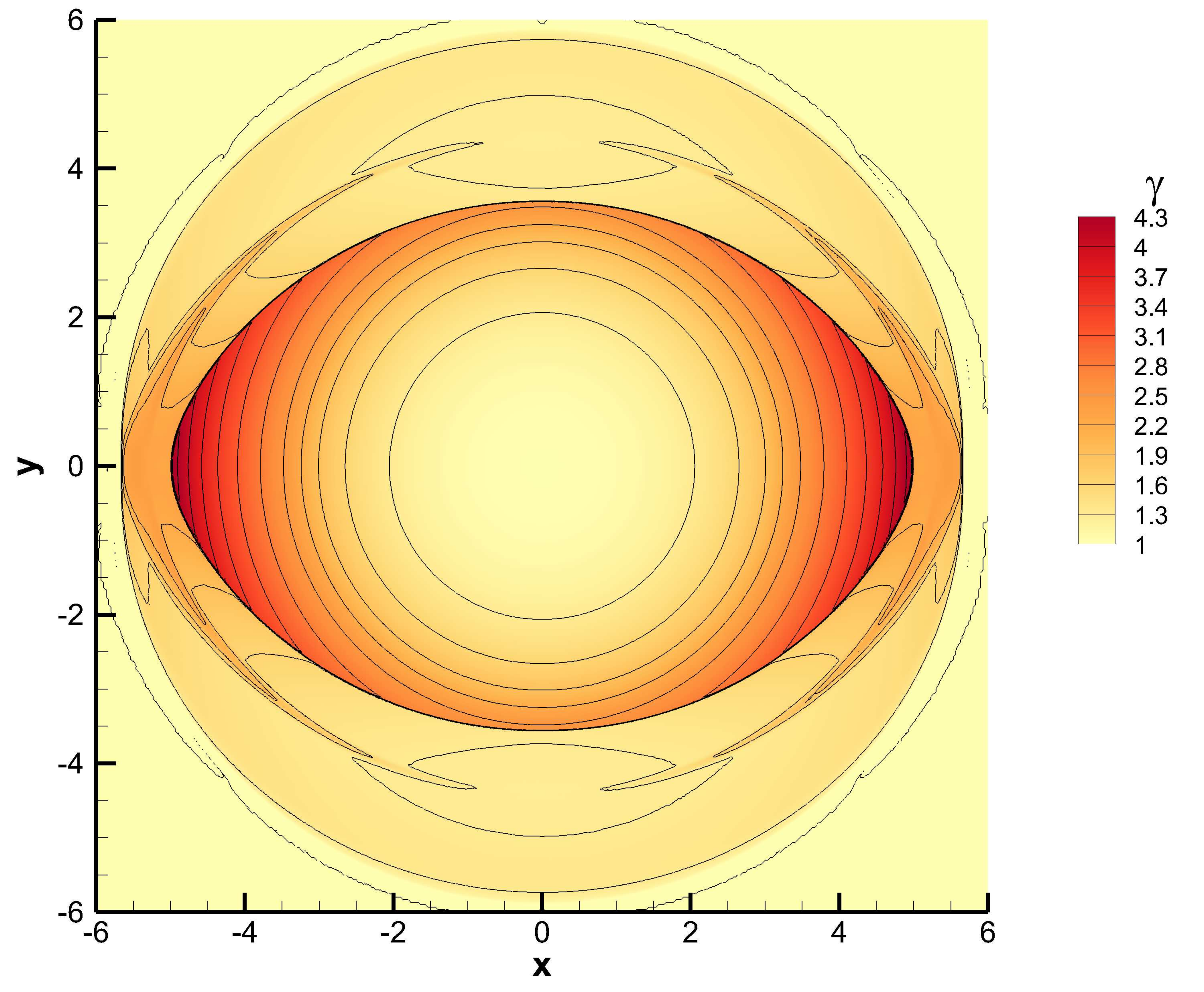}   \\  
\includegraphics[width=0.33\textwidth]{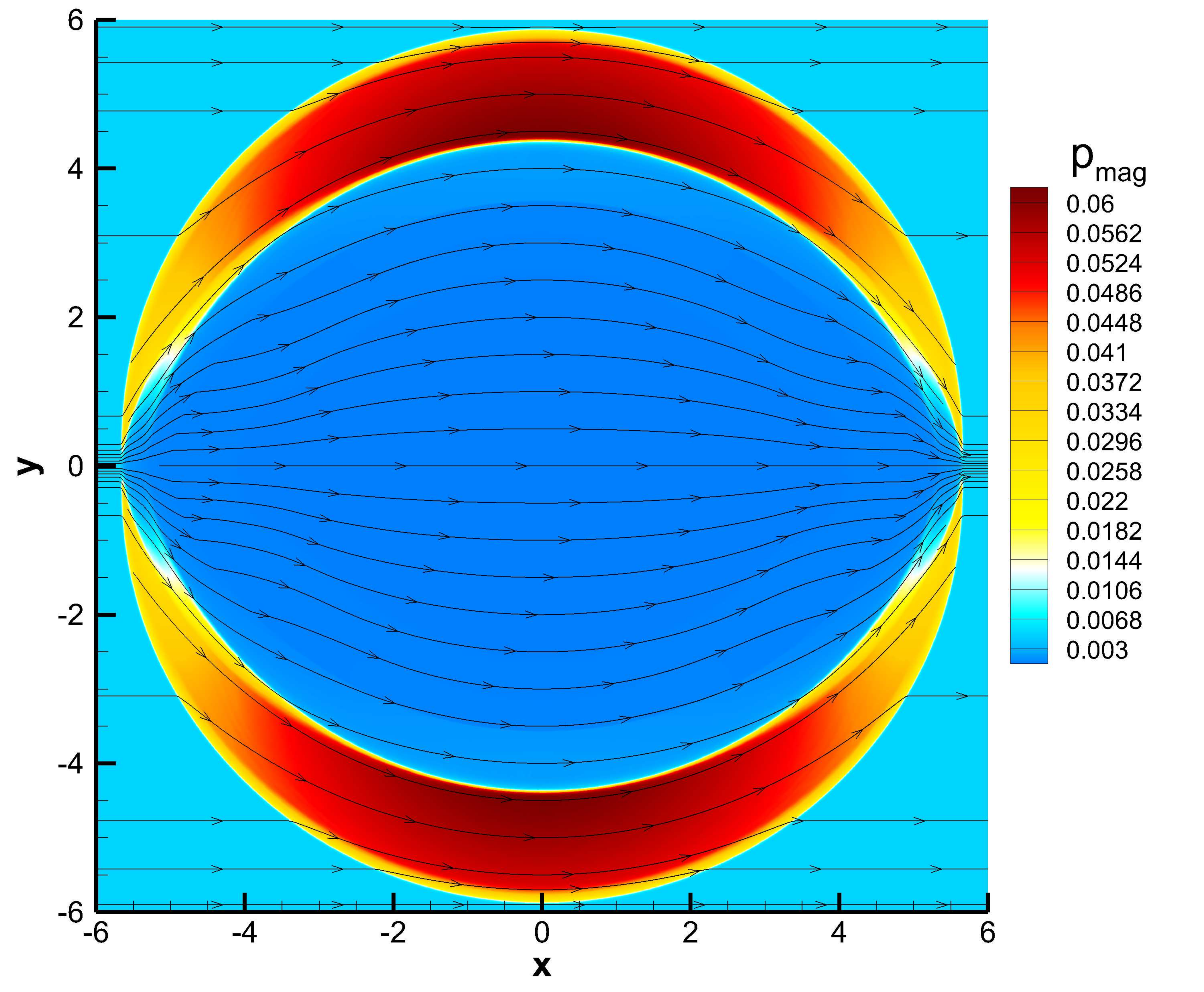}    
\includegraphics[width=0.33\textwidth]{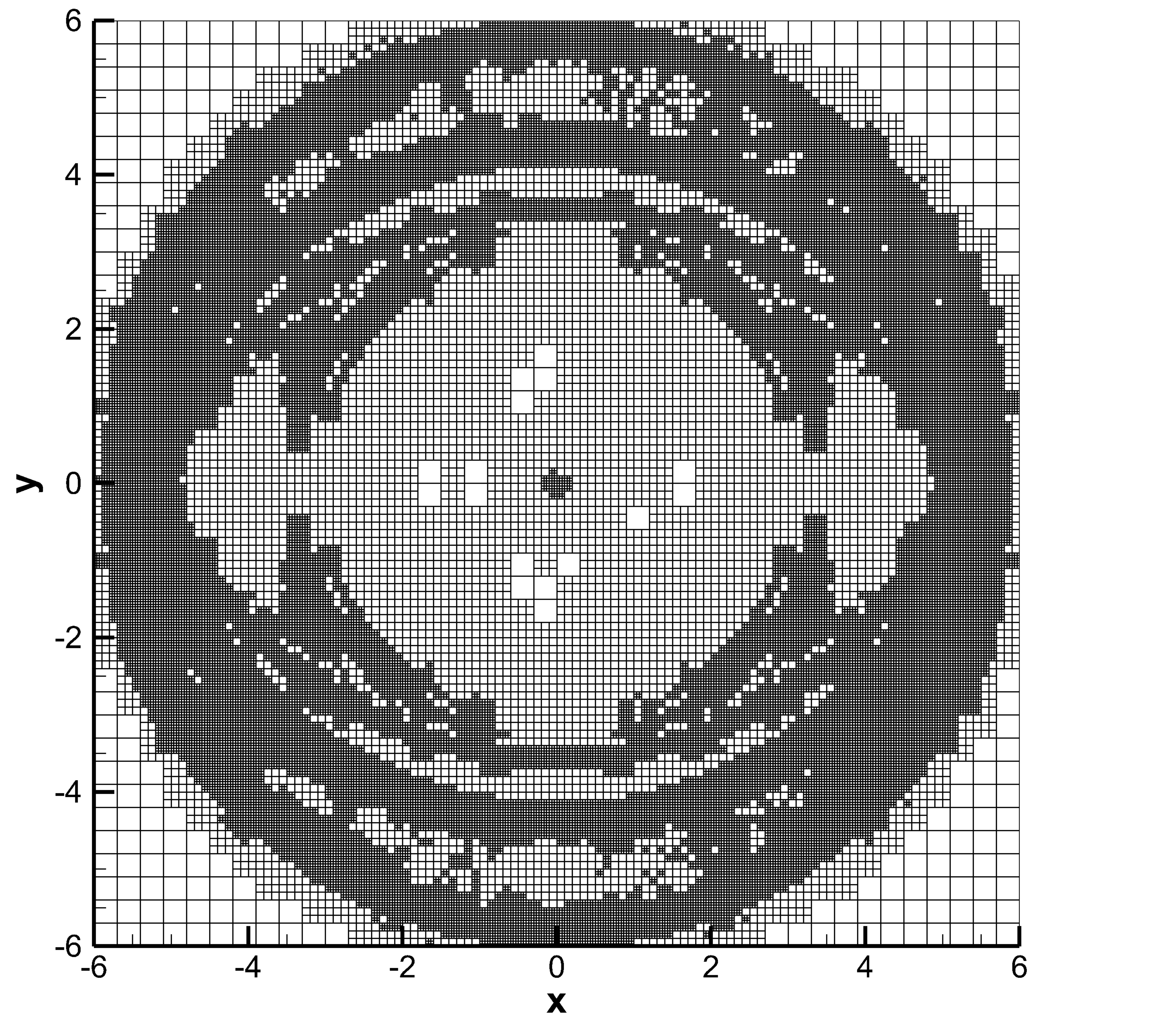}     
\includegraphics[width=0.33\textwidth]{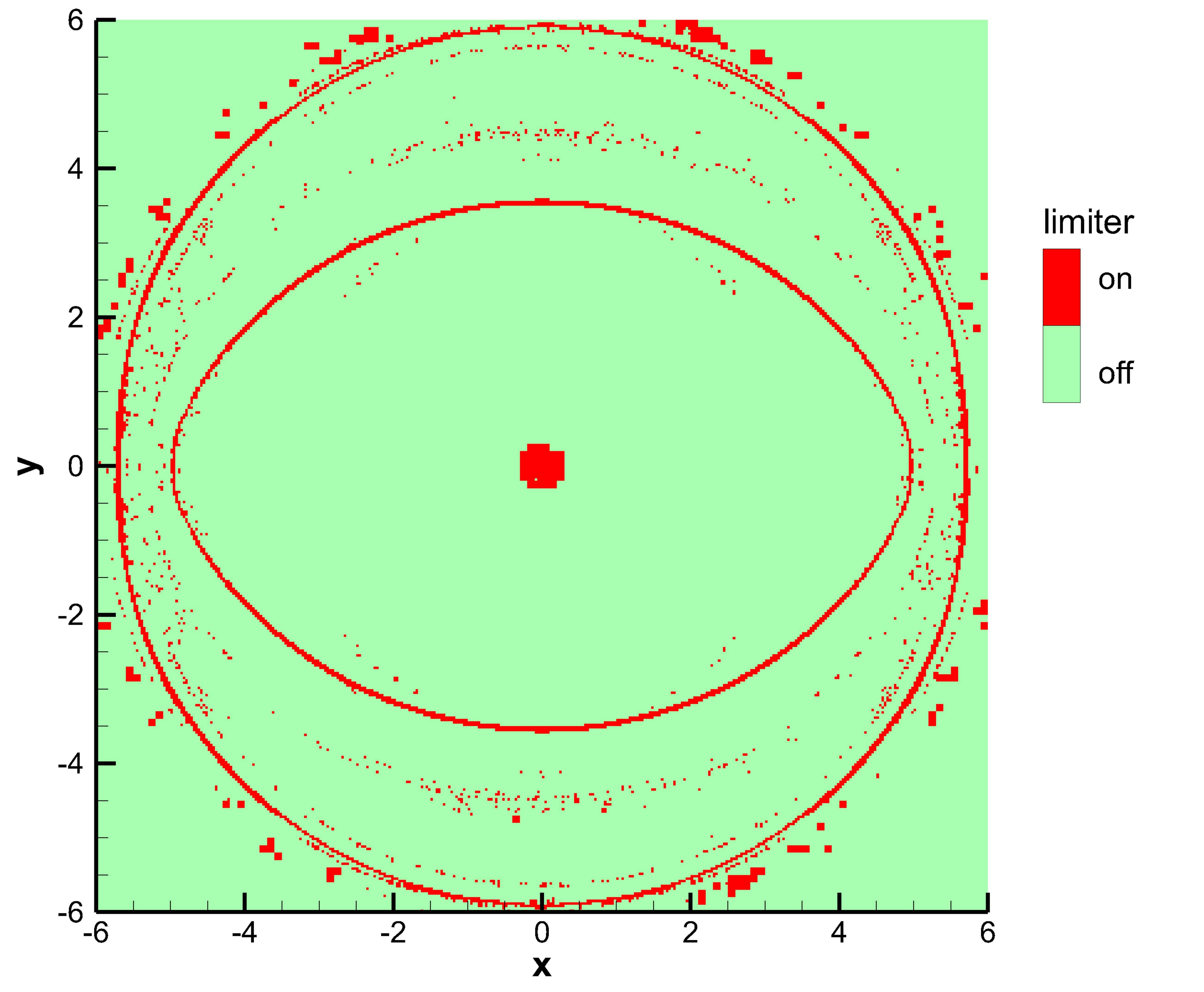}   
\caption{Solution of the SRMHD blast wave with
  $B_x=0.1$ at time $t=4.0$, obtained with the ADER-DG $\mathbb{P}_3$
  scheme supplemented with the \aposteriori second-order TVD subcell
  limiter. 
	Top panels: rest-mass density (left), thermal pressure
  (center) and Lorentz factor (right).
	Bottom panels: magnetic pressure (left) with magnetic field lines reported, AMR grid  (center) and limiter map (right) with troubled cells marked in red and
  regular unlimited cells marked in green.}
\label{fig:BW_01}
\end{center}
\end{figure*}
\begin{figure*}
\begin{center}
  \includegraphics[width=0.33\textwidth]{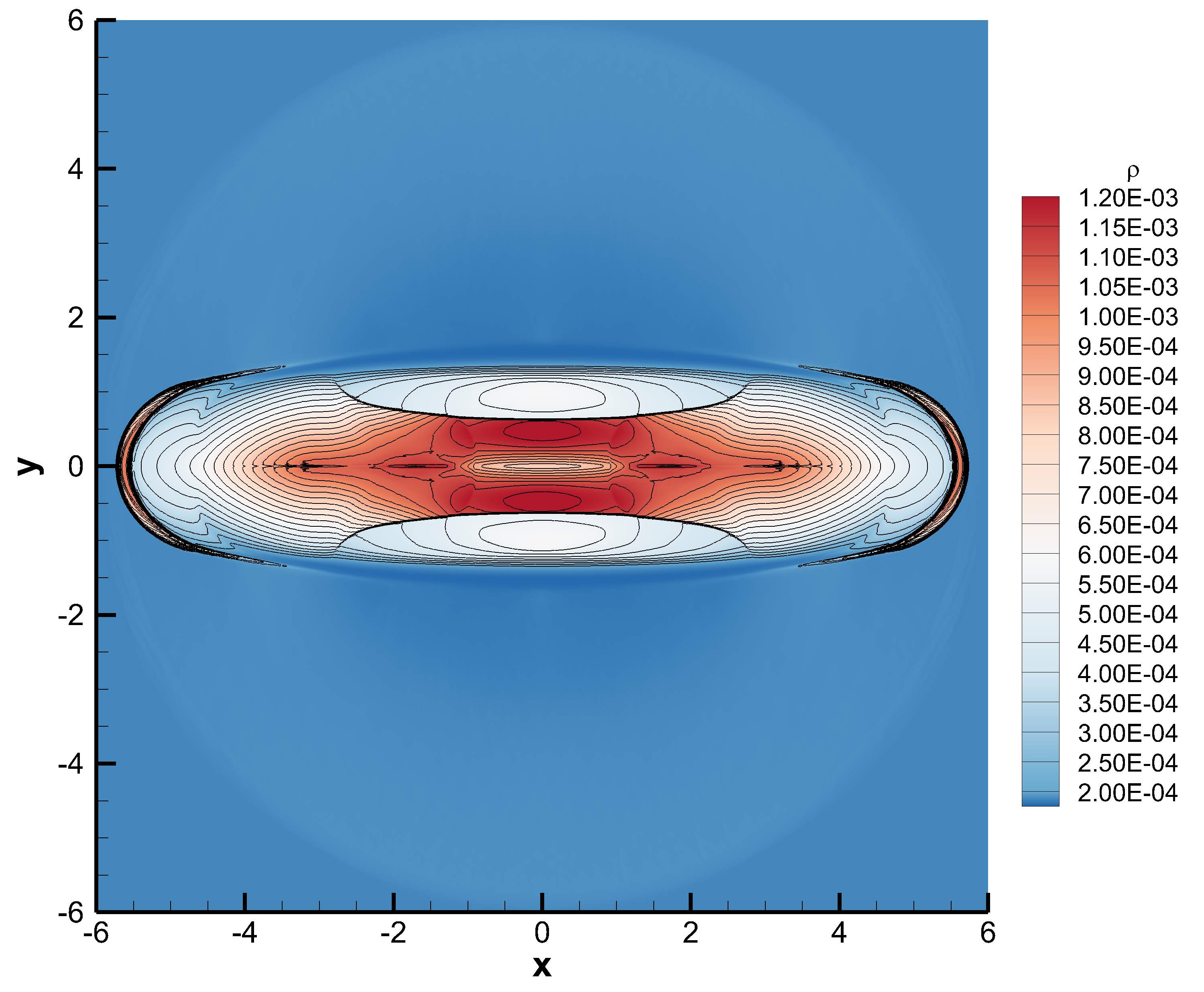}     
  \includegraphics[width=0.33\textwidth]{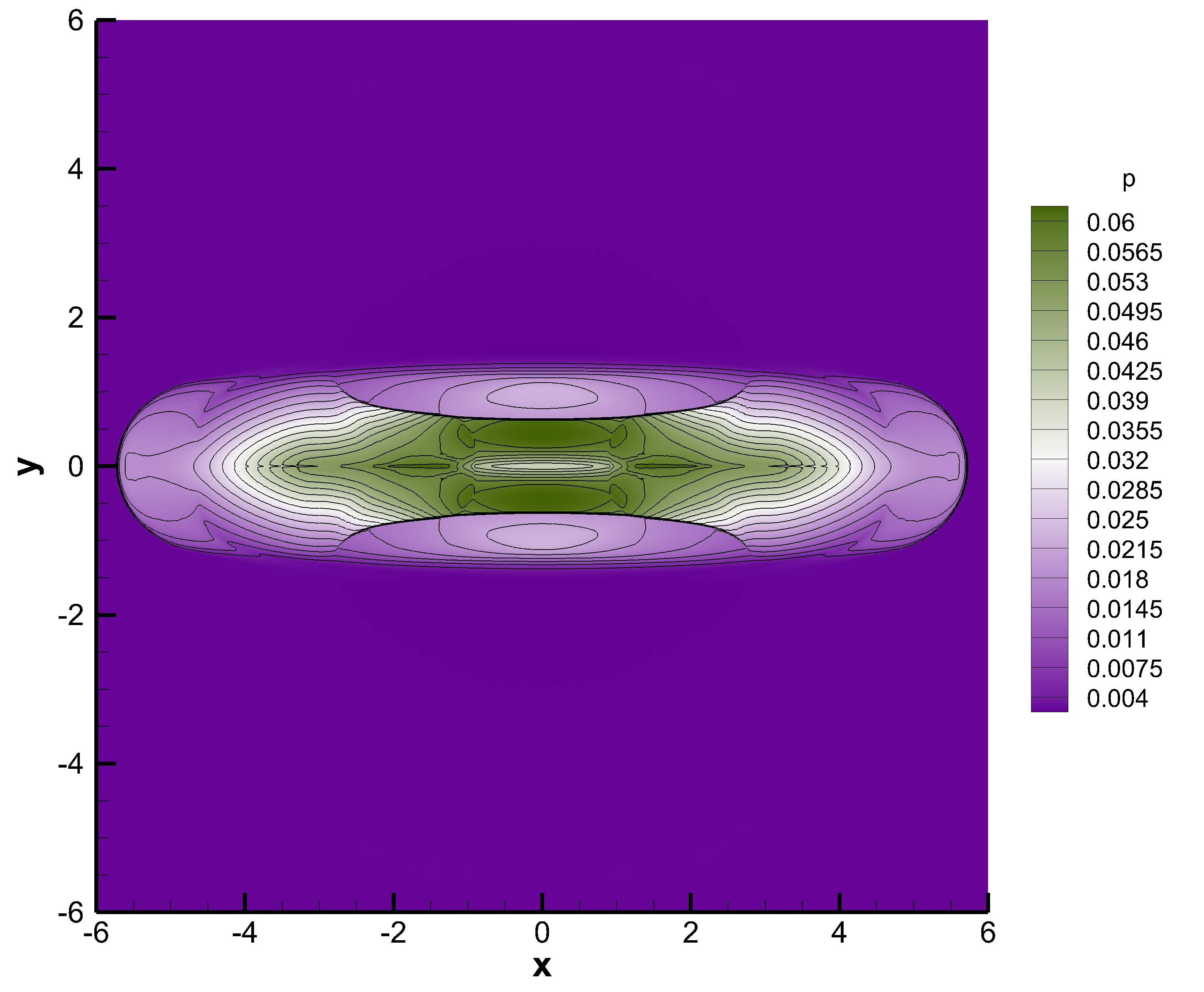}    
  \includegraphics[width=0.33\textwidth]{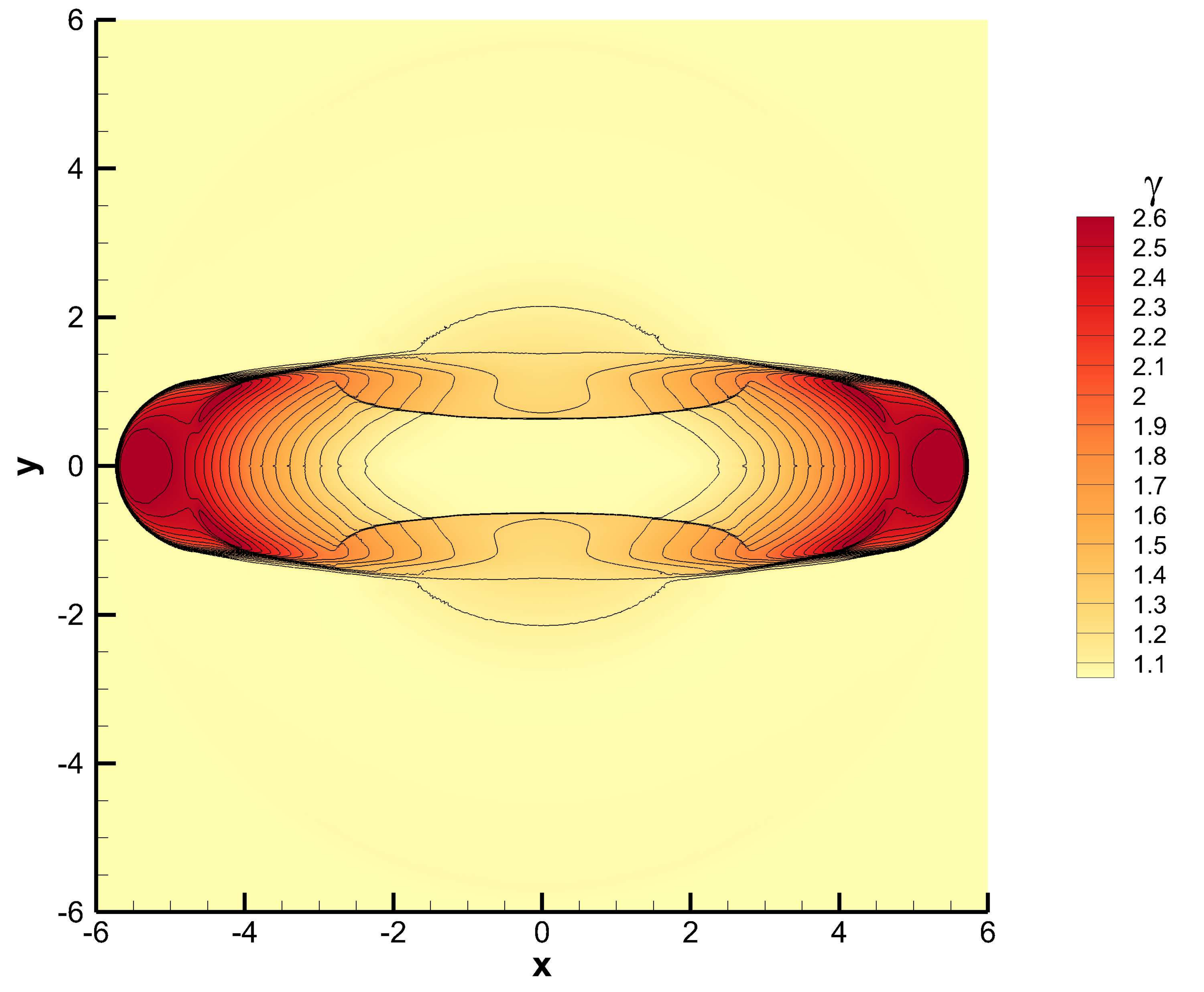}   \\  
  \includegraphics[width=0.33\textwidth]{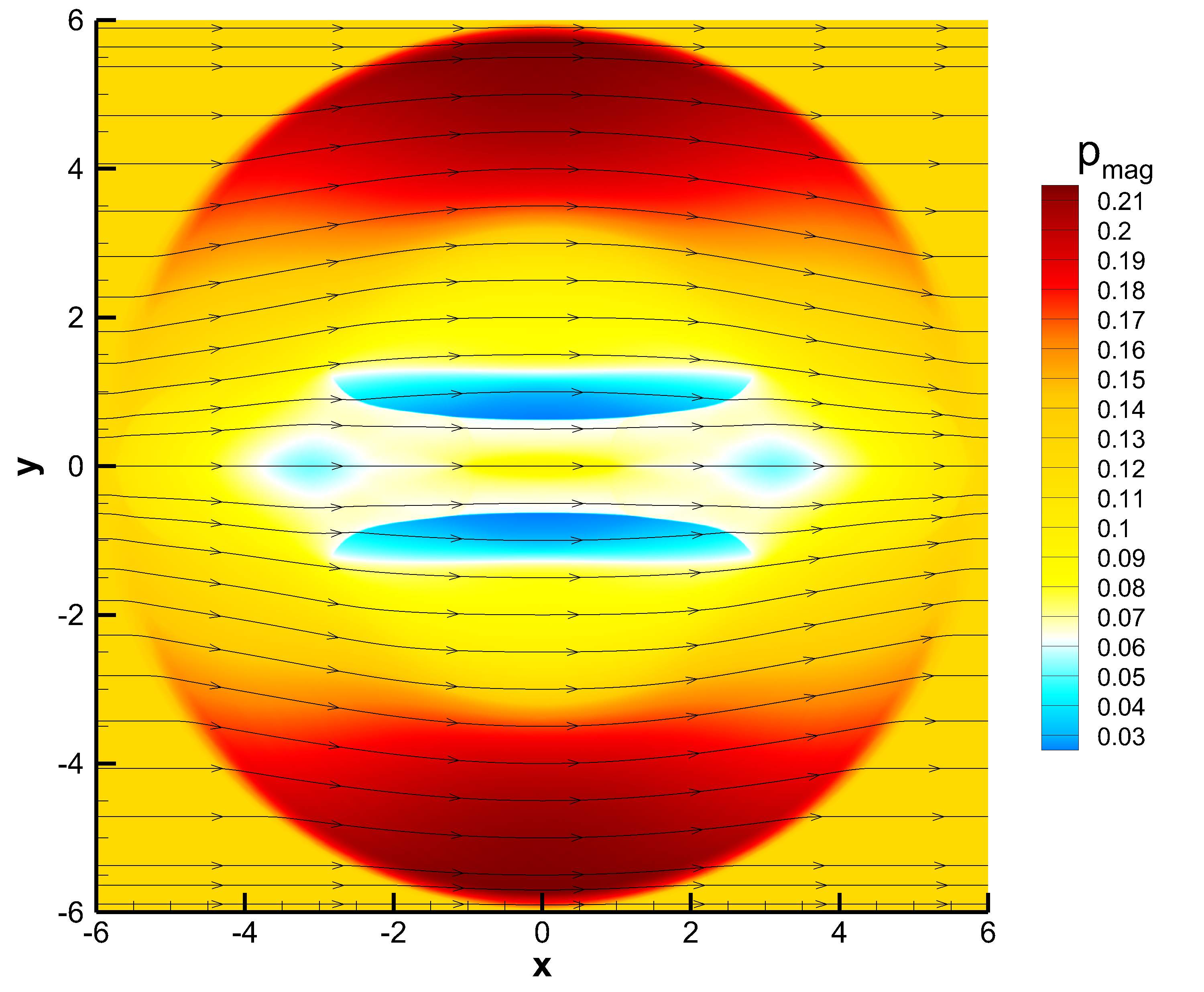}    
  \includegraphics[width=0.33\textwidth]{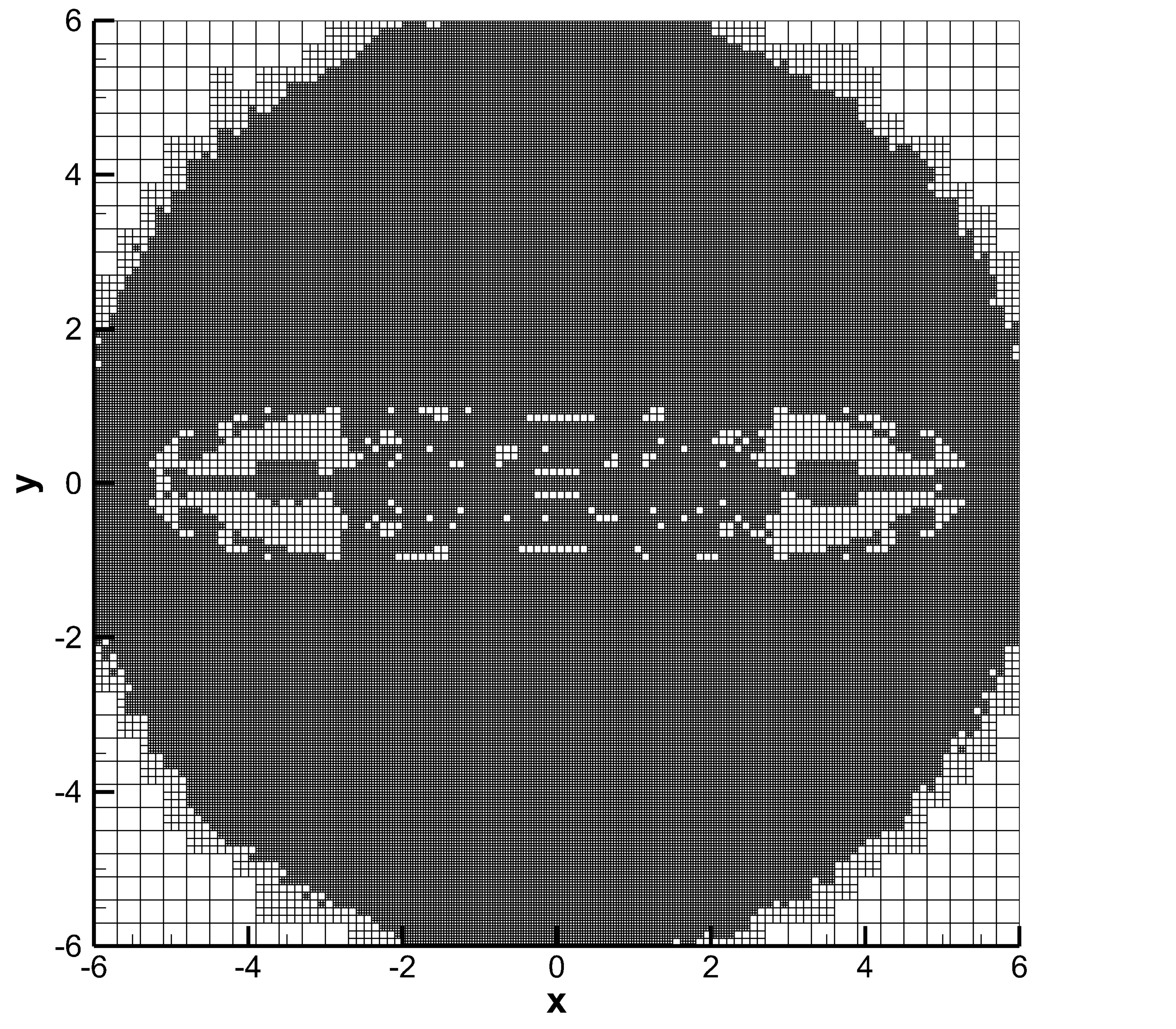}     
  \includegraphics[width=0.33\textwidth]{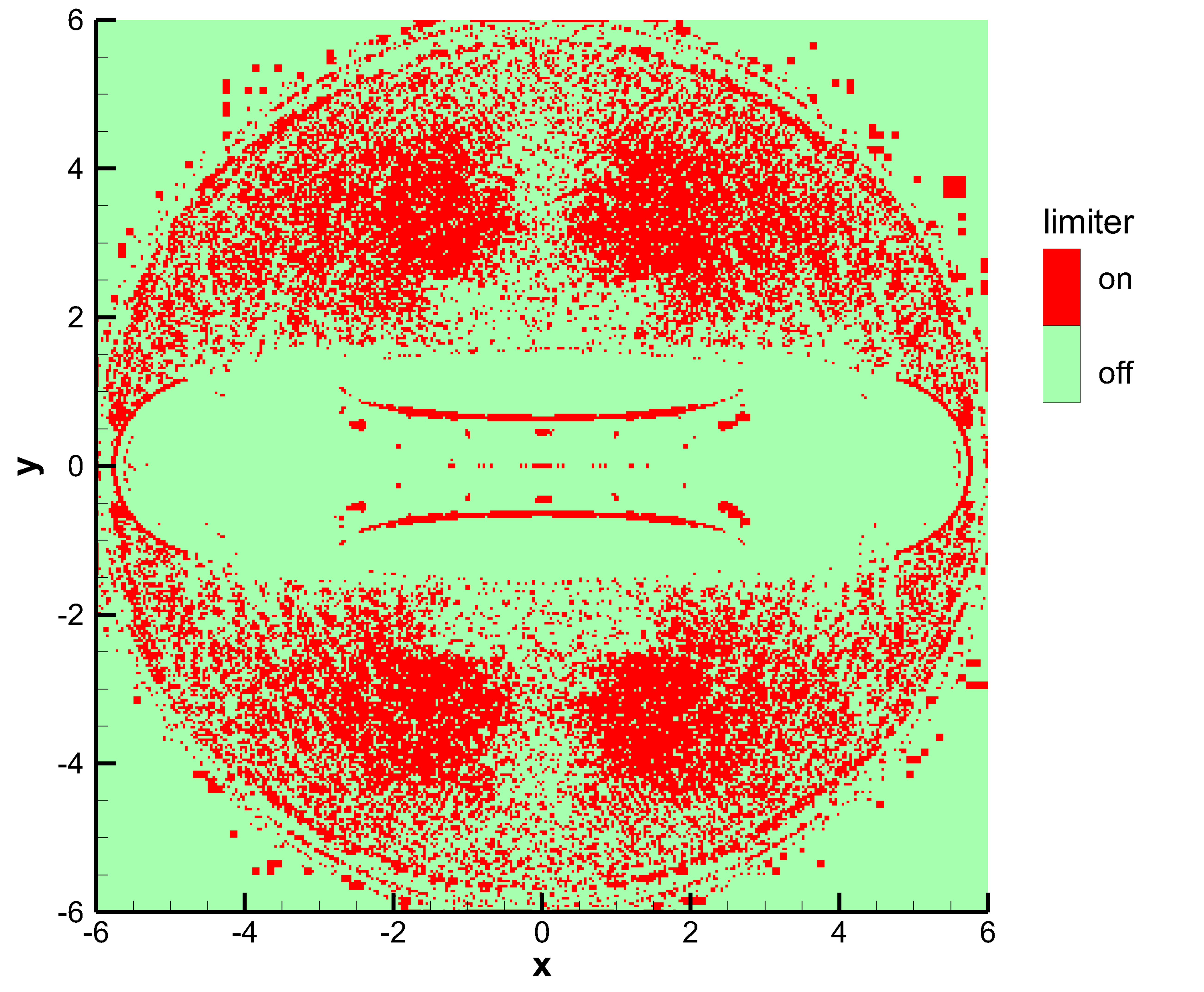}    
  \caption{Solution of the SRMHD blast wave with
    $B_x=0.5$ at time $t=4.0$, obtained with the ADER-DG $\mathbb{P}_3$
    scheme supplemented with the \aposteriori second-order TVD subcell
    limiter. 
	Top panels: rest-mass density (left), thermal pressure
  (center) and Lorentz factor (right).
	Bottom panels: magnetic pressure (left) with magnetic field lines reported, AMR grid  (center) and limiter map (right) with troubled cells marked in red and
  regular unlimited cells marked in green.}
\label{fig:BW_05}
\end{center}
\end{figure*}
\begin{figure*}
\begin{center}
\includegraphics[width=0.33\textwidth]{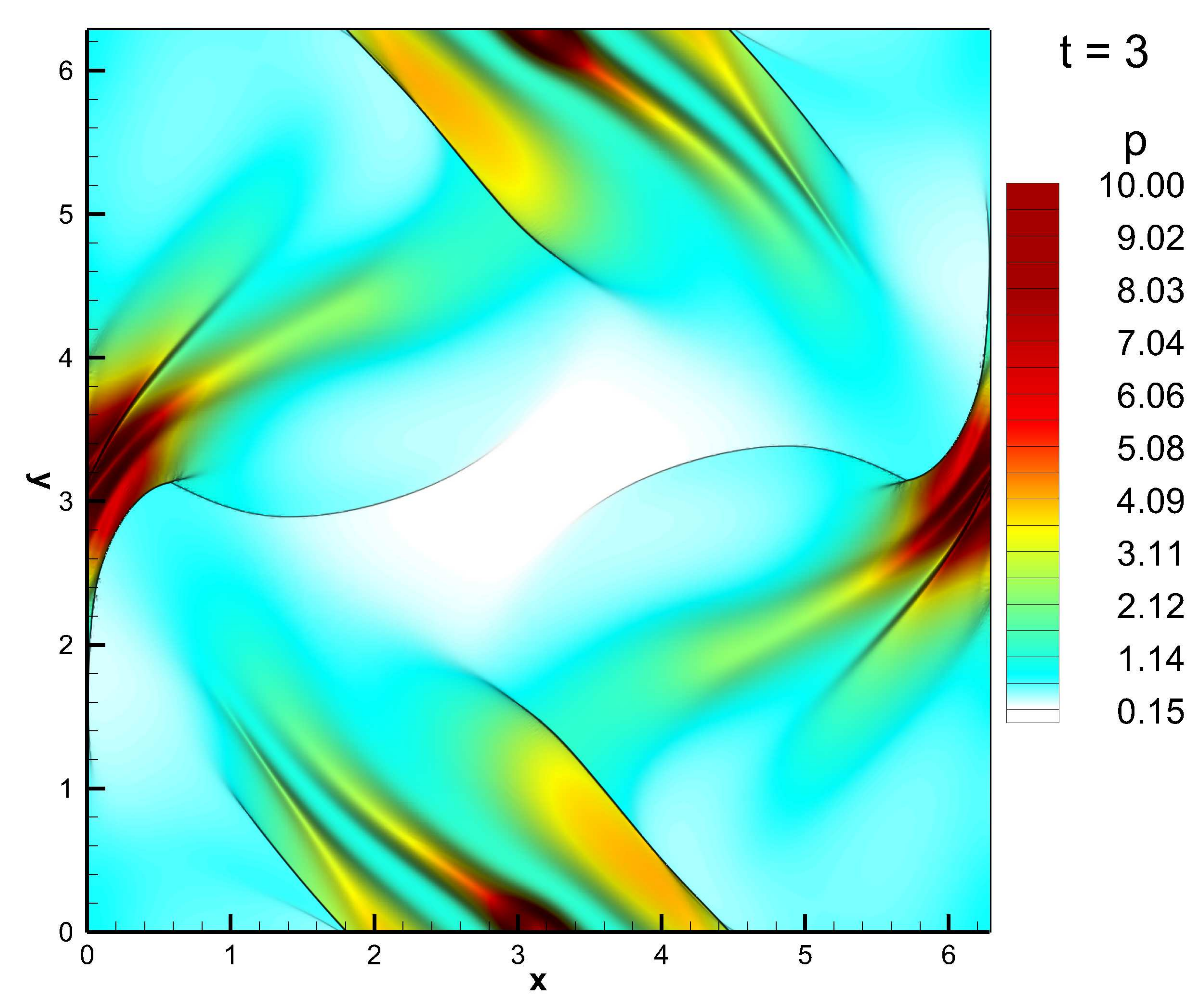}    
\includegraphics[width=0.33\textwidth]{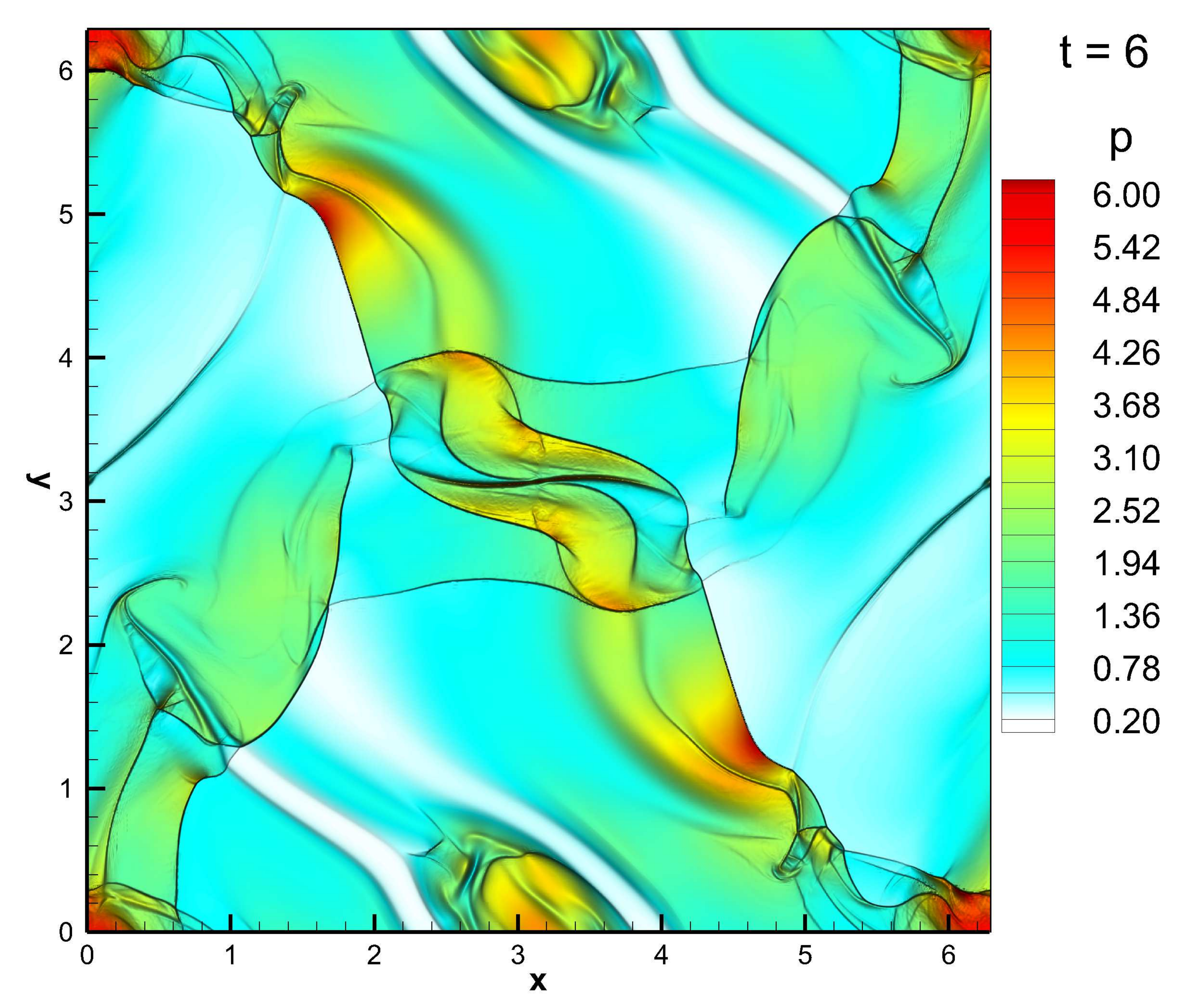} 
\includegraphics[width=0.33\textwidth]{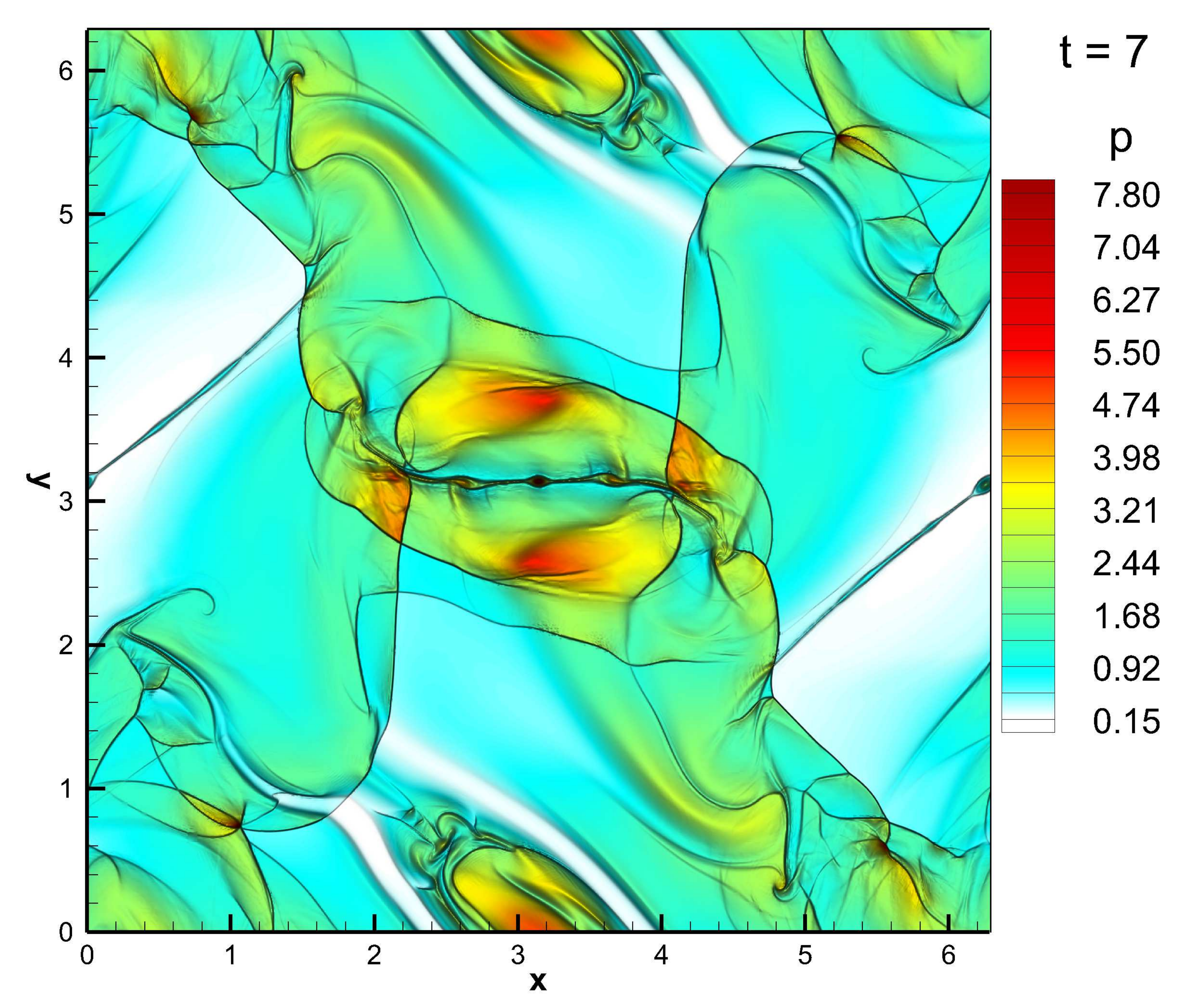}    \\  
\includegraphics[width=0.33\textwidth]{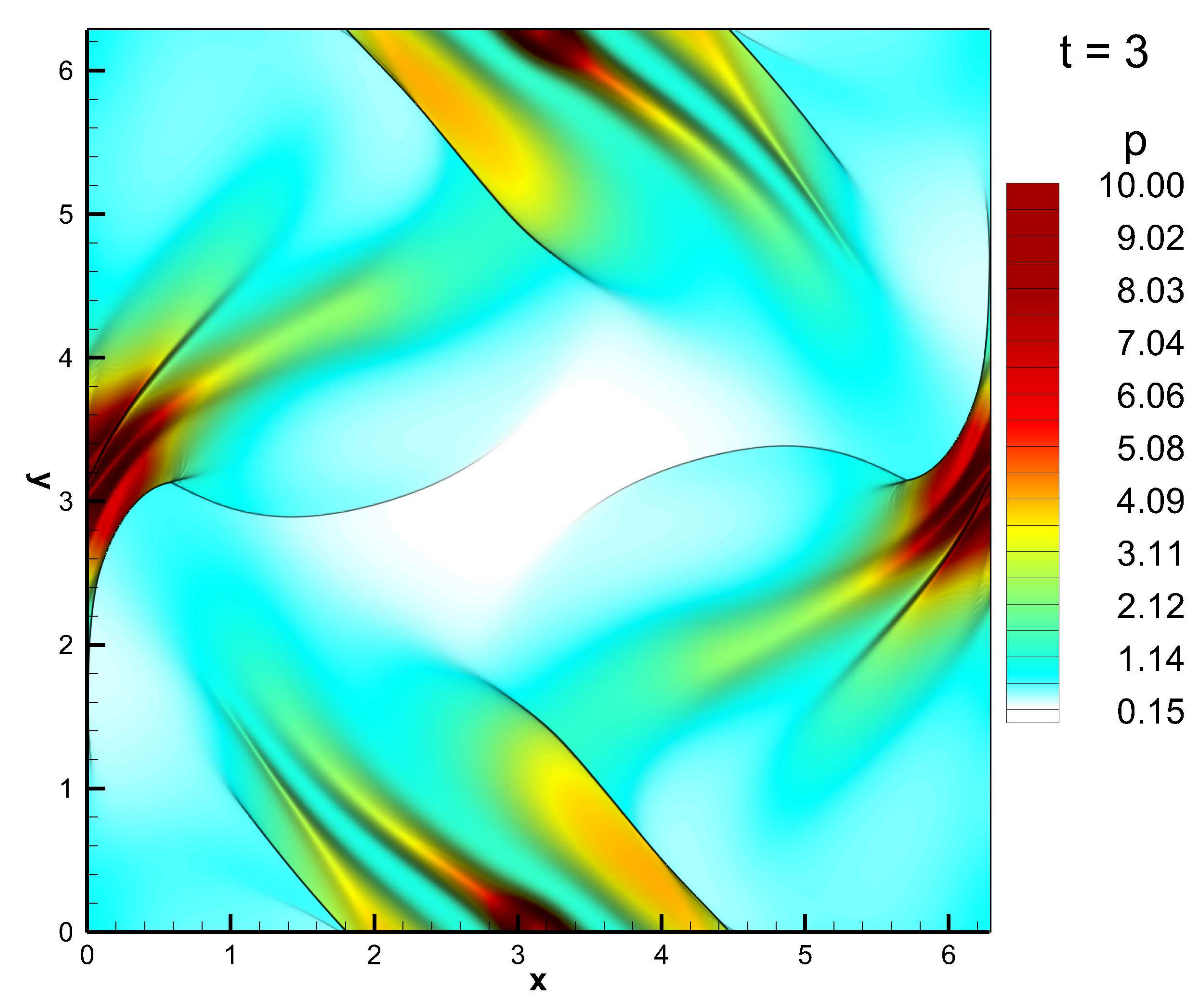}	   
\includegraphics[width=0.33\textwidth]{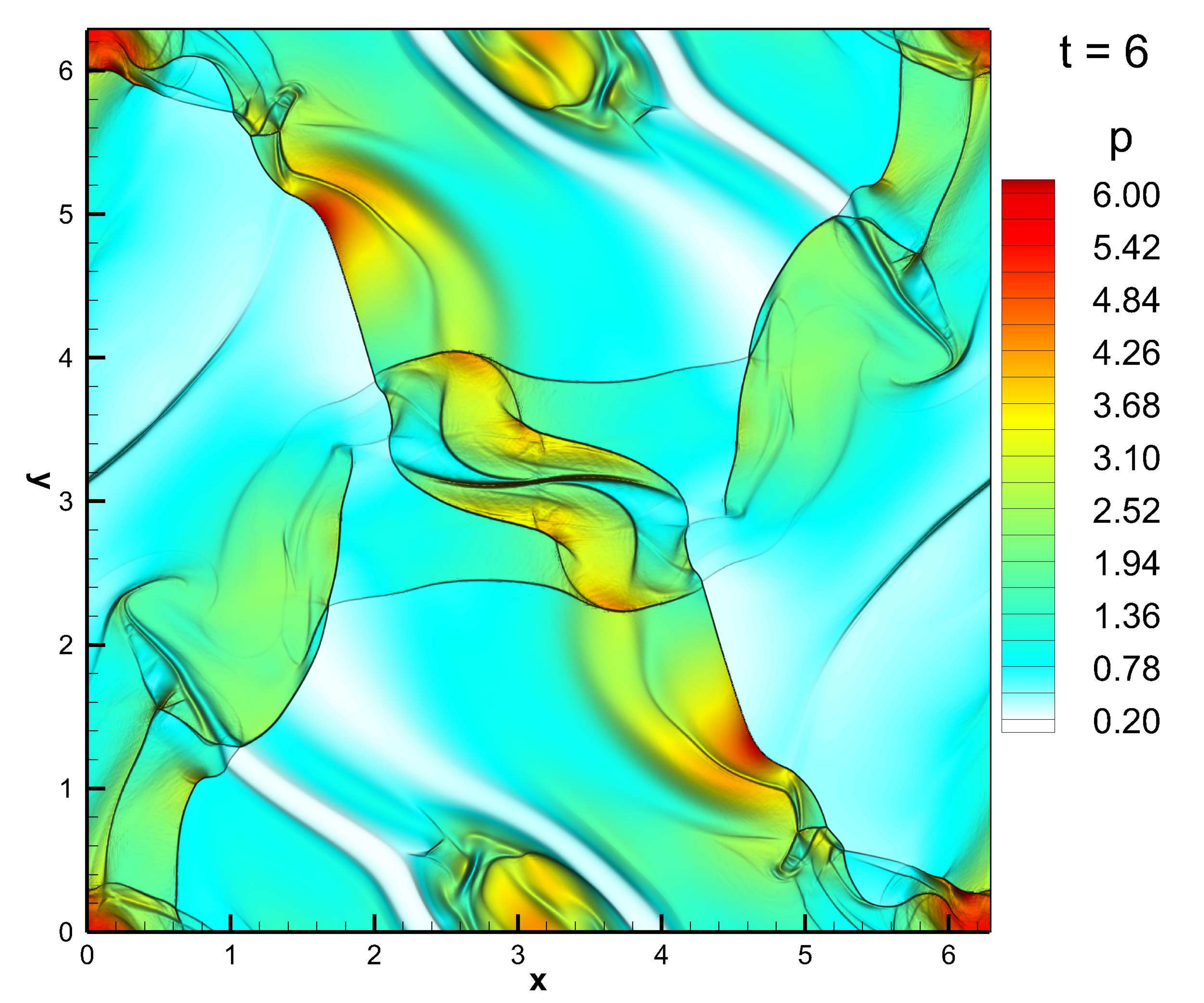} 
\includegraphics[width=0.33\textwidth]{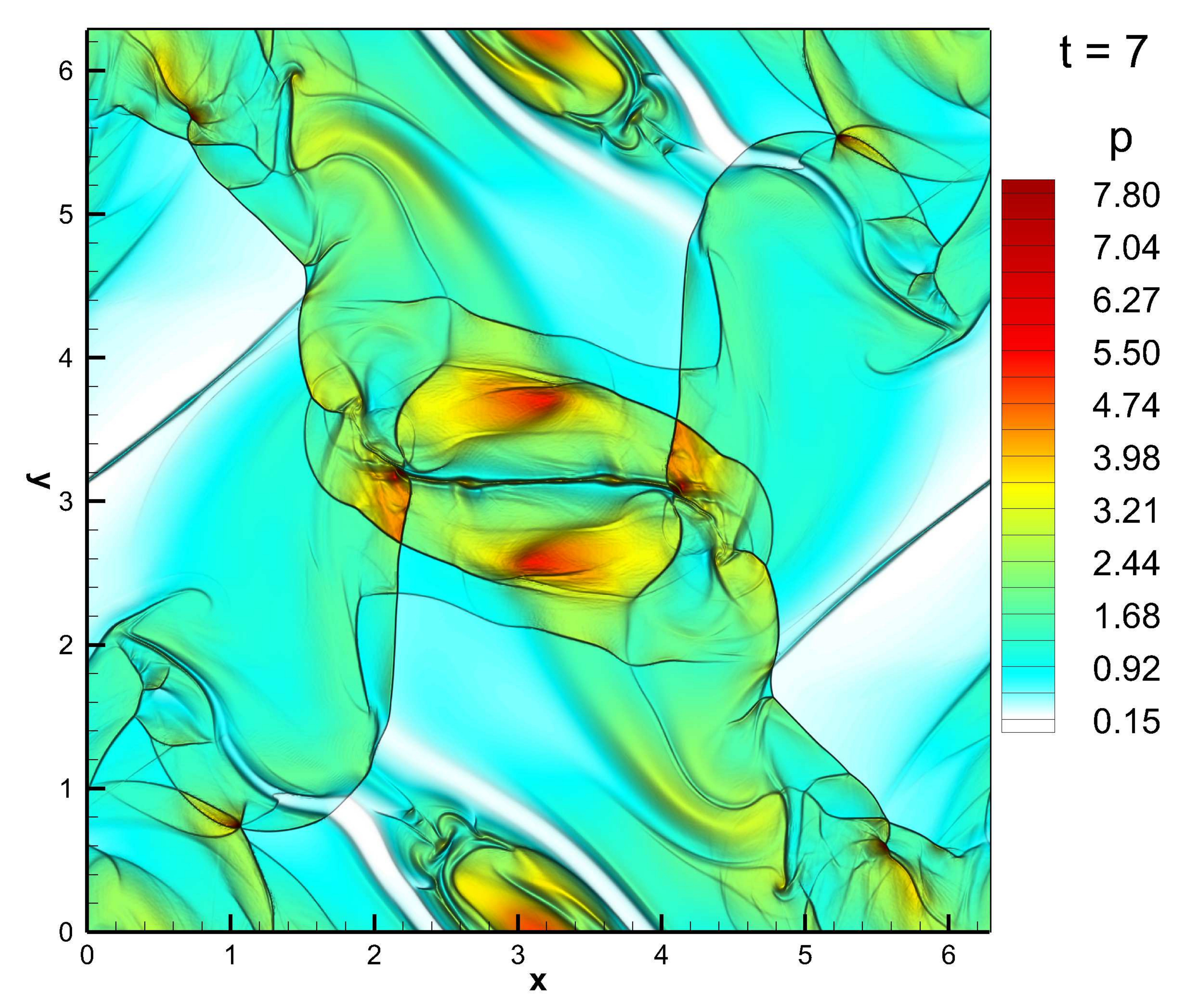}    \\  
\includegraphics[width=0.33\textwidth]{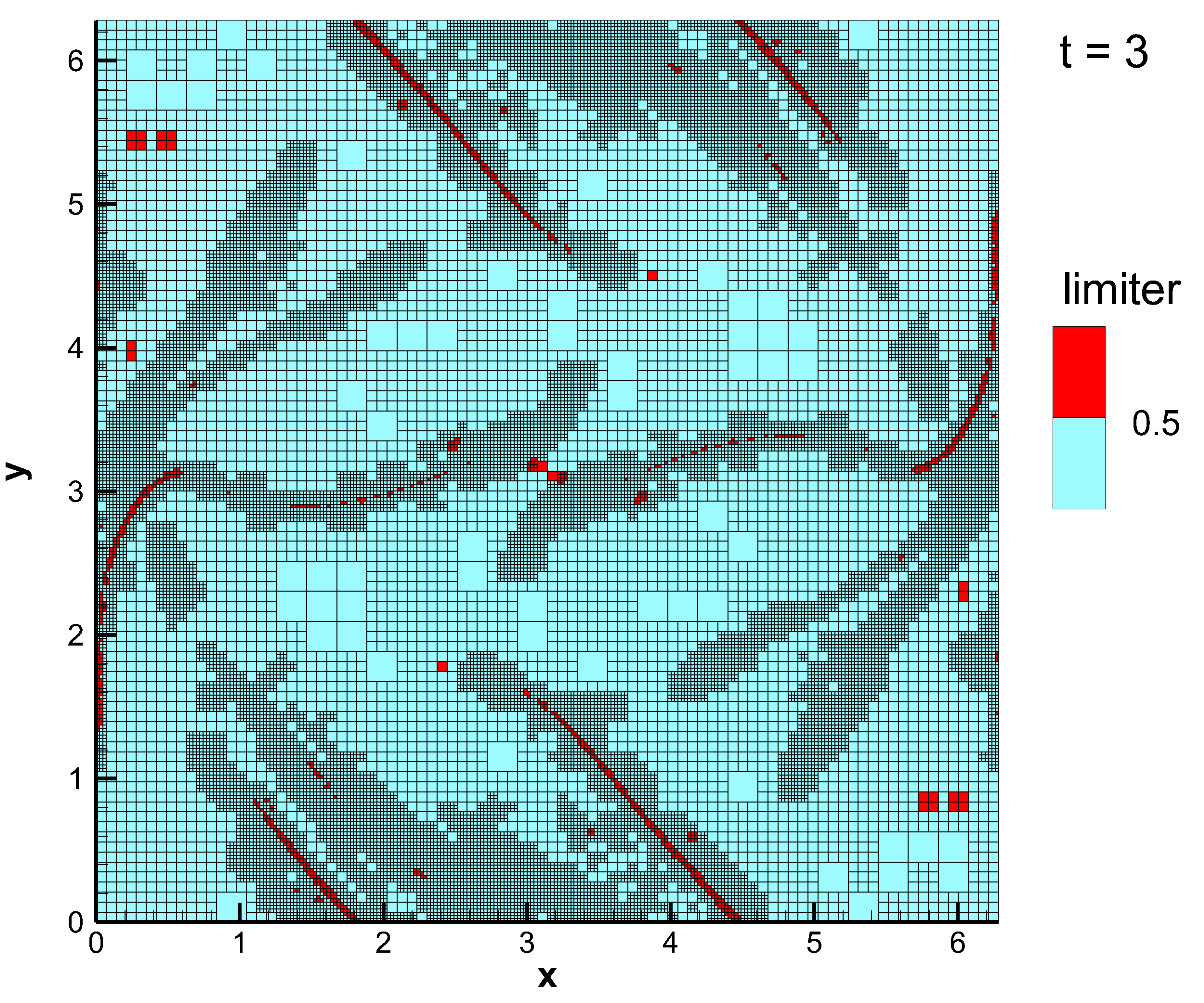}	   
\includegraphics[width=0.33\textwidth]{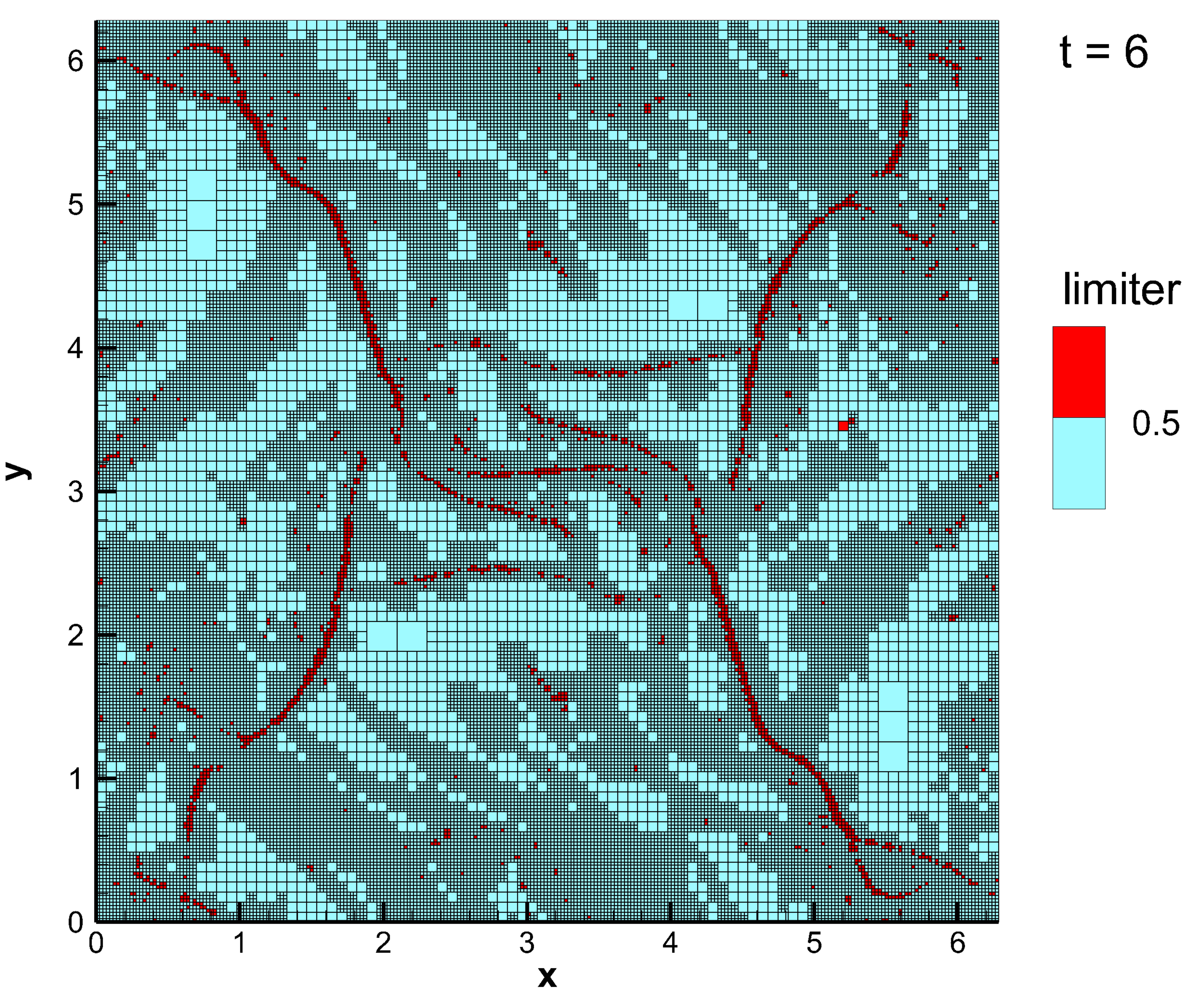} 
\includegraphics[width=0.33\textwidth]{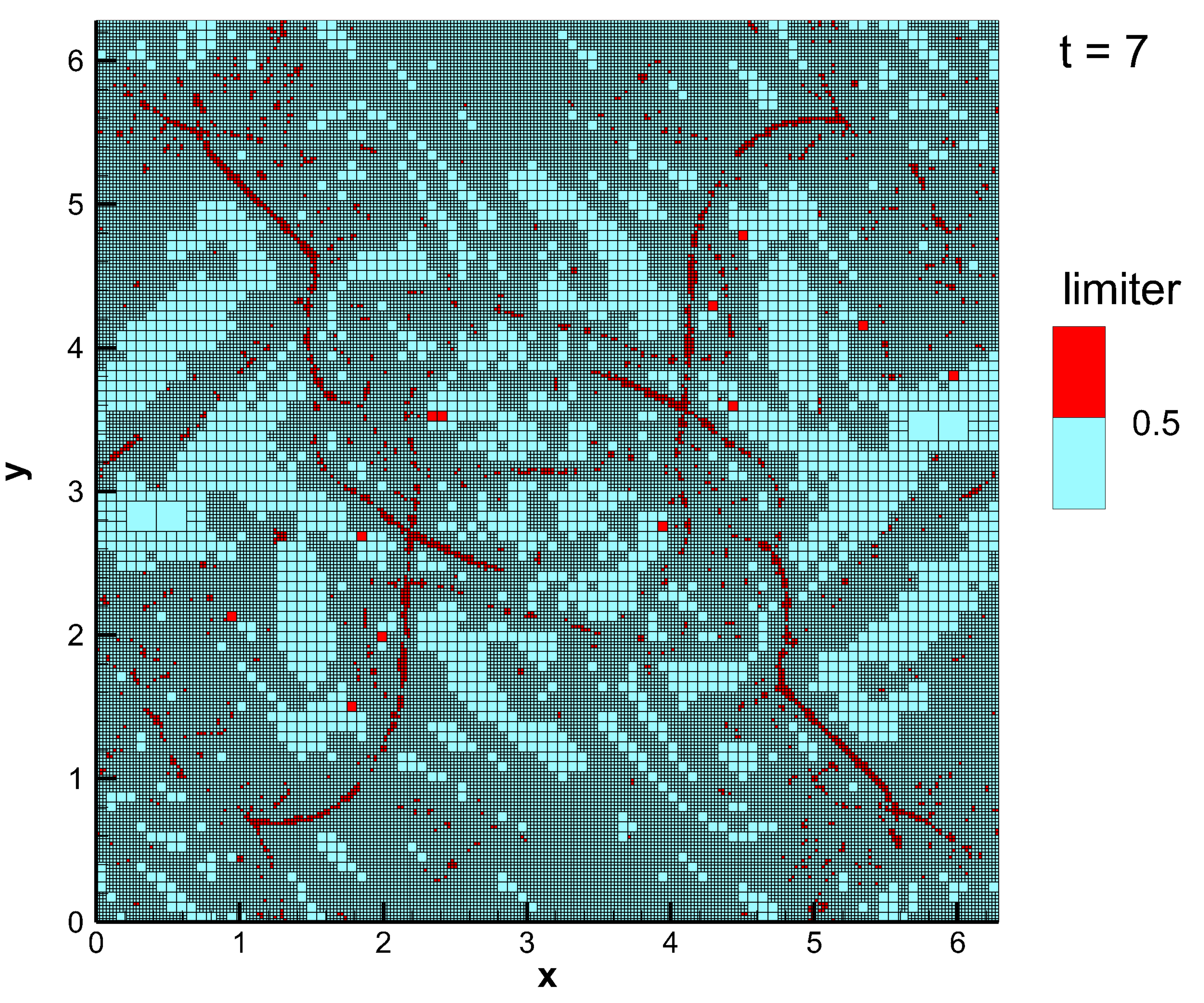}   \\
\includegraphics[width=0.33\textwidth]{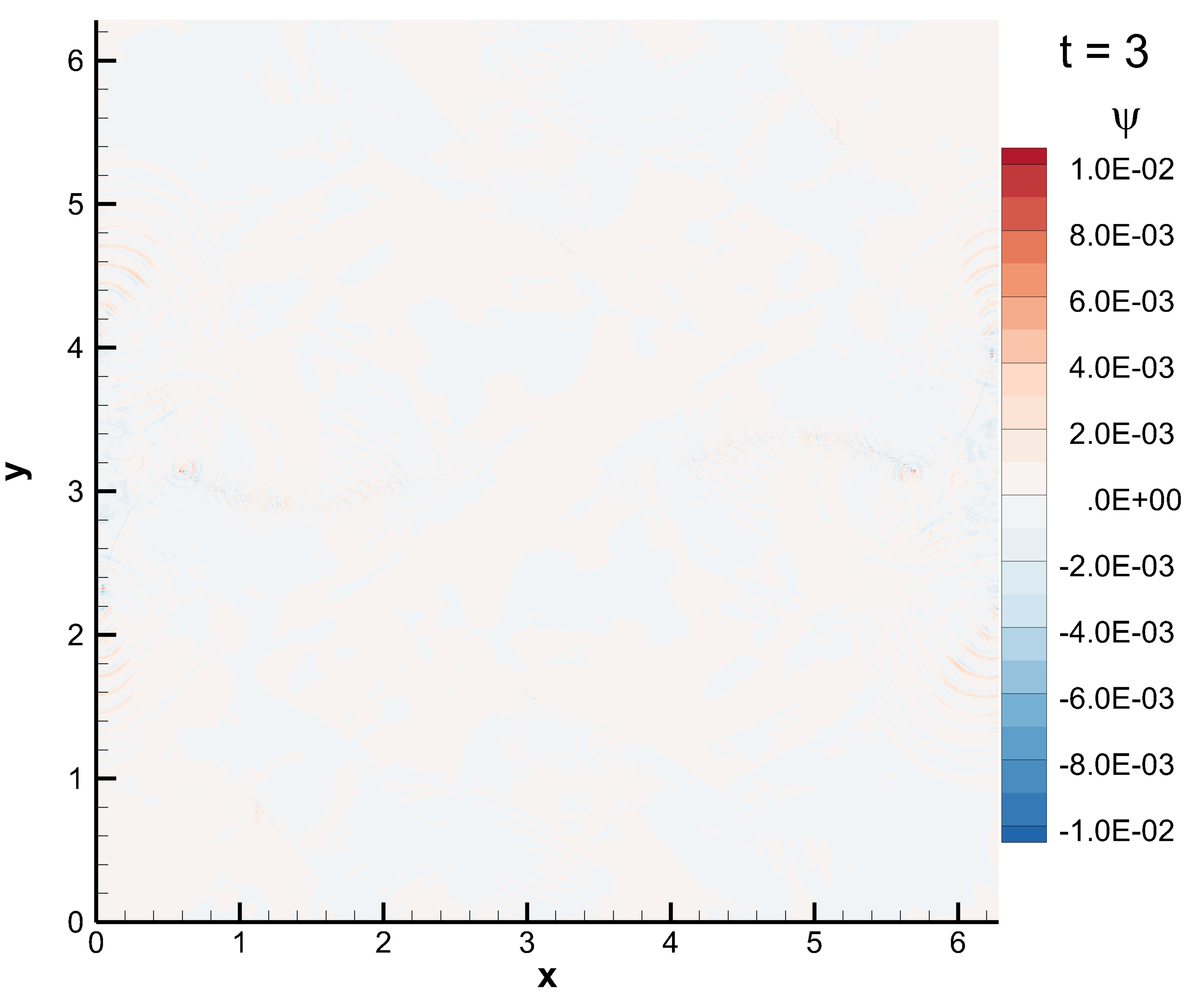}	   
\includegraphics[width=0.33\textwidth]{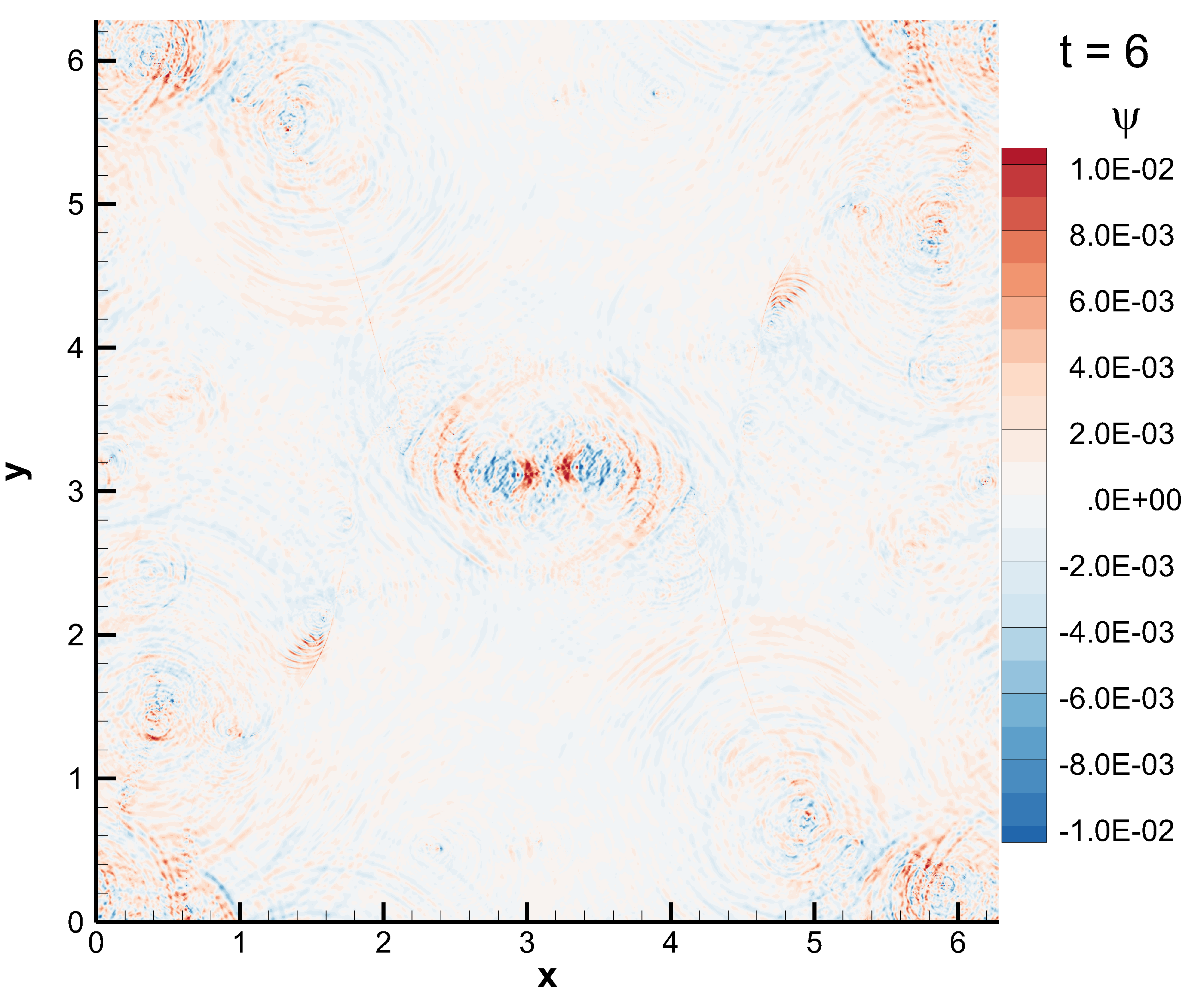} 
\includegraphics[width=0.33\textwidth]{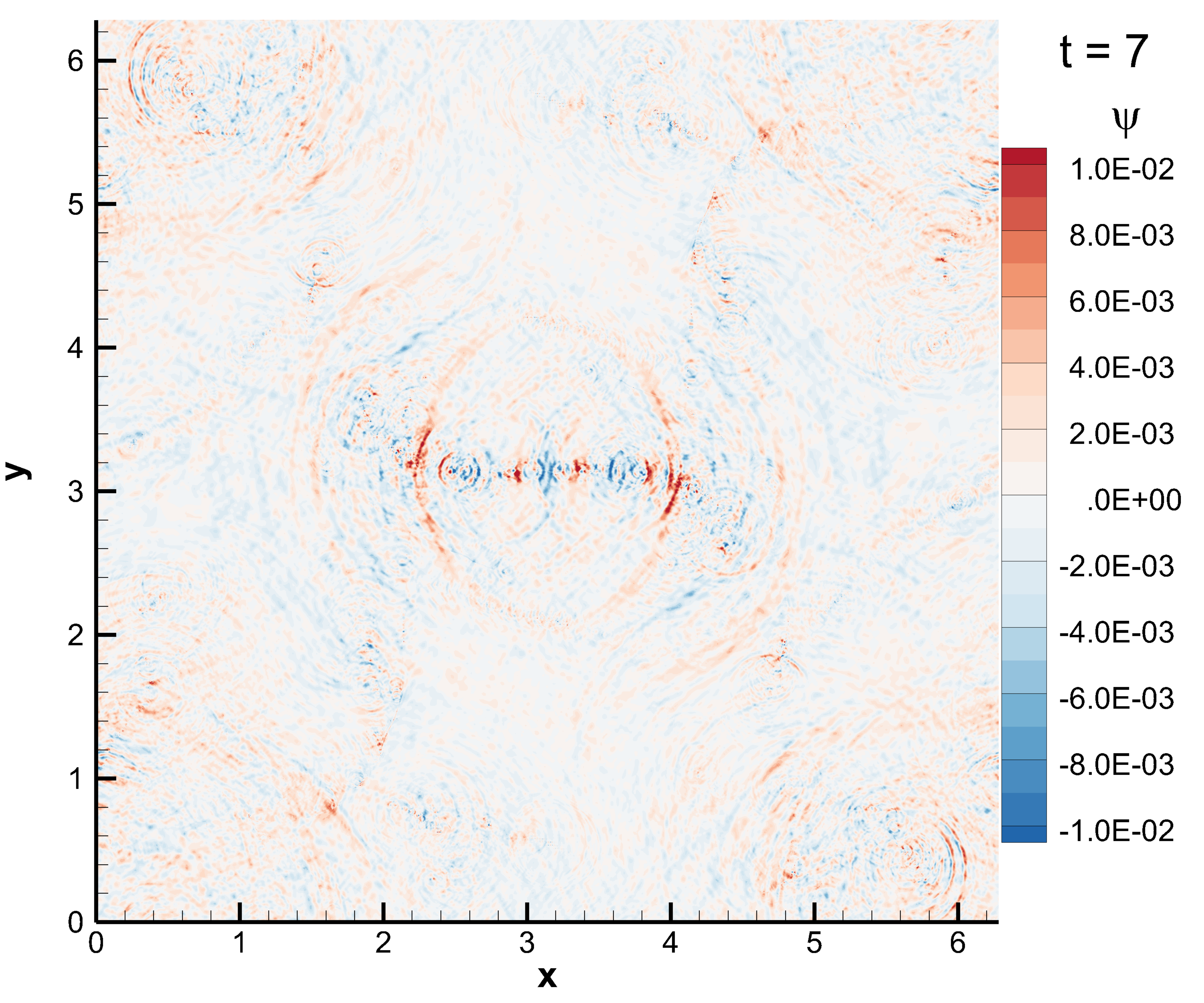}   \\
\caption{SRMHD Orszag-Tang vortex problem at
  times $t=3$, $t=6$, $t=7$, from left to right, obtained through
  the ADER-DG-$\mathbb{P}_5$ scheme supplemented with the \aposteriori
  TVD subcell limiter on a $30^2$ elements on the coarsest grid
  ($\ell=0$), two maximum refinement levels and a refinement factor
  $\mathcal{R}=3$. From the top to the bottom row: $1^{\text{st}}$) $\mathbb{P}_5$-solution
  obtained on the AMR grid; $2^{\text{nd}}$) $\mathbb{P}_5$-solution
  obtained on the corresponding finer uniform grid, \ie $270^2$ space
  elements of the maximum refinement level
  $\ell_{\text{max}}=2$; $3^{\text{rd}}$) AMR-grid, troubled cells (red) and
  unlimited cells (blue); $4^{\text{th}}$) divergence cleaning scalar $\psi$. }
\label{fig:OrszagTang}
\end{center}
\end{figure*}
\begin{figure*}
\begin{center}
\begin{tabular}{cc} 
   \includegraphics[width=0.45\textwidth]{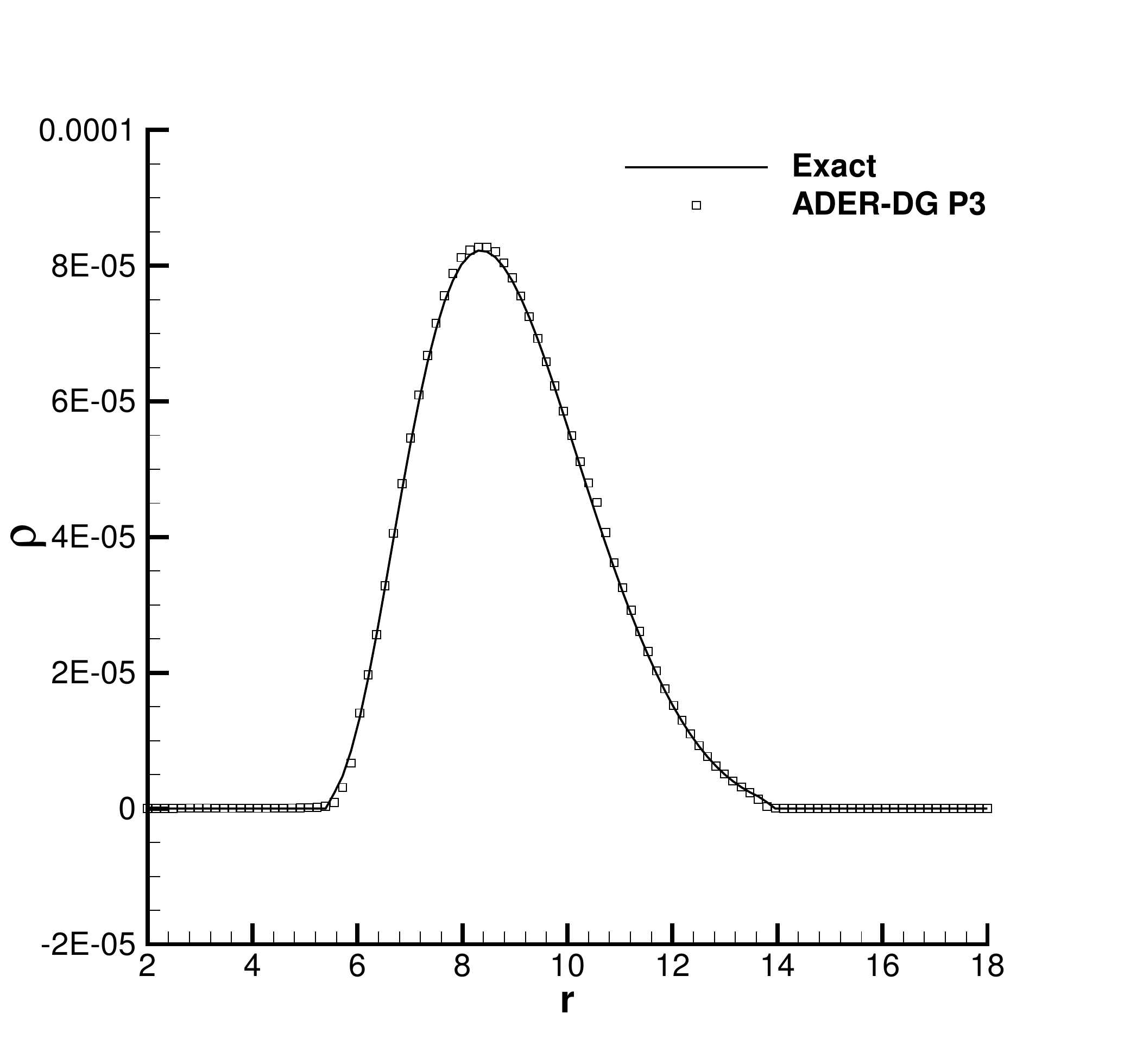}  & 
   \includegraphics[width=0.45\textwidth]{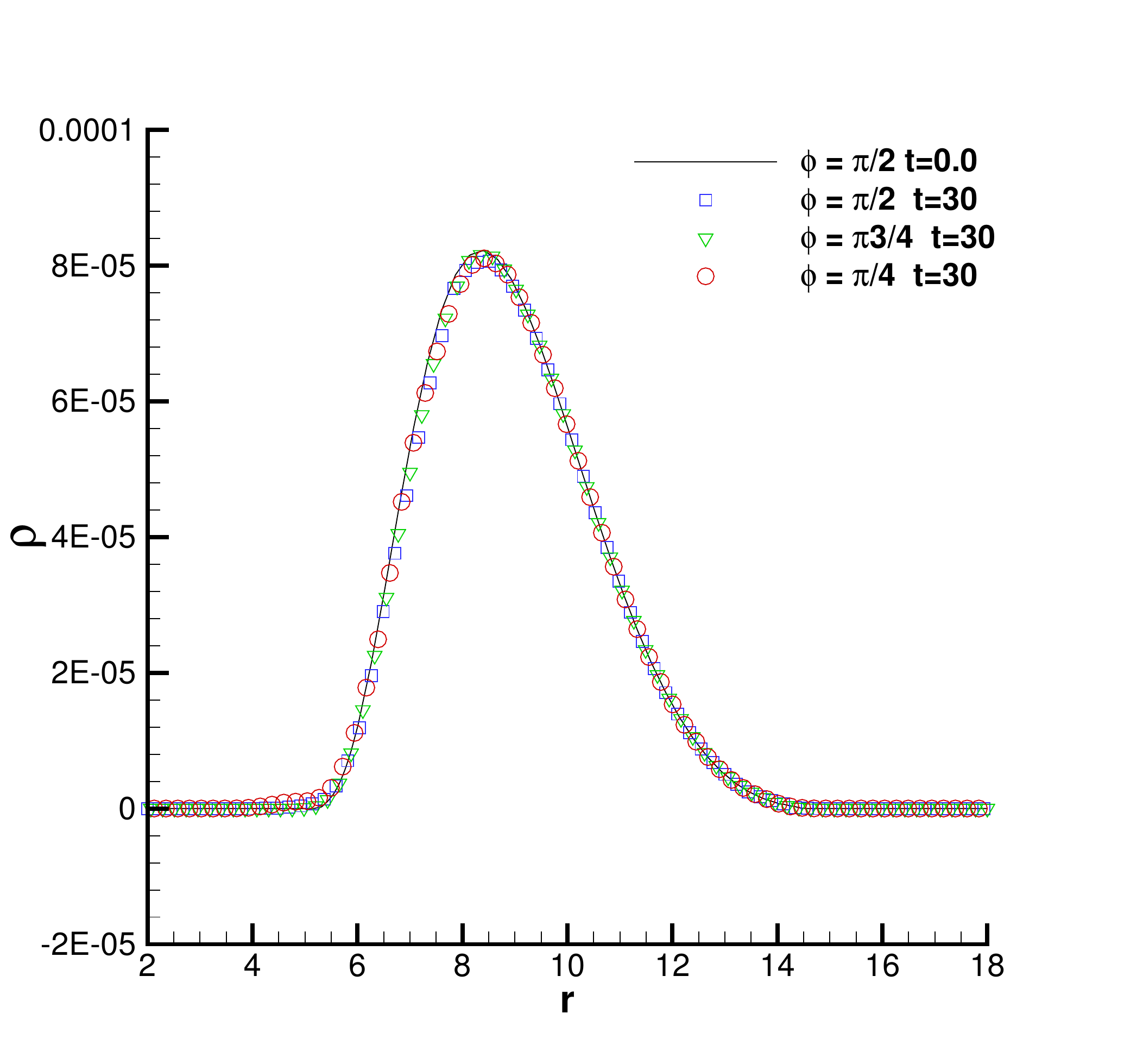} 
\end{tabular} 
\caption{Results obtained with the ADER-DG
  $\mathbb{P}_3$ scheme supplemented with the \aposteriori second-order
  TVD subcell limiter. 1D cut and comparison with the exact solution. 2D
  simulation of the torus in spherical Kerr-Schild coordinates at time
  $t=100$ (left), and 3D simulation in Cartesian Kerr-Schild coordinates
  at time $t=30$ for different azimuthal angles (right).}
\label{fig:Torus.cut}
\end{center}
\end{figure*}

\subsubsection{\textbf{Advection of a 2D magnetic field loop}}

In this special-relativistic 2D problem we advect a loop of magnetic
field which is at a magnetic pressure much smaller than the corresponding
fluid pressure. The computational domain in Cartesian coordinates is
given by is $(x,y)\in \Omega = [-1,+1]\times[-0.5,0.5]$ with periodic
boundary conditions everywhere. Using unitary (dimensionless) rest-mass
density and gas pressure, \ie $\rho=p=1$, the velocity field is set to be
constant with and initialised as $(v_x,v_y)=(2,1)V_0$, where
$V_0=1/5$. The magnetic-field vector is derived from
the magnetic vector potential, which is specified as
\begin{align}
  A_z = \left\{ \begin{array}{lr} A_0 (R - r )
    & \text{for} \ r\leq R\,, \\
    0 
    & \text{otherwise}\,,
\end{array}\right.
\end{align}
where $r$ is the radial coordinate, $R=0.3$ is the radius of the advected
loop and the parameter $A_0=10^{-3}$ modules the magnetic field. The
discontinuity at the loop boundaries has been initially slightly
smoothed, \eg by means of a standard linear smoothing in the form
\begin{align}
  B_x & = \left\{ \begin{array}{cl} A_0 \frac{y}{r}
    & \text{for} \ r\leq R\,, \\s(r) A_0 \frac{y}{r}
    & \text{for} \ R < r\leq R_1\,, \\
    0 
    & \text{otherwise}\,.
\end{array}\right. \\   
B_y& = \left\{ \begin{array}{cl} - A_0 \frac{x}{r}
    & \text{for} \ r\leq R\,, \\ - s(r) A_0 \frac{x}{r}
    & \text{for} \ R < r\leq R_1\,, \\
    0 
    & \text{otherwise}\,.
\end{array}\right.
\end{align}
where $s(r) = 1 - (r-R)/(r-R_1)$ is the adopted linear taper-function,
with $R_1$ chosen to be close to $R$, \eg $R_1=0.315$.

Given the initial conditions and the periodic boundary conditions, the
magnetic loop is advected across the computational
domain and we have performed simulations using the lapse function set
either to $\lapse = 1$ or to $\lapse = 2$, so that the corresponding
simulation times to recover the initial configuration are $t=5$ and $t=2.5$,
respectively; conversely,
the shift vector $\beta^i$ is set to zero.

This test has been solved using a level-zero mesh of $20^2$ space
elements with tho maximum refinement levels $\ell_{\text{max}}=2$ via an
ADER-DG-$\mathbb{P}_4$ scheme, supplemented with the \aposteriori TVD
subcell limiter and by adopting an HLL Riemann solver. At this point we
would like to emphasize that instead of HLL or Rusanov-type Riemann
solvers any other stable and monotone numerical flux could have been used
equally well. The Riemann solver has to be understood as a building block
of the DG scheme, exactly in the same way as it is in the finite-volume
context.  Figure \ref{fig:ML_B} reports the numerical results, which show
a good agreement between the advected solution and the reference one
given by the initial condition (left panels). Furthermore, the limiter is
only rarely activated, as expected for this test case (right panels). {
  The solutions for the divergence cleaning scalar $\psi$ are
  plotted in Fig. \ref{fig:ML_B_psi}.}

\subsubsection{\textbf{2D blast wave}}

\vspace{3mm} 

Another standard test of the RMHD equations is represented by the 
cylindrical blast wave problem. In this benchmark, the plasma is 
initially at rest and subject to a constant magnetic field along the
$x$-direction; we have therefore considered two different configurations
strengths of the magnetic field, \ie $B_x = 0.1$ and $B_x= 0.5$,
representing the case of a moderately and of a highly magnetized plasma,
respectively.

The initial conditions for the rest-mass density and
pressure are given respectively by
\begin{align}
  (\rho, p ) = \left\{\begin{array}{lr} (0.01, 1)
  & \text{if} \  r<R \,, \\
  10^{-4} \times (1,5) & \text{otherwise}\,,
  \end{array}\right.
\end{align}
and together with the magnetic-field strength are sufficient to fully
specify the initial setup. Also in this case, and following see
\citet{Balsara1999b}, a linear smoothing is used in order to avoid sharp  
discontinuities in the initial conditions. 

The computations have been carried out in 2D with a Cartesian coordinate
system over a computational domain given by $\Omega = [-6,6]^2$, with
$40^2$ elements on the coarsest mesh level, and a maximum refinement
level $\ell_\text{max}=2$. We have used the Rusanov Riemann solver with
our ADER-DG-$\mathbb{P}_3$ scheme. The computed results for different
physical quantities, the AMR grid and the limiter status are shown in
Fig. \ref{fig:BW_01} for the moderately magnetized case, and in
Fig. \ref{fig:BW_05} for the highly magnetized case. Note in the
bottom-right panels of figures the map of the ``troubled cells'' and how
these are limited in extent and nicely map the dynamics of the
discontinuities in the magnetic field. Clearly, the fraction of troubled
cells in the case of the low-magnetisation setup represent only a very
small fraction of the evolved cells (see Fig. \ref{fig:BW_01}); this
is to be contrasted with what happens in the case of the much more
challenging case of high magnetisation, where however the troubled cells
still represent less than $50\%$ of the evolved cells (see
Fig. \ref{fig:BW_05}). 

Lacking an analytic solution to compare with, the
assessment of the results in this case is harder, but it is reassuring
that the results match well those presented in other tests in the
literature, \eg by \cite{DelZanna2007,Dionysopoulou:2012pp,Zanotti2015d}.

\subsubsection{\textbf{Orszag-Tang vortex}}

Our final special-relativistic test of non-smooth flows is another classic
benchmark represented by the relativistic version of the Orszag-Tang
vortex system \cite{OrszagTang}. This is a useful application of our
numerical infrastructure as it involves the development of a complex and
non-smooth magnetic-field structure and hence it explores geometries
without trivial symmetries.

The initial conditions in this case are given by the vector of conserved
variables 
\begin{align*}
&\left( \rho, u, v, w, p, B_x ,B_y,B_z \right) = \\ &\left( 1 , -
  \frac{3}{4\sqrt{2}}\sin y\,, \frac{3}{4\sqrt{2}}\sin x \,, 0, 1, - \sin
  y\,, \sin 2x \,, 0 \right)\,,
\end{align*}
with $\Gamma=4/3$. The computational domain is $\Omega = [0,2\pi]^2$,
with $30^2$ elements on the level-zero grid, a maximum refinement level
of $\ell_{\text{max}}=2$, periodic boundary conditions and a Rusanov
Riemann solver for the subcell finite-volume limiter.

Figure \ref{fig:OrszagTang} shows the numerical results for the AMR grid {
with limiter status, the rest-mass density and the divergence-cleaning scalar 
$\psi$ at different times, together}
with the corresponding numerical solution obtained with the same scheme
on a fine uniform $270^2$ mesh, corresponding to the finest mesh
resolution at $\ell=\ell_{\text{max}}$ and which serves here as a
reference. The figure, in particular, refers to simulations in which the
$\mathbb{P}_5$-version of our ADER-DG has been adopted. Also for this
test, a rigorous accuracy analysis is not trivial but we note the very
good agreement between the AMR simulations and the fine uniform-grid
reference solution, as well as with the corresponding solutions that have
been published elsewhere [see, \eg \cite{Zanotti2015d,Porth2017}]. 
Note also how the AMR grid structure and the troubled-cells
patterns closely follow the development of steeper gradients and 
discontinuities.

\subsection{Non-smooth general-relativistic flows}
\label{sec:discontinuous_gr}

In the following two sections we discuss the use of our ADER-DG method in
non-smooth general-relativistic flows, either in 2D and spherical
coordinates or in 3D and Cartesian coordinates. The tests involve the
evolution of non-selfgravitating tori as those presented in
Sec. \ref{sec:discontinuous_sr} with the important difference that the
computational domain here fully contains the torus, whose exterior is
therefore filled with a uniform atmosphere at a rest-mass density of
$\rho_0=10^{-9}$ that is five orders of magnitude smaller than the one at
the torus centre.

\subsubsection{\textbf{2D torus around a Schwarzschild black hole}}
\label{sec:full_torus}

First, we consider a thick torus in equilibrium orbiting around a
black-hole with the parameters previously described in
Sec. \ref{sec:2D_torus_interior} and using horizon-penetrating spherical
KS coordinates in 2D. The computational domain $(r,\theta) \in \Omega =
[2,18] \times[0.5,2.5]$ is discretized with a uniform mesh of $50^2$ 
elements using an ADER-DG-$\mathbb{P}_3$ scheme with TVD subcell 
finite-volume limiter (as a comparison, the torus has an inner radius 
$r_{\rm in} = 5.5$ and an outer radius $r_{\rm out}= 13.8$, so that 
the entire torus is resolved with only 26 elements in radial direction
and 14 elements in angular direction). 
On the outer edge we impose the initial data as boundary condition in all 
variables. 

A 1D cut of the rest-mass density in the radial direction is shown in the
left panel of Fig. \ref{fig:Torus.cut} and is plotted over the analytic
solution at $t=100\,M$.  Note the excellent agreement between the
numerical results and the exact solution, with differences in the central
rest-mass density that are less than $0.7\%$.

It is useful to remark that the low-density atmosphere has been
successfully simulated and robustly evolved in time with a high-order
ADER-DG scheme and that inside the computational domain the limiter is 
activated only on the border of the torus, where 
spurious oscillations may generate possibly negative-valued densities and
pressures in the high-order DG polynomials. However, the {a-posteriori}
subcell finite-volume limiter appears to be robust enough to accurately
treat the atmosphere of the torus. Furthermore, we note that the fluid in
this low-density region is treated so as to be evolved as a standard
fluid, \ie the velocity is not set to zero in a computational cell that
is marked to host the atmosphere. As a result, during the simulations,
the atmosphere the fluid in the atmosphere starts accreting onto the
black hole; in practice the amount of matter accreted in this manner is
minute and does not influence with the dynamics of the much denser matter
lost from the torus.

\subsubsection{\textbf{3D torus around a Schwarzschild black hole}}

The final test considered in this battery is represented by a fully 3D
evolution of the torus considered in the previous section, therefore
adding the azimuthal spatial dimension.

For this, we use a horizon-penetrating Cartesian KS coordinates which
cover a computational domain chosen to be $(x,y,z) \in \Omega = [-18,+18]
\times [2,18]\times[-8,+8]$. The portion of the domain around the origin
is excised following the same logic discussed in sec. \ref{sec:3D_Michel}.
 The solution has been computed using an
ADER-DG-$\mathbb{P}_3$ scheme on a uniform mesh composed of
$40\times20\times20$ elements.  

The 1D cut of the rest-mass density profile on the equatorial plane
$\theta=\pi/2$ and along different angular directions $\phi=\pi/4$,
$\pi/2$ and $3 \pi/4$ at $t\sim 30\,M$.  The various numerical solutions
are overlayed with the corresponding analytic solutions in the right
panel of Fig. \ref{fig:Torus.cut}. Once again, we can observe an
excellent agreement between numerical and exact solution, with
differences in the central rest-mass density that are less than $1.5\%$.

\section{Strong MPI scaling and performance comparison with other schemes}
\label{sec:performance}
\begin{figure}
	\centering
	\includegraphics[width=0.4\textwidth]{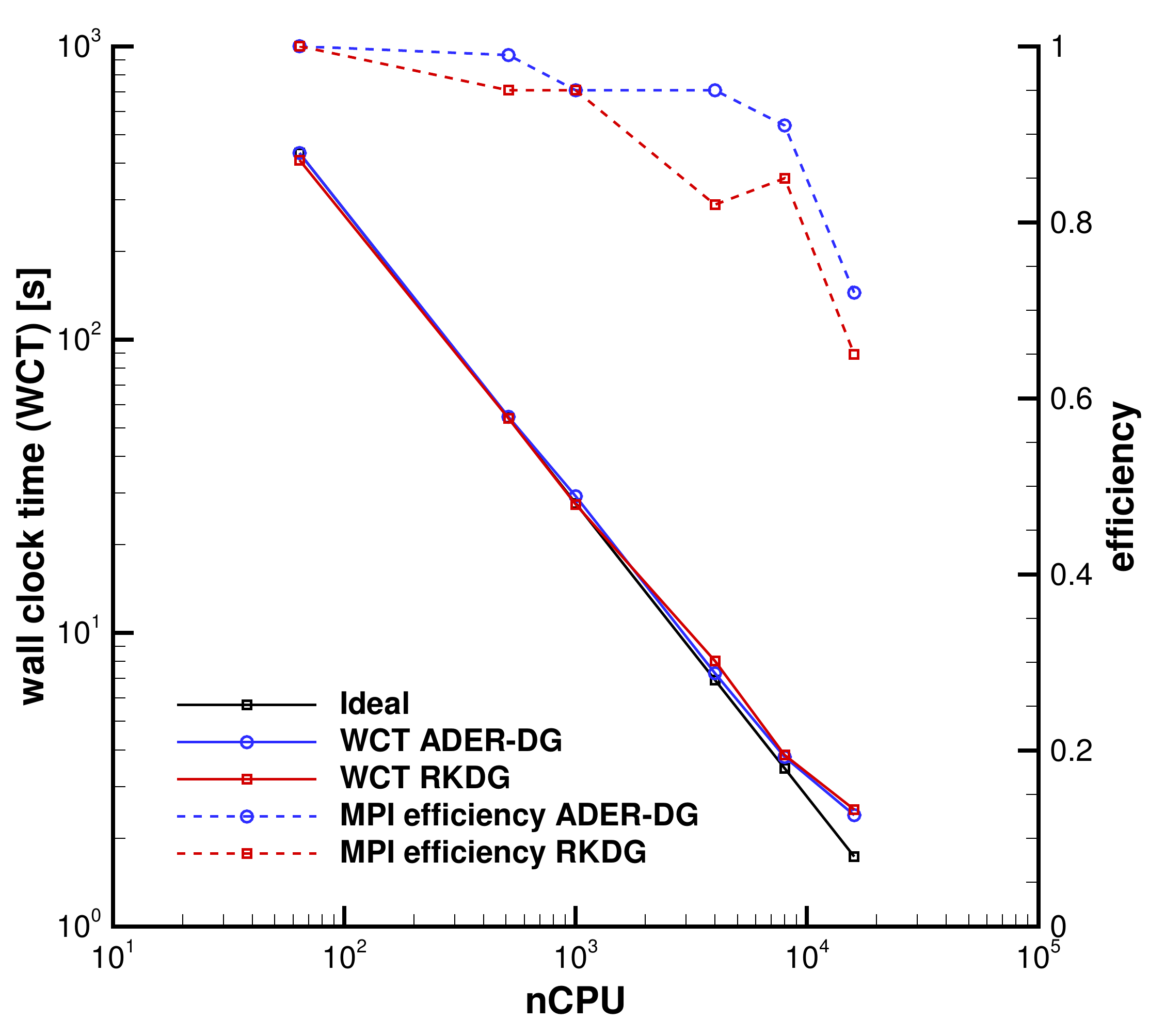}
	\caption{ Strong scaling test for the 3D GRMHD equations and
          performance comparison between fourth-order ADER-DG and RKDG
          schemes ($N=3$).  The test case is the large amplitude Alfv\'en
          wave problem solved in 3D up to $t=1$ on a uniform Cartesian
          mesh composed of $40 \times 40 \times 40$ elements.  The
          results were obtained with a pure MPI implementation on the
          SuperMUC phase I system at the LRZ in Garching, Germany, using
          64 to 16,000 CPU cores. On 16k cores, each MPI rank has only 4
          elements to update.}
	\label{fig:scaling}
\end{figure}

In this section we provide a detailed and quantitative performance analysis of 
the new ADER-DG schemes for the GRMHD equations proposed in this paper. 
We compare CPU times and MPI scaling results for ADER-DG in comparison with 
classical Runge-Kutta DG (RKDG) schemes. We furthermore provide CPU time 
comparisons between ADER-DG and ADER-WENO finite-volume (FV) methods. \\
As first test we run the Michel accretion problem again on the domain 
$\Omega=[3,5.5] \times [1,\pi-1]$ in two space dimensions using a sequence 
of successively refined meshes of $N_x \times N_x$ DG elements and 
$\ N_x (N+1) \times N_x (N+1)$ finite-volume 
zones until a final time of $t=10$. 
We use a third-order ADER-DG scheme ($N=2$) and compare with 
a third-order ADER-WENO finite-volume scheme, see \cite{AMR3DCL,Dumbser2008}. 
In order to make the comparison 
fair, the mesh of the FV scheme is $N+1$ times finer than the one of the DG scheme, 
since the DG method has $N+1$ degrees of freedom per cell and per space dimension. 
The total number of degrees of freedom is therefore the same for both methods. 
We present the $L_2$ and $L_\infty$ errors for the density $\rho$    
obtained with both methods. We also report the wall clock time (WCT) measured in 
seconds and the time needed by the scheme to update one single degree 
of freedom on one single CPU core (DTU), measured also in seconds. The inverse  
of this number represents the number of degrees of freedom that the scheme is 
able to update in one second on one CPU core and can be compared with other 
finite-volume and finite-difference methods. 
As computer hardware for this test we use one single 
CPU core of a workstation with an Intel i7-4770 CPU with 3.4 GHz clock speed 
and 16 GB of RAM. The results are shown in Table \ref{tab:CTimes}, from which
it becomes clear that the ADER-DG scheme is faster and more accurate than the 
ADER finite-volume scheme using the same number of degrees of freedom. Similar 
results have already been reported in \cite{Dumbser2008} and \cite{ADERNSE} for 
the Euler equations of hydrodynamics, the MHD equations and the compressible 
Navier-Stokes equations, using the unified framework of $P_NP_M$ schemes. 
\begin{table}
	\centering
	\begin{tabular}{ c c  cccc  }
		\hline
		& $N_x$ &  $L_2$ error & ${L_\infty}$ error & WCT [s] & TDU [s]  \\ 
		\hline
		\multirow{4}{*}{\rotatebox{90}{{DG $\mathcal{O}3$}}}
		&  6     &    2.53E-05  & 3.26E-05 	    &  15.9     &  1.0470E-04  \\ 
		& 12     &    3.32E-06 	&  4.46E-06	 	& 74.4     &  6.3726E-05  	\\ 
		& 18     &    1.01E-06 	&  1.37E-06	 	& 193.5     & 4.9770E-05 	\\ 
		& 30     &    2.26E-07 	&  3.07E-07	 	& 733.4     & 4.1173E-05 	\\ 
		\cline{2-5}
		\hline
		\multirow{4}{*}{\rotatebox{90}{{FV $\mathcal{O}3$}}}
		& 18    &    2.77E-05 	&  5.99E-05 	 &   37.7     &    5.1765E-04  \\  
		& 36    &   6.40E-06 	&  1.72E-05 	 	&  231.9     &   4.0117E-04 	\\ 
		& 54    &   2.73E-06 	&  8.81E-06 	 	&  694.0     &   3.5679E-04 	\\ 
		& 90    &   9.44E-07 	&  3.78E-06 	 	&  2754.8     &  3.0694E-04  	\\  
		\cline{2-5}
		\hline	
	\end{tabular}
	\caption{ \label{tab:CTimes} Comparison of $L_2$ and $L_\infty$
          errors for the Michel accretion problem in 2D. Wall clock times
          (WCT) and CPU time per degree of freedom update (TDU) in
          seconds for a third-order ADER-DG scheme ($N=2$) compared with
          a third-order ADER-WENO finite-volume (FV) scheme.}
\end{table} 
\\ 
As second test case we take the large amplitude Alfv\'en wave problem in flat
Minkowski spacetime described in \cite{DelZanna2007} and also used later in 
\cite{Dumbser2008} and \cite{Zanotti2015}. 
We use the 3D computational domain $\Omega=[0,2 \pi]^3$, which is discretized 
with ADER-DG schemes of increasing order of accuracy in space and time and 
using a sequence of successively refined meshes of size 
$N_x \times N_x \times N_x$. 
To provide a direct a comparison, we solve the same test problem also with high 
order Runge-Kutta DG schemes \citep{CockburnShu98,cockburn_2001_rkd}. Since 
ADER-DG schemes are uniformly high order accurate in space and time, for the 
RKDG method we use appropriate Runge-Kutta schemes in time whose temporal order 
of accuracy exactly matches the spatial one. In particular, we use the classical 
third and fourth-order RK schemes of \cite{Kutta1901}, the fifth order 
Runge-Kutta scheme of \cite{Fehlberg} and the first one of the sixth order 
Runge-Kutta schemes proposed in \cite{Butcher1964}. Note that due to the 
well-known Butcher barriers that apply to high order RK schemes for nonlinear
ODE systems, the fifth order RK scheme has six stages, and the sixth order RK 
scheme has seven stages. 
We run the test problem with both schemes without any limiter up to a final 
time of $t=1$ and report the errors of the variable $B_y$ measured in $L_2$ 
norm. \\ 
The computational results for ADER-DG and Runge-Kutta DG schemes are
reported in Table \ref{tab.conv.comp}, together with the measured wall
clock times (WCT) in seconds and the time needed by each scheme to update
one single degree of freedom (TDU) in microseconds. Again, the inverse of
TDU in seconds represents the number of degrees of freedom that the
scheme is able to update in one second on one single CPU core and can be
directly compared with existing finite-volume and finite-difference
codes.  We observe that the CPU times and error norms are comparable for
both schemes.
For all mesh sizes $N_x$ and polynomial approximation degrees $N$ we have used  
512 CPU cores of the Phase I system of the SuperMUC of the LRZ in Garching, 
Germany. This means that for the coarsest mesh with $N_x=8$, each MPI rank has 
only one single element to update. The results of Table \ref{tab.conv.comp} 
clearly show that for a small number of elements per MPI rank our 
\textit{communication avoiding} ADER-DG schemes are computationally 
less expensive than RKDG schemes of the same order, since RKDG requires MPI 
communication in each Runge-Kutta stage. We finally run this test problem on 
a fixed grid of 64,000 elements ($N_x=40$) using fourth-order ADER-DG
and RKDG schemes on an increasing number of CPUs, from 64 to 16,000. 
The parallel implementation is based on pure MPI and thus each CPU core 
corresponds to one MPI rank.  
The speedup graph and the parallel efficiency as measured on the Phase I 
system of the SuperMUC supercomputer of the LRZ in Garching, Germany, are  
presented in Fig. \ref{fig:scaling}. 
It shows the better MPI scaling of the communication avoiding ADER-DG schemes 
compared to conventional RKDG methods. 
\begin{table*} 
	\centering
	\begin{tabular}{ccccc|ccccc}
		\hline
		 $N_x$ & $L_2$ error &  $L_2$ order & WCT [s] & TDU  [$\mu$s] & $N_x$ & $L_2$ error &  $L_2$ order & WCT [s] & TDU [$\mu$s] \\ 
		\hline
		\multicolumn{5}{c|}{ADER-DG ($N=3$)} & 	\multicolumn{5}{c}{RKDG ($N=3$)} \\ 
		\hline
        8   & 7.6396E-04	&      	& 0.093 	& 33.8	& 8 	& 8.0909E-04	&     	& 0.107 	& 39.2 \\ 
        16	& 1.7575E-05	&  5.44	& 1.371 	& 31.5	& 16	& 2.2921E-05	& 5.14	& 1.394 	& 32.0 \\ 
        24	& 6.7968E-06	&  2.34	& 6.854 	& 31.0	& 24	& 7.3453E-06	& 2.81	& 6.894 	& 31.2 \\ 
        32	& 1.0537E-06	&  6.48	& 21.642	& 31.1	& 32	& 1.3793E-06	& 5.81	& 21.116	& 30.3 \\  
		\hline
		\multicolumn{5}{c|}{ADER-DG ($N=4$)} & 	\multicolumn{5}{c}{RKDG ($N=4$)} \\ 
		\hline	
        8   & 6.6955E-05	&      	& 0.363 	& 46.8	& 8 	& 6.8104E-05	&     	& 0.456 	& 51.4 \\ 
        16	& 2.2712E-06	&  4.88	& 5.696 	& 45.9	& 16	& 2.3475E-06	& 4.86	& 6.666 	& 51.0 \\ 
        24	& 3.3023E-07	&  4.76	& 28.036	& 44.9	& 24	& 3.3731E-07	& 4.78	& 29.186	& 45.3 \\ 
        32	& 7.4728E-08	&  5.17	& 89.271	& 45.2	& 32	& 7.7084E-08	& 5.13	& 87.115	& 43.4 \\  
		\hline
		\multicolumn{5}{c|}{ADER-DG ($N=5$)} & 	\multicolumn{5}{c}{RKDG ($N=5$)} \\ 
		\hline	
        8   & 5.2967E-07	&      	& 1.090  	& 53.1	& 8 	& 5.7398E-07	&     	& 1.219  	& 55.9 \\ 
        16	& 7.4886E-09	&  6.14	& 16.710 	& 51.2	& 16	& 8.1461E-09	& 6.14	& 17.310 	& 52.5 \\ 
        24	& 7.1879E-10	&  5.78	& 84.425 	& 51.2	& 24	& 7.7634E-10	& 5.80	& 83.777 	& 49.4 \\ 
        32	& 1.2738E-10	&  6.01	& 263.021	& 50.3	& 32	& 1.3924E-10	& 5.97	& 260.859	& 49.5 \\  
		\hline	
    \end{tabular}
	\caption{ \label{tab.conv.comp} Accuracy and cost comparison between ADER-DG and RKDG schemes of different orders for the GRMHD equations in three space dimensions. 
	The test problem is the large amplitude Alfv\'en wave solved in the domain $\Omega=[0,2\pi]^3$ up to $t=1$ on a sequence of  successively refined Cartesian meshes with $N_x^3$ elements. The errors refer to the variable $B_y$. The table also contains  
	total wall clock times (WCT) measured in seconds and the time needed by the scheme to update one single degree of freedom on  one single CPU core  (TDU) measured in microseconds. All simulations have been performed in parallel on 512 MPI ranks of the  SuperMUC phase I system at the LRZ in Garching, Germany. Note that for the coarsest grid with $N_x=8$, each MPI rank has only one single element to update.} 
\end{table*}

\section{Discussion and conclusions}
\label{sec:Conclusions}

We have proposed a new high-order DG scheme for the numerical solution of
the system of the GRMHD equations in the ideal-MHD limit using multiple
spatial dimensions and on spacetime adaptive meshes. An important and
novel aspect of our discretization is that we have made use of
nonconservative products in order to account for the metric terms
directly inside the Riemann solver at the element interfaces instead of
considering them as purely algebraic source terms. While there is no
development yet of exactly well-balanced numerical schemes for GRMHD for
some relevant stationary equilibrium solutions, our approach here is
motivated by the encouraging results already obtained in this respect by
\cite{pares2006,Castro2006,ADERNC}, who have employed the framework of
well-balanced path-conservative finite-volume and DG schemes for the
successful solution of the shallow-water equations. One of the main
feature of our ADER-DG scheme is its ability to reach arbitrary high
order of accuracy in space and time for smooth parts of the solution,
while it falls back to a robust finite-volume scheme at discontinuities
such as shocks and material interfaces, without loosing the subcell
resolution capabilities of the high-order DG scheme.

We have validated the numerical implementation of the novel ADER-DG
scheme with an {a-posteriori} subcell finite-volume limiter by solving
the system of GRHD and GRMHD equations in the ideal-MHD limit for a
number of classical benchmark tests. These tests have been performed both
in 2D and in 3D with either spherical or Cartesian coordinate
mappings. Furthermore, they have involved either smooth relativistic
flows, for which we have been able to compute the convergence order and
compare it with the expected one, or non-smooth relativistic flows, 
for which we have been able to compare our results with exact solutions 
or other reference solutions available in the literature. Overall, the 
benchmarks have shown a very good performance of the new scheme,
exhibiting an excellent agreement between analytical and numerical
solutions and that the latter converge at the expected and high order for
smooth flows.

The developments presented here on the solution of the GRMHD equations is
part of a long-term plan to develop a numerical infrastructure for the
study of problems in relativistic astrophysics in general and to simulate
the merger of binary systems of neutron stars in particular [see, \eg
  \cite{Baiotti2016} for a recent review]. Indeed, another important
development in this respect has been the successful development and
testing of a first-order hyperbolic formulation of the Einstein equations
given by the first-order reduction of the CCZ4 system \citep{Alic:2011a},
which was recently presented by \citet{ADERCCZ4} (FO-CCZ4). These two 
independent but related developments naturally lead to the construction
of a computational framework where the GRMHD equations are evolved
\emph{together} with the Einstein field equations in a fully coupled
manner. This is one of the goals of the ExaHyPE framework
\citep{exahype-review, Koeppel2017, exahype-web} and is part
of our present and future research.  {The key idea here is
  to use our new ADER-DG schemes to solve the GRMHD equations and the
  FO-CCZ4 formulation of the Einstein equations \textit{together}, \ie in
  a \textit{monolithically coupled} way, simulating with the same
  numerical scheme one single evolution system for both matter and
  spacetime.}

We also plan to carry out an extension to full general relativity of the
first-order symmetric hyperbolic model of continuum mechanics recently
proposed by \cite{PeshRom2014,HPRmodel} and by \cite{HPRmodelMHD}, and
which is based on the pioneering work of \cite{GodunovRomenski72} on
nonlinear hyperelasticity in the Newtonian limit. This new unified
formulation of continuum mechanics allows one to deal with {viscous
  fluids} and {elastic solids} within one single and unified system of
symmetric-hyperbolic partial differential equations and has bounded
signal speeds for all involved physical processes, including dissipative
effects. In addition, this mathematical development will be accompanied
by a numerical one, with the implementation of a novel indicator for AMR
and subcell limiting based on the definition of the numerical entropy
density and relative fluxes as done, \eg by \citet{PS:entropy,
  SCR:CWENOquadtree} and \cite{CS:epsweno}.

\section*{Acknowledgements}

We are grateful to Bruno Giacomazzo for the numerical code used for the
exact solution of the Riemann problem in RMHD and to Luca Del Zanna for
the useful suggestions about the initial conditions of the stationary
torus.  {We also thank the anonymous referee for valuable
  suggestions and constructive comments.}  This research was funded by
the European Union's Horizon 2020 Research and Innovation Programme under
the project \textit{ExaHyPE}, grant no. 671698 (call FETHPC-1-2014) and
was also supported by the ERC synergy grant ``BlackHoleCam'' (Grant
No. 610058), by ``NewCompStar'', COST Action MP1304, by the LOEWE-Programme
in the Helmholtz International Center (HIC) for FAIR. The simulations
were performed on the SuperMUC supercomputer at the LRZ in Garching,
Germany, on the LOEWE cluster in CSC in Frankfurt, on the HazelHen
supercomputer at the HLRS in Stuttgart, Germany, as well as on the local
HPC cluster at the University of Trento.

\bibliographystyle{mnras}
\bibliography{references7}

\appendix

\section{Torus initial condition}
\label{sec:TorusIC}
The acceleration experienced by a fluid element rotating around around a
compact object which acts as a source of gravity can be cast into the
following differential equation
\begin{align}
  \label{eq:diff}
  d \log |u_t| - \left(\frac{\Omega}{1-\Omega \ell} \right) d \ell +
  \frac{d p}{\rho h} = 0\,, 
\end{align}
where
\begin{align}
  \label{eq:Omega}
  \ell(r,\theta)&:=-\frac{u_\phi}{u_t} \,, & \Omega(r,\theta)&:=
  \frac{u^\phi}{u^t} \,,
\end{align}
are the so-called {specific angular momentum} and the {coordinate angular
  velocity}, respectively. For {barotropic} fluids the last differential
on the right in Eq. (\ref{eq:diff}) is exact, \ie one can define the
so-called {effective potential} $\mathcal{W}$ as
\begin{align}
  \mathcal{W}-\mathcal{W}_{\text{in}} := - \int_0^p \frac{d \tilde{p}}{\rho
  h}=
\log |u_t| -
\log|(u_t)|_{\text{in}} - \int \limits_{\ell_{\text{in}}}^{\ell}
\frac{\Omega d\tilde{\ell}}{1-\Omega \tilde{\ell}}\,.
\end{align}


In the test-case considered here, the specific angular momentum is assumed
to be constant $\ell=\ell_0=\text{const.}$, so that it is possible to obtain
an explicit and simplified expression for the potential
\begin{align}
  \mathcal{W}(r,\theta) = \log |u_t|\,,
  \label{eq:effPot}
\end{align}
where, for a Schwarzschild black hole, one has
\begin{align}
  u_t = -r \sin \theta \left( \frac{r-2}{r^3 \sin^2 \theta - \ell^2
  (r-2)}\right)^{\frac{1}{2}}\,.
\end{align}

In the axisymmetric equilibrium torus, there are some special radial
positions in the equatorial plane ($\theta = \pi/2$) that are worthwhile
recalling: the {inner} and {outer} edge of the torus $r_{\text{in}}$ and
$r_{\text{out}}$; the radial position of the {cusp}, $r_{\text{cusp}}$;
the radial position of the maximum pressure peak, $r_{\text{c}}$, which
is the {center of the torus}; the radial position of the so-called
``{marginally stable}'' and ``{marginally bound}'' orbit, $r_{\text{ms}}$
and $r_{\text{mb}}$. The cusp position $r_{\text{cusp}}$ and the centre
$r_{\text{c}}$ can be identified as the local extrema of the effective
potential, but also by the condition $\ell_K=\ell_0$, where $\ell_K$ is
the {Keplerian specific angular momentum} which is given by $\ell^2_K(r)
:= M r^3 / (r - 2M)^2$. Similarly, also $r_{\text{ms}}$ and
$r_{\text{mb}}$ are identified by the condition $\ell_K =
\ell_{\text{ms}}$ and $\ell_K=\ell_{\text{mb}}$. For a Schwarzschild
(nonrotating) black-hole: $\ell_{\text{ms}} = (3\sqrt{6}/2) M$ and
$\ell_{\text{mb}} = 4 M$, so that the corresponding to the radial
positions are $r_{\text{ms}} = 6 M$ and $r_{\text{mb}} = 4 M$. Finally,
the inner and outer radial position, $r_{\text{in}}$ and
$r_{\text{out}}$, can be estimated by the condition $\Delta \mathcal{W}
:= \mathcal{W}_{\text{in}} - \mathcal{W}_{\text{cusp}} = 0$. Indeed,
whenever $\Delta \mathcal{W}>0$ the orbit of the corresponding fluid
particle is open, whenever $(\mathcal{W}_{\text{c}} -
\mathcal{W}_{\text{in}}) < \Delta \mathcal{W}<0$ the orbits are
closed. The spatial volume delimited by the widest closed equipotential
surface of the torus, \ie $\mathcal{W}=\mathcal{W}_{\text{cusp}}$ is
named as the ``{Roche lobe}'' of the torus. Using these definitions,
several constraints need to be satisfied: first, the cusp
$r_{\text{cusp}}$ must necessarily be located within $r_{\text{mb}}$ and
$r_{\text{ms}}$, and the inner edge $r_{\text{in}}$ can be located
anywhere within $r_{\text{cusp}}$ and $r_{\text{c}}$. For isentropic
fluids obeying the {polytropic} equation of state
\begin{align}
  p = K \rho^{\gammap} \label{eq:polieos}
\end{align}
$K$ being the polytropic constant, $\gammap$ the polytropic exponent, 
an analytical expression for the rest-mass density exists and takes the
form
\begin{align}
\rho(r,\theta) = \left[\frac{\gammap - 1}{K \gammap} \left( \exp(
  \mathcal{W}_{\text{in}} - \mathcal{W}(r,\theta) ) -
  1\right)\right]^{1/\left(\gammap - 1\right)} \label{eq:rhotorus}
\end{align}


After choosing the value of the polytropic constant $K$, polytropic
exponent $\Gamma$, the specific angular momentum $\ell_0$, and the
potential gap $\Delta \mathcal{W}$, then the Keplerian points are
estimated after ensuring the following scalar equalities: for the radial
cusp position $r_{\text{cusp}}$, 
\begin{align}
  \ell_K(r) = \ell_0\,, \quad \text{with} \quad r_{\text{hor}}< r <
  r_{\text{ms}}\,,
\end{align}
for the center $r_{\text{c}}$,  $r_{\text{hor}}$ being the radial position of the horizon,
\begin{align}
  \ell_K(r) = \ell_0\,, \quad \text{with} \quad r_{\text{ms}} < r\,.
\end{align}
Then, the corresponding potentials $\mathcal{W}_{\text{cusp}}$. and
$\mathcal{W}_{\text{c}}$ are evaluated according to Eq.
(\ref{eq:effPot}). On the other hand, the effective potential at the
inner edge $\mathcal{W}_{\text{in}}$ is computed according to the
prescribed potential gap $\Delta \mathcal{W}$ after estimating
\begin{align}
  (u_{t})_{\text{ in}} = (u_{t})_{\text{ cusp}} \;e^{\Delta \mathcal{W}}\,.
\end{align}
Then, since the fluid distribution is inside the Roche lobe, the inner
and outer edge positions $r_{\text{in}}$ and $r_{\text{out}}$ are
computed through the conditions
\begin{align}
  u_t(r) = (u_{t})_{\text{ in}} \quad \text{with} \quad r_{\text{cusp}} <
  r < r_{\text{c}}\,,
\end{align}
and
\begin{align}
  u_t(r) = (u_{t})_{\text{ in}} \quad \text{with} \quad r_{\text{c}} <
  r\,,
\end{align}
respectively. The rest-mass density at the center $\rho_{\text{c}}$ is
provided directly by the analytical solution (\ref{eq:rhotorus}), the
corresponding pressure $p_{\text{c}}$ through the polytropic equation of
state (\ref{eq:polieos}). Finally, for every spatial position
$(r,\theta)$ within the torus, \ie which fulfils the condition
\begin{align}
  r>r_{\text{in}} \quad \text{and} \quad
  \mathcal{W}<\mathcal{W}_{\text{in}}\,,
\end{align}
the angular velocity $\Omega(r,\theta)$ is computed through the
definition (\ref{eq:Omega}), the rest-mass density $\rho$ directly from
(\ref{eq:rhotorus}), the pressure $p$ from the polytropic equation of
state (\ref{eq:polieos}), and the velocity is given by
\begin{align}
&(v^r ,v^\theta ,v^{\phi} ) = \left( \frac{\shift^r}{\lapse} , 0 ,
  \frac{1}{\lapse}(\Omega + \shift^\phi) \right)\,.
\end{align}
%

\bsp	
\label{lastpage}
\end{document}